\documentclass[pra,aps,twocolumn, nobalancelastpage, superscriptaddress]{revtex4-1}
\usepackage[utf8]{inputenc}
\usepackage{bbold}
\usepackage{geometry}
\usepackage{float}
\usepackage{graphicx}
\graphicspath{ {images/} }
\usepackage{braket}
\usepackage{subfigure}% in preamble
\usepackage[normalem]{ulem}
\usepackage[T1]{fontenc}
\usepackage{gensymb}
\usepackage{amsmath,amsfonts,amssymb}
\usepackage{amsmath}
\usepackage{amsthm}
\usepackage[]{natbib}
\usepackage{comment}
\newcommand{\be}{\begin{equation}}
\newcommand{\ee}{\end{equation}}
\newcommand{\ba}{\begin{aligned}}
\newcommand{\ea}{\end{aligned}}

\newcommand{\bc}{\begin{center}}
\newcommand{\ec}{\end{center}}
\newcommand{\beq}{\begin{equation}}
\newcommand{\eeq}{\end{equation}}
\newcommand{\beqq}{\begin{equation*}}
\newcommand{\eeqq}{\end{equation*}}
\newcommand{\beqa}{\begin{align}}
\newcommand{\eeqa}{\end{align}}
\newcommand{\barr}{\begin{array}}
\newcommand{\earr}{\end{array}}
\newcommand{\bi}{\begin{itemize}}
\newcommand{\ei}{\end{itemize}}

\newcommand{\norm}[1]{\ensuremath{\vert\vert#1\vert\vert}}

\newcommand{\s}{\ensuremath{\sigma}}
\newcommand{\w}{\ensuremath{\omega}}

\newcommand{\Tr}{\ensuremath{\,\mathrm{Tr}}}

\usepackage[T1]{fontenc} %fontenc makes the pdf ugly
\usepackage{lmodern} % Nicer font fith T1 fontenc
\usepackage{microtype} % Even nicer typography
\usepackage{graphicx}
\usepackage{dcolumn}
\usepackage{enumerate,amsthm,amssymb,color}
\usepackage{amsfonts}
\usepackage{amsmath}
\usepackage{mathtools}
\usepackage{dsfont}
\usepackage{siunitx}
\usepackage{tabularx}

\usepackage{ragged2e}
\usepackage{bbm}
\usepackage{bm}
\usepackage{synttree}
\usepackage{float}
\usepackage{bbold}
\usepackage{braket}
\usepackage{xcolor}
\usepackage{gensymb}

\usepackage{booktabs}
\usepackage{upgreek, eurosym}
\usepackage[colorlinks=true]{hyperref}

\newcommand{\ketbra}[2]{|#1\rangle\!\langle#2|}

\newcommand{\mbf}[1]{\mathbf{#1}}
\renewcommand{\t}[1]{\textrm{#1}}
\newcommand{\q}{\mbf{q}}

\newcommand{\g}{\gamma}

\renewcommand{\r}{\rho}

\newcommand{\D}{\Delta}

\renewcommand{\L}{\mathcal{L}}

\renewcommand{\P}{\mathcal{P}}

\newcommand{\+}{^\dagger}
\renewcommand{\>}{\rangle}

\newcommand{\XX}{\vert X\>\<X\vert}

\newcommand{\BB}{\vert B\>\<B\vert}

\newcommand{\GX}{\vert G\>\<X\vert}

\newcommand{\XB}{\vert X\>\<B\vert}

\graphicspath{{images/}}

\begin{document}

%-------------------------------------------------------

\bibliographystyle{apsrev}

%-------------------------------------------------------

%\title{Tuning quantum dot sources for quantum cryptography}
\title{Enhancing quantum cryptography with quantum dot single-photon sources}

\author{Mathieu Bozzio}\thanks{Both authors contributed equally.\\ mathieu.bozzio@univie.ac.at /  michal.vyvlecka@univie.ac.at}
\affiliation{University of Vienna, Faculty of Physics, Vienna Center for Quantum Science and Technology (VCQ), 1090 Vienna, Austria}

\author{Michal Vyvlecka}\thanks{Both authors contributed equally.\\ mathieu.bozzio@univie.ac.at /  michal.vyvlecka@univie.ac.at}
\affiliation{University of Vienna, Faculty of Physics, Vienna Center for Quantum Science and Technology (VCQ), 1090 Vienna, Austria}

\author{Michael Cosacchi}
\affiliation{Theoretische Physik III, Universität Bayreuth, 95440 Bayreuth, Germany}

\author{Cornelius Nawrath}
\affiliation{Institut für Halbleiteroptik und Funktionelle Grenzflächen (IHFG), Center for Integrated Quantum Science and Technology (IQST) and SCoPE, University of Stuttgart, 70569 Stuttgart, Germany}

\author{Tim Seidelmann}
\affiliation{Theoretische Physik III, Universität Bayreuth, 95440 Bayreuth, Germany}

\author{Juan C. Loredo}
\affiliation{University of Vienna, Faculty of Physics, Vienna Center for Quantum Science and Technology (VCQ), 1090 Vienna, Austria}\affiliation{Christian Doppler Laboratory for Photonic Quantum Computer, Faculty of Physics, University of Vienna, 1090 Vienna, Austria}

\author{Simone L. Portalupi}
\affiliation{Institut für Halbleiteroptik und Funktionelle Grenzflächen (IHFG), Center for Integrated Quantum Science and Technology (IQST) and SCoPE, University of Stuttgart, 70569 Stuttgart, Germany}

\author{Vollrath M. Axt}
\affiliation{Theoretische Physik III, Universität Bayreuth, 95440 Bayreuth, Germany}

\author{Peter Michler}
\affiliation{Institut für Halbleiteroptik und Funktionelle Grenzflächen (IHFG), Center for Integrated Quantum Science and Technology (IQST) and SCoPE, University of Stuttgart, 70569 Stuttgart, Germany}

\author{Philip Walther}
\affiliation{University of Vienna, Faculty of Physics, Vienna Center for Quantum Science and Technology (VCQ), 1090 Vienna, Austria}\affiliation{Christian Doppler Laboratory for Photonic Quantum Computer, Faculty of Physics, University of Vienna, 1090 Vienna, Austria}

%

%-------------------------------------------------------

%

%-------------------------------------------------------

\begin{abstract}

%Quantum dot single-photon sources are remarkable candidates for quantum cryptography: they can generate single photons on-demand, with high brightness and low multiphoton emission. In this work, we show how recently demonstrated properties of such sources, such as the presence of photon number coherence, can be tuned to significantly improve the security of four main quantum cryptography protocols: quantum key distribution (decoy and twin-field), quantum coin flipping, quantum bit commitment and unforgeable quantum tokens. We identify the optimal pumping scheme for each protocol, benchmark its performance with respect to Poisson-distributed sources such as attenuated laser states and down-conversion sources, and establish a friendly cookbook to encourage further interactions between the solid state and quantum cryptography communities. We hope that the presented tools, simulations, and arguments, will guide future developments in the emerging quantum dot-based communication efforts. 

Quantum cryptography harnesses quantum light, in particular single photons, to provide security guarantees that cannot be reached by classical means. For each cryptographic task, the security feature of interest is directly related to the photons' non-classical properties. Quantum dot-based single-photon sources are remarkable candidates, as they can in principle emit deterministically, with high brightness and low multiphoton contribution. Here, we show that these sources provide additional security benefits, thanks to the tunability of coherence in the emitted photon-number states. We identify the optimal optical pumping scheme for the main quantum-cryptographic primitives, and benchmark their performance with respect to Poisson-distributed sources such as attenuated laser states and down-conversion sources. In particular, we elaborate on the advantage of using phonon-assisted and two-photon excitation rather than resonant excitation for quantum key distribution and other primitives. The presented results will guide future developments in solid-state and quantum information science for photon sources that are tailored to quantum communication tasks.

\end{abstract}

%-------------------------------------------------------

\maketitle

\noindent With the rise of quantum algorithms capable of breaking modern encryption schemes, there follows a global response to search for stronger security levels \cite{GS:PRL21,MLL:NatPhot12,S:SIAM97}. While the security of most current schemes relies on the complexity of solving difficult mathematical problems, quantum-mechanical laws can provide security against adversaries endowed with unlimited computational power for some tasks \cite{BB84,Pan:RevMod20}. This type of security, known as information-theoretic security, motivates research towards a quantum internet \cite{WEH:Sci18}.

Modern communication networks rely on a handful of fundamental building blocks, known as cryptographic primitives \cite{BS:dcc16,BCD:PRX21}. These can be combined with one another to provide security in various applications such as message encryption, electronic voting, digital signatures, online banking, anonymous messaging, and software licensing, to name a few. In order to reach information-theoretic security through quantum primitives, information is typically encoded onto quantum properties of light, such as photonic path, time-bin, polarization, and photon number \cite{Pan:RevMod20}. In the quantum realm, the uncertainty principle then ensures that any eavesdropper attempting to access quantum-encoded information, while unaware of the preparation basis, will alter the quantum states in a way that is detectable by the honest parties \cite{Wie:acm83,BB84}.

For such quantum primitives, it is expected that quantum dot-based single-photon sources (QDS) can excel by generating photons on-demand, with high brightness and low multiphoton contribution \cite{Michler17,SSW:NNT17}. In fact, source brightness is crucial in achieving high-speed quantum communication \cite{BBR:PRL18,AMS:PRApp20}, while low multiphoton contribution minimizes information leakage to a malicious eavesdropper \cite{BLM:PRL00}. In contrast to these on-demand single-photon sources, widely used Poisson-distributed sources (PDS), such as attenuated laser states \cite{Pan:RevMod20} and spontaneous parametric down-conversion \cite{SKB:JO19}, suffer from a stringent trade-off between high brightness and low multiphoton emission. Despite elaborate countermeasures proposed to overcome this trade-off \cite{LMC:PRL05,LP:QIC07}, the distance and rate of secure quantum communication can be increased using on-demand single-photon sources.

Some pioneering works have already implemented instances of quantum key distribution (QKD) employing QDS \cite{VRG:AQT22,BVM:SciAdv21,SRH:SciAdv21,TNM:SciRep15,HKR:NJP12,CCF:JAP10,IWK:JOA09,Y:Nat02} or other single-photon sources \cite{WCX:PRL08,MCH:Arx22}, comparing their performance to PDS in terms of secret key rate. In these works, brightness and purity are the sole figures of merit used to establish a comparison, while the additional tuning capabilities of QDS and their role in quantum cryptography have not yet been investigated.

In this work, we explore features of QDS to enhance the performance and security of quantum-cryptographic primitives, with an emphasis on telecom wavelengths. We first optimize and compare the brightness and single-photon purity of three main optical pumping schemes, using realistic intra-cavity simulations of quantum dot dynamics. We then show how photon-number coherence generated from QDS, experimentally demonstrated in \cite{L:NatPhot19}, can be erased or preserved to boost the performance of practical QKD, and match its fundamental security requirements \cite{LP:QIC07,CZL:NJP15}. We further explain how the field of mistrustful quantum cryptography \cite{BS:dcc16,BCD:PRX21}, not yet implemented with QDS, can significantly benefit from this feature. Our findings are designed to bridge the gap between the quantum dot and quantum cryptography communities: we optimize and benchmark QDS optical pumping schemes for four main quantum cryptographic primitives, exploiting the combined advantage of brightness, single-photon purity, and photon-number coherence. The studied primitives include quantum key distribution (standard BB84, decoy and twin-field) \cite{Pan:RevMod20,LMC:PRL05,WYH:NatPhot21,LYD:Nature18}, unforgeable quantum tokens \cite{BOV:npj18,GAA:pra18,Kent:npjQI22}, quantum strong coin flipping \cite{PJ+:natcomm14,BBB:NC11,BCK:PRA20}, and quantum bit commitment \cite{NJC:NatComms12,Zbind:PRL13,Pan:PRL14} under storage assumptions.

%-------------------------------------------------------

\medskip

\large

\noindent\textbf{Results}

\normalsize

\begin{figure*}
	\begin{center}
		\includegraphics[width=180mm]{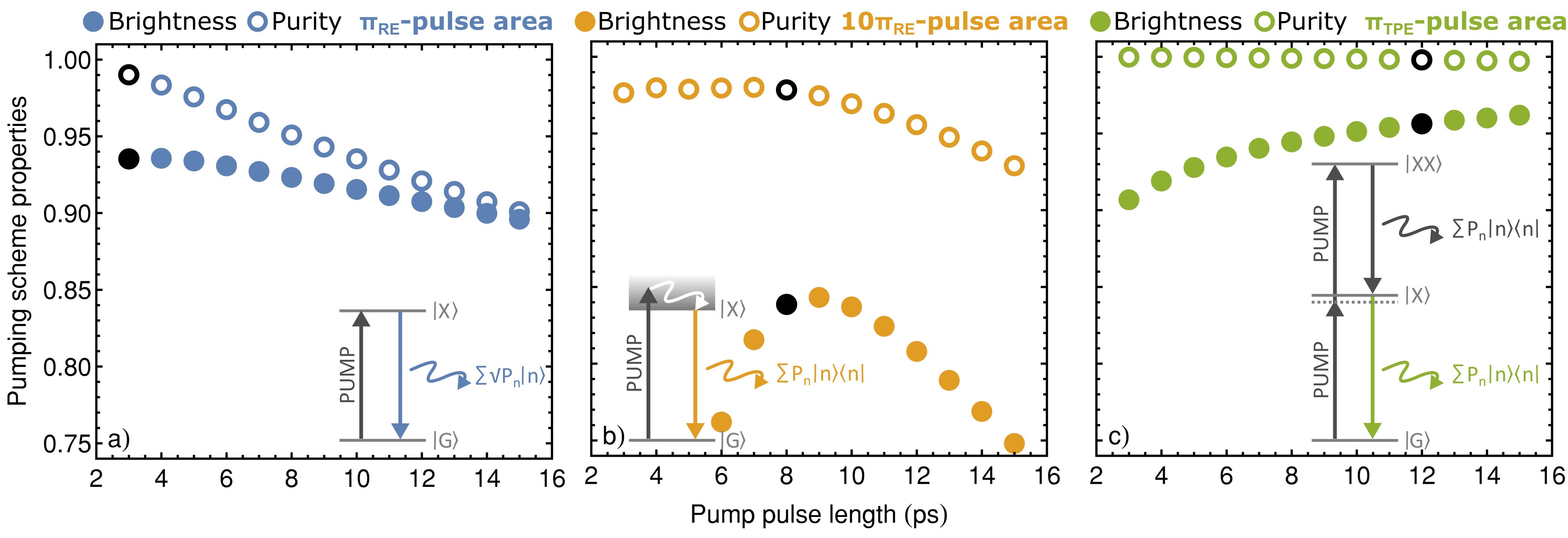}
		\caption{\textbf{Simulation of emission properties under different pumping schemes.} Simulated brightness and single-photon purity for individual pumping schemes calculated from our numerical model (see Methods) for (a) resonant excitation (RE) between ground $\ket{G}$ and exciton $\ket{X}$ states , (b) longitudinal phonon-assisted excitation (LA) (c) two-photon excitation (TPE) between ground $\ket{G}$ and biexciton $\ket{XX}$ states. Insets show a schematic representation of each pumping scheme (the upper grey area in the LA sketch represents the vibrational quasi-continuum of the exciton state). Optimal FWHM pulse lengths are marked with black disks and circles, chosen pulse areas for maximum population inversion are $\pi_\text{RE}$ for RE and $\pi_\text{TPE}$ for TPE, while we choose $10\pi_\text{RE}$ for LA. Simulation parameters are: dot-cavity coupling $\hbar g = 50$ µeV, radiative decay rate $\hbar\gamma = 0.66$ µeV, cavity loss rate $\hbar\kappa = 379$ µeV (yielding a Purcell factor of $P = 10$), initial system temperature $T = 4.2$K, electron confinement length $3$ nm, and material properties typical for GaAs.  Throughout the paper, we assume that photonic states are maximally pure in the photon-number basis for RE, i.e. expressed as $\sum_{n=0}^\infty \sqrt{p_n}\ket{n}$, while they are expressed as diagonal states for LA and TPE, i.e. as $\sum_{n=0}^\infty p_n\ket{n}\bra{n}$.}
		\label{fig:pumping}
	\end{center}
\end{figure*}

\noindent\textbf{Comparison of pumping schemes.} Solid-state single-photon sources can be excited under different optical pumping schemes, and we aim to provide a fair comparison of their performance for quantum cryptography. Using realistic intra-cavity simulations for GaAs-based QDS, we calculate the emitted photon-number occupations up to three photons for resonant excitation (RE), longitudinal phonon-assisted (LA) excitation and two-photon excitation (TPE). In Fig.~\ref{fig:pumping}, we then compare each scheme's brightness and single-photon purity, and estimate the full-width-at-half-maximum excitation pulse length which maximizes both properties (marked with black symbols).

RE schemes are based on resonant excitation of a two-level system~\cite{UPW:SciAdv20,SGS:NP16,Pan:NNano2013}. The spectral degeneracy of the excitation laser and the emitted photons usually imposes separation based on polarization filtering, which may cause significant collection losses of around $50\%$~\cite{SGS:NP16,Lodahl:PRL2014}. Other methods, exploiting dichromatic pumping or trion recombination in asymmetric cavities however, can overcome such limitations \cite{Pan:NatPhys19,TJA:NatNan21}. Since RE exhibits Rabi oscillations of the excitonic state populations, its brightness and single-photon purity are susceptible to pump power fluctuations---thus presenting challenges for quantum network applications~\cite{Armando:ACS17}. As we show in Fig. \ref{fig:pumping}.a., for a fixed RE $\pi$-pulse area, single-photon purity decreases with pulse length due to re-excitation processes, while brightness decreases as the emission statistics tend to a Poisson distribution \cite{Fischer:IOP17,Fisher:NPJ18}.

The main limitations of RE may be overcome by using LA excitation schemes. Here, the pump energy is slightly higher than the relevant excitonic transition, and the fast emission of a longitudinal-acoustic (LA) phonon precedes the population of the excited state. Due to this additional incoherent step, Rabi oscillations vanish, and the purity of the emitted single photons becomes less sensitive to small pump power fluctuations \cite{LR:IOP19}. Recently, it was shown that LA excitation can reach even smaller multiphoton components than its RE counterpart \cite{Axt:PRL19}, while still enabling spectral filtering of the pump \cite{LAParis:21, Ding2016}. As regards to brightness, Fig. \ref{fig:pumping}.b. displays an increase with pulse length, as was experimentally demonstrated in \cite{LAParis:21}. With longer pulses however, the peak intensity decreases for a fixed pulse area, thus lowering the efficiency of the phonon excitation process.

RE and LA multiphoton contributions can be greatly reduced by addressing the exciton-biexciton cascade through TPE schemes, usually employed to generate spectrally-separated entangled photon pairs. Here, the re-excitation probability scales quadratically with the pulse length, as opposed to linearly in the resonantly-driven RE scheme \cite{Fisher:NPJ18}, which can reduce multiphoton emission by several orders of magnitude \cite{SJZ:APL18}. Moreover, TPE offers the possibility to overcome the collection efficiency limitations of RE: the spectral separation of the generated photons allows for frequency filtering of the pump laser \cite{Mueller2014}, avoiding the polarisation filtering losses. We show in Fig. \ref{fig:pumping}.c. that the brightness is low for short pulse lengths, due to a remaining overlap with the exciton transition, causing the biexciton level to be only partially populated. Although the emitted exciton polarization is random, which would limit the brightness to half the values of Fig. \ref{fig:pumping}.c., recent works have shown the possibility of deterministically preparing the exciton polarization with near-unit brightness by adding a stimulated biexciton excitation after the original two-photon excitation \cite{SHS:PRL22,WLL:NatNano22,YLL:NanoLett22}. 

\begin{figure*}
	\begin{center}
		\includegraphics[width=180mm]{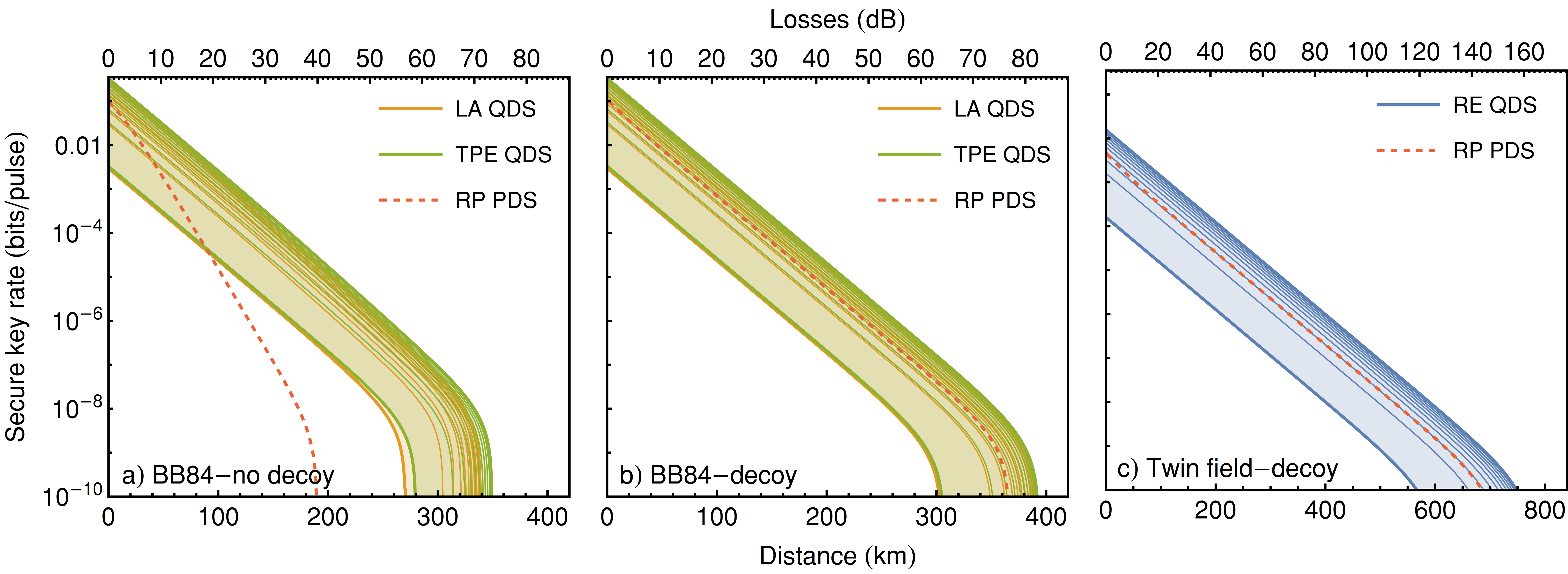}
		\caption{\textbf{Source comparison for three main QKD schemes.} Simulated secret key rates using one-way classical post-processing \cite{GLLP04,KGR:PRL05} for (a) BB84 without decoy states, (b) BB84 with infinite decoy states, (c) twin-field with infinite decoy states. The continuum in each plot shows the attainable key rates for QDS ranging from $1$ to $100\%$ collection efficiencies, with intermediate curves showing steps of $10\%$. The dashed line indicates the optimal performance of randomized-phase (RP) PDS. For QDS, pulse lengths were chosen to simultaneously maximize the brightness and single-photon purities from Fig. \ref{fig:pumping}: $3$ ps for RE, $8$ ps for LA, $12$ ps for TPE. Chosen pulse areas:  $\pi_\text{RE}$ for RE, $10\pi_\text{RE}$ for LA, $\pi_\text{TPE}$ for TPE. photon-number populations $\{p_n\}$ up to $n=3$ were inferred from the subsequent values of brightness and single-photon purity. Parameters for all plots are: $e = 2\%$ single photon error rate, $0.21$ dB/km single mode telecom fiber losses, $Y_0 = 10^{-9}$ dark count probability, unit detection efficiency, and error-correcting code inefficiency $f=1.2$.}
		\label{fig:QKD}
	\end{center}
\end{figure*}

\medskip

\noindent\textbf{Photon-number coherence.} We now discuss the presence of coherence in the photon-number basis, a usually disregarded feature of interest, under each excitation scheme. For PDS such as attenuated laser states, this quantity refers to a fixed phase relationship between the various Poisson-distributed number states. For QDS, this will materialize as a coherent superposition of vacuum, single and two-photon states. 

In RE, it was experimentally demonstrated in \cite{L:NatPhot19} that the coherently-driven Rabi oscillations translate into emitted photon-number coherence: values of coherence purity as high as $96\%$ for $\pi$-pulse areas were measured. On the other hand, this coherence can gradually vanish as the pump is detuned from resonance in LA schemes, along with the vanishing of Rabi oscillations \cite{Axt:PRL19}. Our quantum dot dynamics simulations support these findings: for the optimal pulse lengths of Fig. \ref{fig:pumping}, the normalized off-diagonal density matrix elements of LA between the vacuum and single-photon components are around $10$ times smaller than the RE ones. Accordingly, we will assume in this work that states emitted under RE are pure in photon-number basis, while those emitted under LA present vanishing off-diagonal elements. 

In TPE, our simulation results display off-diagonal elements around $20$ times smaller than the RE ones. Although the biexciton level follows Rabi oscillations under resonant TPE \cite{SJZ:APL18,Mueller2014}, loss of coherence arises from a radiative decay between the biexciton to exciton state, which creates a timing jitter similar to the phonon-induced jitter in LA schemes. We will therefore also assume that states emitted under TPE present vanishing off-diagonal elements. Note that this assumption is expected to hold for the stimulated schemes discussed in \cite{SHS:PRL22,WLL:NatNano22,YLL:NanoLett22}, since the remaining jitter due to the biexciton transition is significantly larger than the jitter responsible for coherence erasure.

\medskip

\noindent\textbf{Practical sources and security.} We now discuss the role of brightness, single-photon purity and photon-number coherence in quantum primitives involving two parties, exchanging a sequence of classical and quantum (photonic) messages that do not rely on quantum entanglement. Each of these primitives achieves a different functionality within quantum networks, and thus also requires its own security figure of merit. 

%Since ideal single-photon sources are still under development, the majority of quantum-cryptographic implementations so far have made use of either PDS (mainly SPDC sources and attenuated laser states) \cite{Pan:RevMod20,BOV:npj18,GAA:pra18,PJ+:natcomm14,NJC:NatComms12,Zbind:PRL13}, or imperfect quantum dots \cite{TNM:SciRep15,CCF:JAP10,IWK:JOA09,Y:Nat02} as studied here.

\begin{figure*}
	\begin{center}
		\includegraphics[width=180mm]{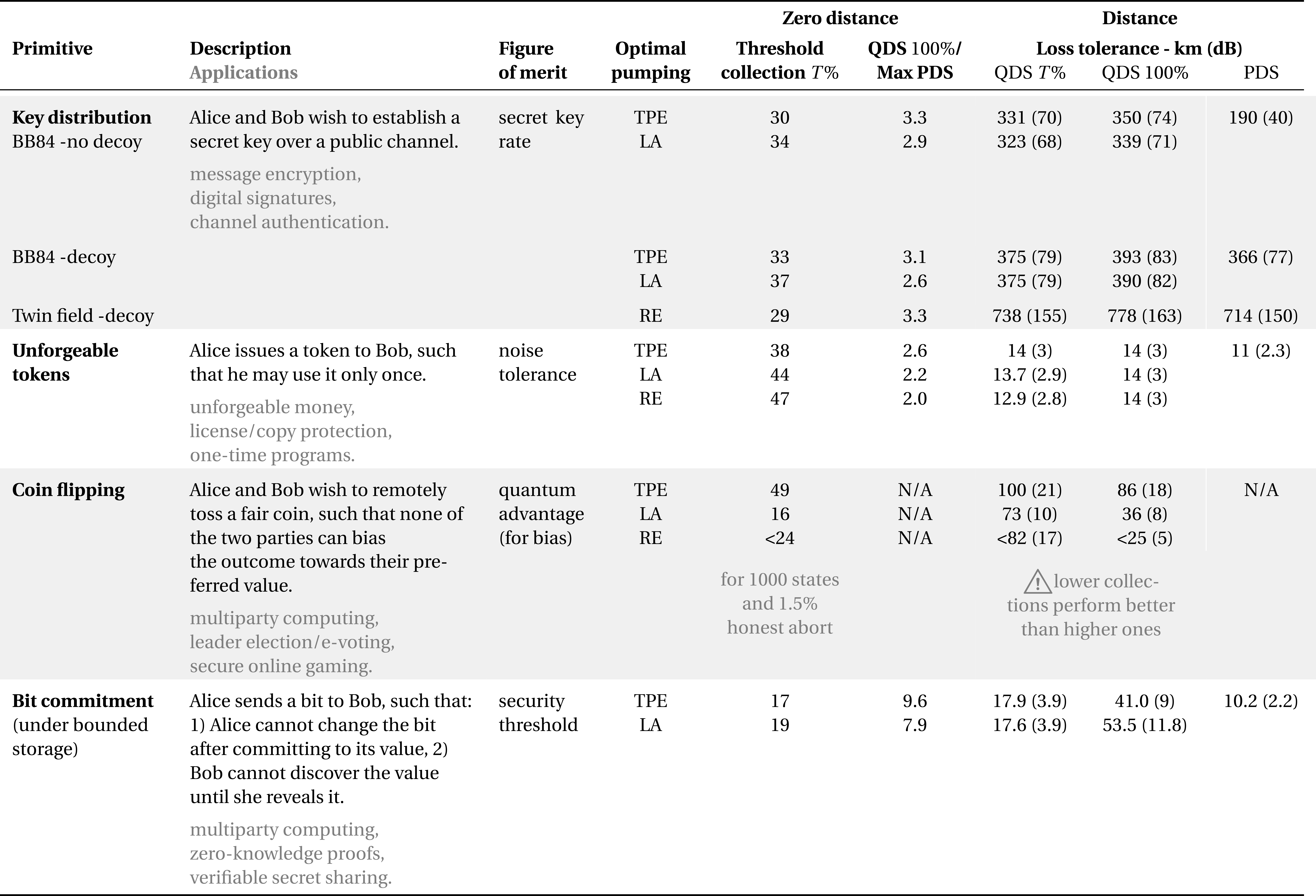}
		\caption{\textbf{QDS performance for the main quantum primitives.} A description of each primitive is provided, along with its main network applications, and our chosen security figure of merit. In each case, we summarize the various QDS pumping scheme performances and the threshold collection efficiencies $T$ required to outperform PDS in a lossless setting. We then display the performance ratio of QDS with $100\%$ collection over the best PDS at zero distance (note that this ratio has to be scaled for QDS achieving higher repetition rates than PDS). We then calculate the loss tolerance, both in terms of distance (in km) assuming single-mode telecom fiber losses of $0.21$ dB/km, and in terms of absolute losses (in dB) for QDS reaching $T\%$ collection, QDS reaching $100\%$ collection, and randomized-phase PDS.}
		\label{fig:protocols}
	\end{center}
\end{figure*}
The main efficiency limitations of PDS may be understood upon inspection of the generated state $\sum_{n=0}^{\infty} C_\mu\left(n\right) \ket{n}$, where the $P_\mu\left(n\right) = \left|C_\mu\left(n\right)\right|^2$ coefficients follow a Poisson distribution with average photon number $\mu$, and $\{\ket{n}\}$ span the photon-number basis. Increasing the source brightness (i.e., increasing $\mu$) comes at the cost of increasing the multiphoton components $n\geqslant 2$, which renders the respective quantum primitive vulnerable to attacks involving photon number splitting on lossy channels \cite{BLM:PRL00}. Thus, $\mu$ is typically kept very low in quantum-cryptographic implementations, in the range $\mu{\sim}0.005{-}0.5$ \cite{Pan:RevMod20,Zbind:PRL13,PJ+:natcomm14,BOV:npj18}, which limits the communication rate. On the other hand, single-photon purity in QDS can be increased without an intrinsic penalty on the multiphoton component. Achieving higher QDS brightness is then ultimately a technological challenge, limited by the collection efficiency of the source \cite{SSW:NNT17,TJA:NatNan21}, and not a fundamental limitation as in the case of PDS.

\begin{figure*}
	\begin{center}
		\includegraphics[width=180mm]{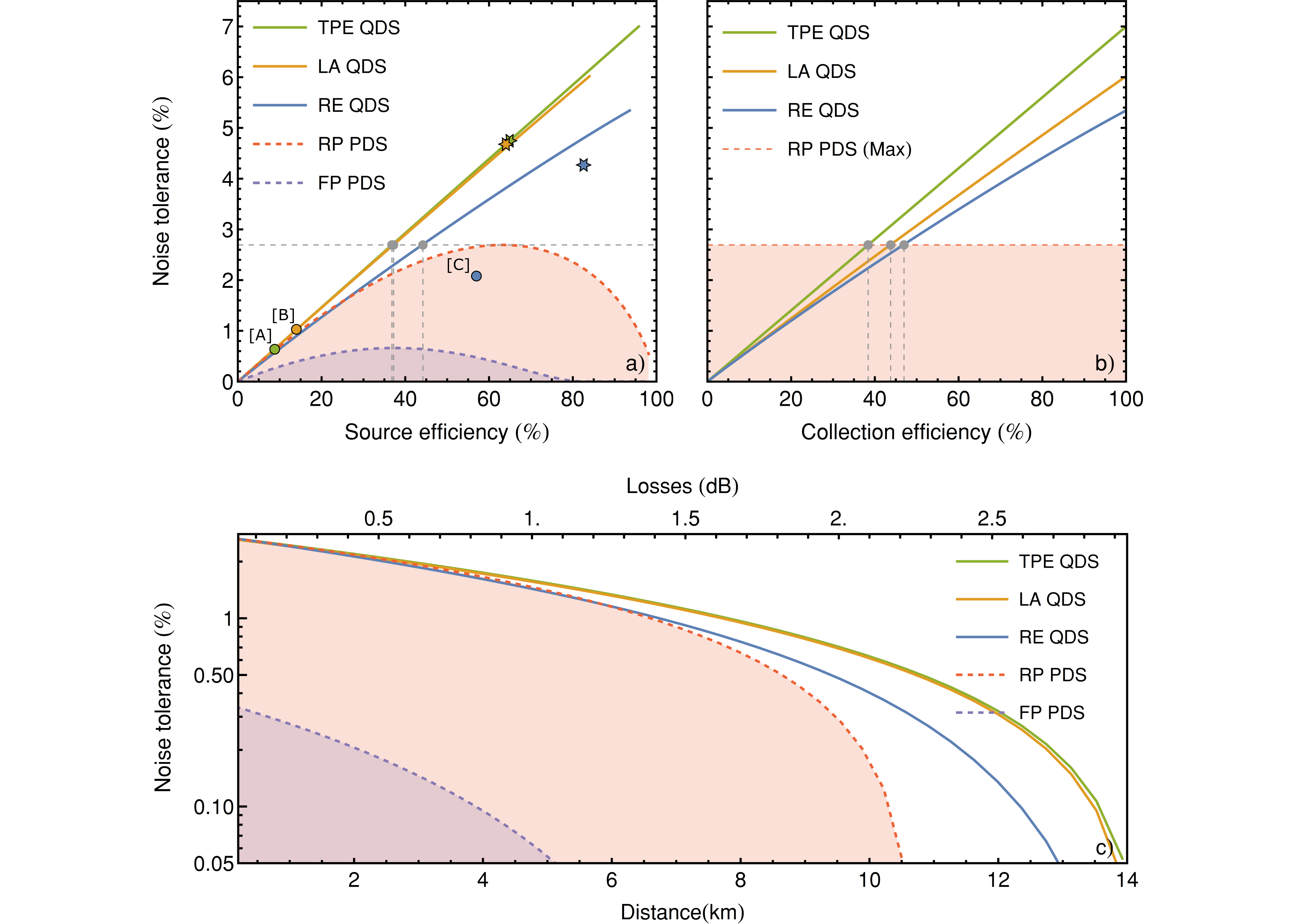}
		\caption{\textbf{Source comparison for unforgeable quantum tokens (quantum verification protocol from \cite{BDG:PRA19})}. (a) Numerical noise tolerance as a function of source efficiency for RE, LA and TPE QDS, along with fixed-phase (FP) and randomized-phase (RP) PDS. Source efficiency is defined as $1-e^{-\mu}$ for PDS and $1-\sum_{n=0}^\infty p_n(1-\eta)^n$ for QDS, where $\eta$ is the QDS collection efficiency. Photon-number populations $\{p_n\}$ emitted under different QDS excitation schemes were obtained from the brightness and single-photon purity results of Fig. \ref{fig:pumping}, assuming the optimal pulse lengths marked in black. RE photonic states are assumed to be maximally pure in number basis, expressed as $\sum_{n=0}^\infty \sqrt{p_n}\ket{n}$, while LA states were expressed as diagonal states $\sum_{n=0}^\infty p_n\ket{n}\bra{n}$. Colored circles indicate the performance of three state-of-the-art quantum dots, inferred from the experimental brightness and purity values reported in [A]=\cite{Wang:2019XX}, [B]=\cite{Ding:2016}, [C]=\cite{TJA:NatNan21}, corrected to unit detector efficiency. Stars indicate the potential of these quantum dots assuming the reported collection losses, but no setup losses. (b) Numerical noise tolerance as a function of QDS collection efficiency, compared to the best performance of PDS  sources (dashed line). (c) Numerical noise tolerance plotted as a function of distance, assuming single mode telecom fiber losses of $0.21$ dB/km. The QDS collection efficiencies were chosen as the intersection points from (b), also summarized in Fig. \ref{fig:protocols}.}
		\label{fig:sdpresults}
	\end{center}
\end{figure*}

In contrast to their PDS counterparts, QDS have not yet been optimized to suit the security requirements of quantum primitives. Most importantly, a main assumption behind the implementation of decoy QKD and other primitives is that the global phase of PDS must be actively scrambled, to effectively destroy the coherence in the number basis \cite{LP:QIC07,CZL:NJP15}:

\begin{equation}
    \sum_{n=0}^{\infty}  C_\mu\left(n\right)\ket{n} \xrightarrow[\text{scrambling}]{\text{phase}} \sum_{n=0}^{\infty} P_\mu\left(n\right)\ket{n}\bra{n}.
    \label{eq:scramble}
\end{equation}
Under this assumption, the adversary's cheating strategy is restricted to performing an attack conditioned on the photon-number content of each pulse. Many works rely on this feature to prove the security of quantum primitive implementations \cite{Pan:RevMod20,LMC:PRL05,LYD:Nature18,PJ+:natcomm14,BOV:npj18}. 

Achieving phase randomization with active phase modulation or laser gain switching imposes practical limitations of a few GHz on repetition rates \cite{KTO:PRA14,BBR:PRL18}. These limitations, combined with the low values of $\mu$ required due to fundamental PDS source statistics, can bring effective communication rates down to a few MHz. Unwanted remnants of coherence in the number basis, furthermore, can be exploited for a large spectrum of attacks, using unambiguous state discrimination for instance \cite{TYM:PRA13,DJL:pra00}. In contrast, as discussed in this work, QDS can be excited in such a way that this coherence is intrinsically suppressed, thus circumventing the need for active phase scrambling. With demonstrated Purcell-enhanced photon lifetimes of tens of picoseconds \cite{LBO:NatNanotech18,TJA:NatNan21} and source efficiencies now beyond the $50\%$ level~\cite{TJA:NatNan21}, QDS have the potential to enable secure communication rates of tens of GHz \cite{AMS:PRApp20}, i.e. around $3$ orders of magnitude higher than effective PDS communication rates. This is not only true for QKD but for many quantum primitives as shown in the following sections. 

\medskip

\noindent\textbf{Quantum key distribution.} A few decades after the birth of quantum key distribution (QKD) \cite{BB84}, experimentalists started demonstrating that QDS with low collection efficiencies can already outperform PDS in terms of secret key rate \cite{HKR:NJP12,CCF:JAP10,IWK:JOA09,Y:Nat02}. We first show that, while this is true for standard QKD implemented without the decoy-state countermeasure, beating PDS with decoy states \cite{LMC:PRL05} requires much higher QDS collection efficiencies at an equal repetition rate. Our results, based on the optimal performance of pumping schemes in Fig. \ref{fig:pumping}, are displayed in Fig. \ref{fig:QKD}: without decoy states $(a)$, QDS with collection efficiency $1\%$ are enough to outperform PDS after $100$ km, while infinite decoy schemes $(b)$ require at least $30\%$. We should emphasize, however, that this benchmark must be scaled by a repetition rate factor for QDS which could achieve considerably higher repetition rates than phase-randomized PDS. Any state preparation losses, including modulator losses, can be absorbed in the QDS collection efficiency, and the resulting key rate inferred from Fig. \ref{fig:QKD}. Comparisons can also be established using finite (and different) number of decoy intensities for QDS and PDS.

The optimal pumping scheme for QKD then follows from Eq. (\ref{eq:scramble}): the states' global phase must be uniformly randomized \cite{LP:QIC07,CZL:NJP15}, which implies that standard and decoy-state QKD should only be implemented with LA and TPE. Any remaining photon-number coherence will lead to a decrease in the secure key rate, as shown in \cite{CZL:NJP15}.

Twin-field QKD, on the other hand, requires two states, generated by Alice and Bob, to interfere on an untrusted party's beamsplitter \cite{LYD:Nature18,WYH:NatPhot21}. This forces Alice and Bob to scramble the global phase of their states in an active manner (using a modulator for instance), such that they can record their original fixed phase encoding. We therefore argue that twin-field schemes must be implemented with RE QDS, in order to provide the two parties with a shared phase reference before the scrambling. We simulate the protocol performance of \cite{LYD:Nature18} under these conditions in Fig. \ref{fig:QKD}.c, assuming the implementation of decoy states. We note that performing TF-QKD with a perfect single photon source (RE $\pi$-pulse and unit single-photon purity) is impossible, since there is no accessible phase to encode the key information. However, by decreasing the pulse area a little below $\pi$, in the same manner as \cite{L:NatPhot19}, it is possible to create a coherent superposition of vacuum and single-photon, thus providing an accessible phase to perform the protocol. 

Our results are focused on QKD implemented with discrete degrees of freedom, such as polarization or time-bin. For completeness, we note the existence of QKD protocols optimized for continuous degrees of freedom of PDS, such as phase and amplitude \cite{GG:PRL02}. These can actually compete with discrete-variable QKD over metropolitan distances, using simpler technology, but are sensitive to larger attenuations due to heavier post-selection and noise estimation techniques \cite{JKL:NatPhot13,VEA:OptExp20}.

\medskip 

\noindent\textbf{Quantum primitives beyond QKD.} Quantum cryptography presents a broad spectrum of other primitives, many of which belong to the branch of mistrustful cryptography \cite{BCD:PRX21,BS:dcc16}: unlike in QKD, Alice and Bob are not collaborators, but adversaries wishing to compute a common function. Decoy-state methods are then more challenging to apply (although not impossible for all protocols), since Alice and Bob do not trust each other.

Remarkably, the desired security properties for such primitives can be very sensitive to photon-number coherence. In this instance, substituting PDS by appropriately optimized QDS can yield even more benefits than in QKD. To show this, we extend the practical security analyses of three mistrustful quantum primitives \cite{BOV:npj18,PJ+:natcomm14,NJC:NatComms12} to the QDS framework, and estimate the QDS collection efficiencies required to outperform PDS for the relevant security figures of merit. Our performance results are summarized in Fig. \ref{fig:protocols} for all primitives. 

We display the performance of QDS and PDS for one example primitive in Fig. \ref{fig:sdpresults}: unforgeable quantum tokens. This primitive allows a central authority to issue tokens, comprised of quantum states, whose unforgeability is intrinsically guaranteed by the no-cloning theorem, thus requiring no hardware assumptions. One famous application is quantum money, which, in its private-key form, can prevent banknote forgery \cite{Wie:acm83}, double-spending with credit cards \cite{BDG:PRA19,BOV:npj18} and guarantee features such as user privacy \cite{Kent:npjQI22}. 

In Fig. \ref{fig:sdpresults}.a., we compare the noise tolerance of the quantum token scheme from \cite{BDG:PRA19} for PDS and QDS as a function of source efficiency. Noise tolerance indicates how much experimental error rate can be tolerated such that the unforgeability property holds, while source efficiency is the probability that a threshold single-photon detector will click in a lossless setting. Naturally, PDS reach a maximal noise tolerance for source efficiencies around $63\%$, corresponding to $\mu\approx 1$, before dropping again when the multiphoton contribution becomes too significant. For QDS, we notice a striking difference between schemes with coherence (RE) and those without (LA and TPE): the latters give an overhead of almost $2\%$ on the noise tolerance with respect to RE at high source efficiencies. This difference is crucial in making implementations feasible, since boosting the fidelity of quantum state preparation and quantum storage by a few percent can be extremely challenging. These differences are also reflected in Fig. \ref{fig:sdpresults}.b., which identifies the collection efficiencies at which QDS can outperform the best PDS performance: while LA and TPE require $44\%$ and $38\%$, respectively, RE must be pushed to $47\%$ to beat PDS. For information purposes, we also select three state-of-the art experimental QDS, and show how they would perform in such a beyond-QKD protocol with their reported values of brightness and single-photon purity. 

Fig. \ref{fig:sdpresults}.c. finally compares the performance of each source as a function of distance. Once again, the difference between LA/TPE and RE is significant due to the coherence feature. We notice here that the maximal distance for all sources is much shorter than in QKD schemes, since our selected quantum token scheme bears a maximal loss tolerance of $50\%$: above this limit, an adversary can clone the quantum token without introducing any errors \cite{BDG:PRA19}. 

Our work showcases the importance of engineering optical pumping schemes towards specific primitives. Fig. \ref{fig:protocols} displays non-trivial requirements for quantum strong coin flipping for instance: unlike with QKD and quantum tokens, QDS perform better at lower collection efficiencies, thus rendering state-of-the art QDS already capable of providing quantum advantage in such mistrustful primitives. Furthermore, the absence of photon number coherence in LA and TPE allow these schemes to reach quantum advantage over significantly longer distances than RE schemes: for the selected protocol from \cite{PJ+:natcomm14}, these values read $86$ km and $36$ km for TPE and LA, respectively, vs. $25$ km for RE.

\medskip

\large

\noindent\textbf{Discussion}

\normalsize

\noindent We have estimated threshold collection efficiencies for which GaAs-based quantum-dot photon sources can outperform Poisson-distributed-based implementations in four main quantum-cryptographic primitives. The estimations include the combined effect of brightness, single-photon purity, and photon-number coherence. We have shown in particular that resonant excitation should be used for twin-field QKD, but not for decoy QKD and other quantum primitives due to the unwanted presence of photon-number coherence that violates practical security assumptions.

We believe that these results will provide a benchmark for future achievements in quantum dot cavity structures, especially at telecom wavelengths \cite{KNB:Nano21,NVF:APL21,AMS:PRApp20,TTH:JAP07}. Although state-of-the-art dot-cavity simulation frameworks cannot account for all characteristics of quantum dots emitting in the telecom range, current telecom performance \cite{AMS:PRApp20,SNK:Nano22,MTN:APL16} shows good agreement with our $900$nm framework: the brightness and purities are not strongly dependent on the emission wavelength. Furthermore, frequency conversion of $900$nm photons to telecom photons can currently reach efficiencies up to $57\%$ \cite{VBG:PRL20}. These extra losses can be absorbed in our collection efficiency quantity.

We wish to encourage future quantum key distribution experiments with optimal pumping schemes, taking into account the security assumptions provided by the quantum cryptography community. Finally, we hope to stimulate experiments that explore the full potential of quantum dot-based single-photon sources for other quantum network primitives like unforgeable tokens \cite{BOV:npj18,GAA:pra18,Kent:npjQI22}, coin flipping \cite{BBB:NC11,PJ+:natcomm14} and bit commitment \cite{NJC:NatComms12,Zbind:PRL13,Pan:PRL14}. We believe our analysis can be extended in future works to multipartite entanglement-based quantum network primitives, such as secret sharing \cite{BMH:NatComms14} and anonymous messaging \cite{UMY:PRL19}.

\medskip

\large

\noindent\textbf{Methods}

\normalsize

\noindent\textbf{Quantum dot dynamics.} Our GaAs-based quantum dot, driven by a pulsed pump laser, is modelled either as a two- or a three-level system coupled to a single-mode microcavity in the Jaynes-Cummings manner. The interaction of the quantum dot with phonons is treated by the standard pure-dephasing Hamiltonian \cite{Besombes2001,Borri2001,Axt2005,Reiter2019}. In this way, we solve for the dynamics of the dot-cavity system by employing a numerically exact path-integral formalism \cite{Vagov2011,Cygorek2017,Cosacchi2018}. The detailed intracavity simulations, along with the derivation of the photon number populations, are presented in Appendix A.

\medskip

\noindent\textbf{Practical security analyses.} Appendices B and C show how the collection efficiencies and state encodings are modelled, both in the presence and absence of photon number coherence, for PDS and QDS, respectively. Appendix D provides some high-level descriptions of the four main quantum primitives, and displays all results justifying our claims. Appendix E briefly introduces mathematical tools required to understand the security analyses, namely semidefinite programs and Choi's theorem on completely positive maps. Appendices F to J provide the practical security analyses of all protocols, and the extensions to account for the presence of coherence in the QDS framework.

\medskip

\large

\noindent\textbf{Acknowledgments}

\normalsize

\noindent M.B., M.V, J.C.L. and P.W. acknowledge support from the European Commission through UNIQORN (no. 820474), the AFOSR via Q-TRUST (FA9550-21-
1-0355), the Austrian Science Fund (FWF) through BeyondC (F7113) and Reseach Group (FG5), and from the Austrian Federal Ministry for Digital and Economic Affairs, the National Foundation for Research, Technology and Development and the Christian Doppler Research Association. M.C., T.S. and V.M.A. are grateful for funding by the Deutsche Forschungsgemeinschaft (DFG, German Research Foundation) under project No. 419036043. C.N., S.L.P. and P.M. gratefully acknowledge the funding by the German Federal Ministry of Education and Research (BMBF) via the project QR.X (16KISQ013) and the European Union’s Horizon 2020 research and innovation program under Grant Agreement No. 899814 (Qurope). The work reported in this paper was partially funded by Project No. EMPIR 20FUN05 SEQUME.

%\makeatletter

\onecolumngrid

\newpage

\begin{center}
  \Large
  \textbf{Appendix}
\end{center}

\begin{appendix}

\noindent Section \ref{sec:dynamics} details the theoretical framework used in our intra-cavity simulations of quantum dot dynamics. The brightness and correlation functions are derived, from which the emitted photon number populations $\{p_n\}$, used in our security analyses, are inferred. Sections \ref{sec:PDS} and \ref{sec:QDS} show how the collection efficiencies and state encodings are modelled, both in the presence and absence of photon number coherence, for PDS and QDS, respectively. Section \ref{sec:resultsanddescriptions} provides some high-level descriptions of the four main quantum primitives, and displays all results justifying the Main Text claims. Section \ref{sec:tools} briefly introduces mathematical tools required to understand the security analyses, namely semidefinite programs and Choi's theorem on completely positive maps. Sections \ref{sec:decoy} to \ref{sec:bitcommitment} provide the practical security analyses of all protocols, and the extensions to account for the presence of coherence in the QDS framework.

\medskip

\textit{Note on security analyses: For all quantum primitives, we make the standard quantum-cryptographic assumption that a dishonest party can replace their lossy channel and detectors by ideal ones, as this only increases their power. Although we take into account experimental imperfections such as channel losses and detector dark counts, we perform all analyses in the asymptotic regime. Our results are designed to illustrate the claims of the Main Text for a handful of protocols, and not to provide a full analysis of quantum primitives with finite-size effects.}   

\section{Quantum dot dynamics and simulations}\label{sec:dynamics}

\subsection{Model and dynamical equation}
\label{app:model}

\noindent We model a quantum-dot--cavity system (QDC) consisting of a laser-driven strongly-confined self-assembled semiconductor quantum dot (QD) coupled to a single-mode microcavity influenced by an environment of longitudinal acoustic phonons (Ph) by the Hamiltonian
\begin{align}
H=\,&H_{\t{QDC}}+H_{\t{Ph}}\, .
\end{align}

\subsubsection{Two-level model}

QDs can often be described as two-level systems, e.g. when one is dealing with trions in charged QDs or, when a circularly polarized laser excites degenerate excitons with vanishing fine-structure splitting \cite{Reiter2019}.
%When the fine-structure splitting between the two oppositely circularly polarized bright exciton states vanishes and the external laser has a defined circular polarization, the quantum dot (QD) can be modeled as a two-level system \cite{Cosacchi2020b}.
The considered two-level system has an excited state $\vert X\>$ at energy $\hbar\w_{\t{X}}$ and the energy of the ground state $\vert G\>$ is chosen to be zero.
Assuming that the cavity supports only a single mode with nearly resonant coupling to the two-level system, the Jaynes-Cummings Hamiltonian can be used.
Denoting the frequency of the microcavity mode by $\w_{\t{C}}$ and the coupling strength by $\hbar g$, the QDC Hamiltonian in a frame co-rotating with the laser frequency $\w_{\t{L}}$ reads
%The coupling to the quantized electromagnetic field mode of the microcavity at $\hbar\w_{\t{C}}$ with strength $\hbar g$ is described by the Jaynes-Cummings Hamiltonian.
%In a frame co-rotating with the laser frequency $\w_{L}$ the QDC Hamiltonian thus reads as
\begin{align}
\label{eq:2LS}
H_{\t{QDC}}=\,&\hbar\D\w_{\t{XL}} \XX
+\hbar\D\w_{\t{CL}} a\+ a
+ \hbar g \left(a \s\+ + a\+ \s\right)
- \hbar \frac{f(t)}{2}\left(\s\+ + \s\right)\, .
\end{align}
$\D\w_{\t{XL}}=\w_{\t{X}}-\w_{\t{L}}$ is the exciton-laser detuning, $\D\w_{\t{CL}}=\w_{\t{C}}-\w_{\t{L}}$ is the cavity-laser detuning, $a$ ($a\+$) annihilates (creates) a cavity photon, $\s:=\GX$ is the QD transition operator, and $f(t)$ is the real envelope of the driving laser.
The cavity mode is assumed to be in resonance with the ground state-to-exciton transition, i.e. the cavity-exciton detuning $\D\w_{\t{CX}}=\w_{\t{C}}-\w_{\t{X}}$ is zero.
We consider pulsed excitations with a train of Gaussian pulses, each of which has the form
\begin{align}
f(t)=\frac{\mathcal{A}}{\sqrt{2\pi}\tau_{\t{G}}} e^{-\frac{t^2}{2{\tau_{\t{G}}}^2}}
\end{align}
with the pulse area $\mathcal{A}$ and width $\tau_{\t{G}}$, which is connected to the full width at half maximum by $T_\text{FWHM}=2\sqrt{2\ln{2}}\,\tau_{\t{G}}$.

The QD interacts with an environment of longitudinal acoustic (LA) phonons.
This interaction is modeled by the Hamiltonian \cite{Besombes2001,Borri2001,Krummheuer2002,Axt2005,Reiter2019}
\begin{align}
\label{eq:H_Ph}
H_{\t{Ph}}=\hbar\sum_\q \w_\q b_\q\+ b_\q+\hbar\sum_\q \left(\g_\q^{\t{X}}b_\q\+ +\g_\q^{\t{X}*}b_\q\right)\XX\, ,
\end{align}
where $b_\q$ ($b_\q\+$) annihilates (creates) a phonon of energy $\hbar\w_\q$ in the mode $\q$.
$\g_\q^{\t{X}}$ is the coupling constant between the QD exciton and the LA phonons.
It fully determines the phonon spectral density $J(\w)=\sum_{\q} \vert\g_{\q}\vert^2 \delta(\w-\w_{\q})$.
Assuming harmonic confinement and a linear dispersion $\w_{\q}=c_s|\q|$ with sound velocity $c_s$, it becomes
\begin{align}
J(\w)=\frac{\w^3}{4\pi^2\r_D\hbar c_s^5}
\left(D_e e^{-\w^2a_e^2/(4c_s^2)}-D_h e^{-\w^2a_h^2/(4c_s^2)}\right)^2\, ,
\end{align}
where we have considered deformation potential coupling which is usually the dominant coupling mechanism in the type of QDs considered here \cite{Krummheuer2002}.
$\r_D$ is the density of the material, $D_e$ ($D_h$) the electron (hole) deformation potential, and $a_e$ ($a_h$) the electron (hole) confinement length.
We use standard GaAs material parameters listed in \cite{Krummheuer2005,Cygorek2017}.
Assuming identical potentials for electrons and holes, the confinement ratio is fixed by the effective masses as $a_h=a_e/1.15$.
The electron confinement length $a_e$ as the only free parameter thus becomes a measure for the size of the QD.
Choosing $a_e$ between $3\,$nm and $5\,$nm has produced results in good agreement with experiment \cite{Bounouar2015,Quilter2015,kaldewey2017demonstrating}.
Indeed, both the confinement ratio and the electron confinement length can be considered as fitting parameters when modeling specific samples in experiment.
Here, we choose $a_e=3\,$nm.
Note that the electronic density matrix is affected by phonons only via the phonon spectral density $J(\w)$.
For QDs of any shape, it is always possible to obtain a spherical dot model, which generates the identical $J(\w)$ \cite{Luker2017}.
Then, the smallest dimension of the nonspherical QD has the largest influence on the phonon coupling.

Furthermore, we account for the radiative decay of the QD exciton to the free field outside the cavity with rate $\gamma$ and cavity losses with rate $\kappa$.
Both processes are well approximated by a phenomenological Markovian description using Lindblad superoperators acting on a density matrix $\rho$
\begin{align}
\mathcal{L}_{O,\Gamma}\rho=\Gamma\left(O\rho O\+ -\frac{1}{2}\left\lbrace\rho,O\+ O\right\rbrace_+\right)\, ,
\end{align}
where $\Gamma$ is the decay rate associated with a process described by the operator $O$.
$\left\lbrace A,B\right\rbrace_+$ denotes the anti-commutator of operators $A$ and $B$.

The system dynamics is described by the Liouville-von Neumann equation for the density matrix $\r$.
\begin{align}
  \label{eq:Liouville-von Neumann}
  \frac{\partial}{\partial t} \r =\,&
-\frac{i}{\hbar}\{H,\r\}_-
+\L_{\s,\g}\r
+\L_{a,\kappa}\r
\end{align}
with the commutator $\{\cdot,\cdot\}_-$.

\subsubsection{Three-level model}

%\noindent When the two oppositely polarized modes in the microcavity have a large energetic mismatch and the external driving has a defined linear polarization, one of the two oppositely polarized exciton states is decoupled from the external laser.
Considering external driving by lasers with well defined linear polarization and again assuming that there is just a single nearly resonant cavity mode, one of the linearly polarized excitons is decoupled from the dynamics.
Therefore, including the biexciton state $|B\>$ in this situation amounts to the addition of only one further electronic level.
The QDC Hamiltonian becomes
\begin{align}
\label{eq:3LS}
H_{\t{QDC}}=\,&\hbar\D\w_{\t{XL}} \XX
+(2\hbar\D\w_{\t{XL}}-E_{\t{B}}) \BB
+\hbar\D\w_{\t{CL}} a\+ a
+ \hbar g \left(a \s\+ + a\+ \s\right)
- \hbar \frac{f(t)}{2}\left(\s\+ + \s\right)\, ,
\end{align}
where the biexciton binding energy $E_{\t{B}}$ has been introduced.
We assume a value of $E_{\t{B}}=4\,$meV (cf., Table~\ref{tab:par}), which is large for typical QDs, but even if it may be hard to find a naturally grown QD with this value, it can be achieved by applying biaxial stress \cite{Ding2010}.
Having such a rather large value for $E_{\t{B}}$ means that essentially all collected photons originate from the exciton-to-ground state transition.
$\s:=\GX+\XB$ now contains both transitions which are optically excited in the QD.
The phonon coupling strength of the biexciton state is assumed to be twice the one of the single exciton, i.e.
\begin{align}
\label{eq:H_Ph_3LS}
H_{\t{Ph}}=\hbar\sum_\q \w_\q b_\q\+ b_\q+\hbar\sum_\q \left(\g_\q^{\t{X}}b_\q\+ +\g_\q^{\t{X}*}b_\q\right)\left(\XX+2\BB\right)\, .
\end{align}
Finally, the biexciton is assumed to radiatively decay with twice the exciton decay rate $\g$.
Therefore, the dynamical equation becomes
\begin{align}
  \label{eq:Liouville-von Neumann_3LS}
  \frac{\partial}{\partial t} \r =\,&
-\frac{i}{\hbar}\{H,\r\}_-
+\L_{\GX,\g}\r
+\L_{\XB,2\g}\r
+\L_{a,\kappa}\r
\end{align}

\subsubsection{Method and parameters}

\noindent We employ a numerically exact iterative real-time path integral method to solve the Liouville-von Neumann equation for the QDC's reduced density matrix $\overline{\rho}:=\Tr_{\t{Ph}}[\r]$, where the phonon subspace is traced out.
Details of the method are explained in \cite{Vagov2011,Barth2016,Cygorek2017}.
This path integral formalism is exact up to the time discretization and the memory truncation length.
We call a solution numerically exact, when it does not change noticeably when making the discretization finer or the truncation length longer.
All relevant system parameters used for the calculations are listed in Tab.~\ref{tab:par}.

In the two-level model, we consider two different excitation conditions: (i) resonant $\pi$-pulse and (ii) off-resonant phonon-assisted excitation.
In the former, the laser is on resonance with the exciton energy, i.e. $\D\w_{\t{XL}}=0$ and the excitation pulse has the area $\mathcal{A}=\pi$.
In the latter, the laser is detuned above the exciton energy by $\D\w_{\t{XL}}=-0.9\,$meV and the pulse has an area of $\mathcal{A}=10\pi$.

In the three-level model, we only consider the two-photon resonant excitation of the biexciton state, i.e. $2\hbar\D\w_{\t{XL}}-E_{\t{B}}=0$.
For every $T_\text{FWHM}$ chosen for the simulation, first, the area $\mathcal{A}$ has to be found, for which the occupation of the biexciton state $
|B\>$ is unity after the pulse.
This calibration is done for a standalone QD, i.e. $g=0$, and without any losses, i.e. $\g=\kappa=0$.

\begin{table}[t]
\begin{center}
\caption{Phyiscal parameters used in the simulations.}
  \begin{tabularx}{90mm} {>{\hsize=50mm}X >{\hsize=15mm}X  >{\hsize=20mm}X}

\toprule

\textbf{Quantum dot-cavity coupling} & \Centering $\hbar g$ & \Centering 0.05 meV \\

\textbf{Biexciton binding energy} & \Centering $E_{\t{B}}$ & \Centering 4 meV \\

\textbf{Radiative decay rate} & \Centering $\g$ & \Centering 1 ns$^{-1}$ \\

\textbf{Cavity loss rate} & \Centering $\kappa$ & \Centering 0.577 ps$^{-1}$ \\

\textbf{Temperature} & \Centering $T$ & \Centering 4.2 K \\

\textbf{Cavity-exciton detuning} & \Centering $\hbar\Delta\omega_{CX}$ & \Centering 0 meV \\

\bottomrule
  \end{tabularx}
  \label{tab:par}
\end{center}
\end{table}

\subsection{Derivation of photon number populations $p_n$}\label{sec:probabilities}

\subsubsection{Correlation functions}\label{sec:correlations}

\noindent We calculate second-order two-time correlation functions
\begin{align}
G^{(2)}(t,\tau)=\<a\+(t)a\+(t+\tau)a(t+\tau)a(t)\>
\end{align}
to obtain the multiphoton component of the cavity state.
First, we average over the entire pulse sequence to obtain a function of the delay time argument $\tau$ only:
\begin{align}
G^{(2)}(\tau):=\,&\lim_{T\to\infty}\frac{1}{T}\int_{0}^T dt\, G^{(2)}(t,\tau)
\end{align}
Then, we integrate the peak-like structure around $\tau=0$, which corresponds to the amount of multiphoton contribution within a single pulse of the pulse train, and normalize it to the first uncorrelated side peak.
This results in the probability $\P_2$ of having two or more photons during one pulse
\begin{equation}
\begin{aligned}
\P_2=\frac{\int_{-T_{\t{Pulse}}/2}^{T_{\t{Pulse}}/2}d\tau\,G^{(2)}(\tau)}{\int_{T_{\t{Pulse}}/2}^{3 T_{\t{Pulse}}/2}d\tau\,G^{(2)}(\tau)}\, .
\end{aligned}\label{eq:p2}
\end{equation}
Here, $T_{\t{Pulse}}$ is the separation of two subsequent pulse maxima within the excitation pulse train.

To estimate the probability to obtain three or more photons during one pulse, we evaluate the third-order three-time correlation function
\begin{align}
G^{(3)}(t,\tau_1,\tau_2)=\<a\+(t)a\+(t+\tau_1)a\+(t+\tau_1+\tau_2)a(t+\tau_1+\tau_2)a(t+\tau_1)a(t)\>\, .
\end{align}
Again, it is averaged over the pulse sequence as
\begin{align}
G^{(3)}(\tau_1,\tau_2):=\,&\lim_{T\to\infty}\frac{1}{T}\int_{0}^T dt\, G^{(3)}(t,\tau_1,\tau_2)\, .
\end{align}
In close analogy to the case above, the probability $\P_3$ of three-photon (or more) coincidence within a single pulse is obtained as
\begin{equation}
\begin{aligned}
\P_3=\frac{\int_{-T_{\t{Pulse}}/2}^{T_{\t{Pulse}}/2}d\tau_1\int_{-T_{\t{Pulse}}/2}^{T_{\t{Pulse}}/2}d\tau_2\,G^{(3)}(\tau_1,\tau_2)}{\int_{T_{\t{Pulse}}/2}^{3 T_{\t{Pulse}}/2}d\tau_1\int_{T_{\t{Pulse}}/2}^{3 T_{\t{Pulse}}/2}d\tau_2\,G^{(3)}(\tau_1,\tau_2)}\, .
\end{aligned}\label{eq:p3}
\end{equation}
Both $G^{(2)}$ and $G^{(3)}$ are then calculated in a numerically exact way following \cite{Cosacchi2018}. Finally, note that we define the unnormalized brightness of the QD source as
\begin{equation}
\begin{aligned}
\widetilde{\mathcal{B}}=\,&\kappa \int_{t_0-T_{\t{Pulse}}/2}^{t_0+T_{\t{Pulse}}/2} dt\, \<a\+(t) a(t)\>\, ,
\end{aligned}\label{eq:brightnesss}
\end{equation}
where $t_0$ is the center time of the pulse.

\subsubsection{Population extraction}

To calculate the probabilities of $n$-photon emission, we used the following assumptions: $\P_{\geqslant4}=0$ for two-level systems and $\P_{\geqslant3}=0$ for three-level systems. These assumptions are based on $n$-photon generation probabilities which scale as $(\gamma T_\text{FWHM})^{(n-1)}$ for two-level systems \cite{Fischer:IOP17} and $(\gamma T_\text{FWHM})^{2(n-1)}$ for three level systems \cite{Fisher:NPJ18}. Using the correlation functions derived in Eqs. (\ref{eq:p2}) and (\ref{eq:p3}), along with unnormalized brightness from Eq. (\ref{eq:brightnesss}), we inferred the emitted photon number populations $\{p_n\}$ as: 

\begin{equation}
\begin{aligned}
&p_1=\widetilde{\mathcal{B}}-2\P_2-3\P_3,\\
&p_2=\P_2-\P_3,\\
&p_3=\P_3,\\
&p_{\geqslant4}=0\\
&p_0 = 1-p_1-p_2-p_3-p_{\geqslant4}.
\end{aligned}
\label{eq:probabilities}
\end{equation}
The normalized brightness can now be expressed as:

\begin{equation}
\begin{aligned}
\mathcal{B}&=\sum_{n\geqslant 1}p_n.
\end{aligned}
\label{eq:ourbrightness}
\end{equation}
The single-photon purity may now be written as:

\begin{equation}
\begin{aligned}
\P&=\frac{p_1}{\mathcal{B}}.
\end{aligned}
\label{eq:purityz}
\end{equation}

\begin{table}[t]
\begin{center}
\caption{Optimal photon number populations derived from Eq. (\ref{eq:probabilities}).}
\begin{tabularx}{145mm} {>{\hsize=50mm}X >{\hsize=10mm}X  >{\hsize=25mm}X >{\hsize=25mm}X >{\hsize=25mm}X}
\toprule
& &  \Centering \textbf{RE} & \Centering \textbf{LA-assisted
excitation} & \Centering \textbf{TPE}  \\
\midrule
\textbf{Pump pulse length} & \Centering $T_\text{FWHM}$ & \Centering 3 ps & \Centering 8 ps & \Centering 12 ps \\
\textbf{Pump pulse area} & \Centering  $\mathcal{A}$  & \Centering $\pi_\text{RE}$ & \Centering $10\pi_\text{RE}$ & \Centering $\pi_\text{TPE}\approx6.8\pi_\text{RE}$ \\
\textbf{Normalized brightness} &  \Centering  $\mathcal{B}$ & \Centering 0.9366 & \Centering 0.8399 & \Centering 0.9526 \\
\textbf{Single-photon purity} &  \Centering  $\mathcal{P}$ & \Centering 0.9903 & \Centering 0.9785 & \Centering 0.9988 \\
\textbf{Single-photon population} &  \Centering  $p_1$ & \Centering 0.9275 & \Centering 0.8219 & \Centering 0.9514 \\
\textbf{Two-photon population} &  \Centering  $p_2$ & \Centering 0.0091 & \Centering 0.0180 & \Centering 0.0012 \\
\textbf{Three-photon population} &  \Centering  $p_3$ & \Centering $10^{-8}$ & \Centering $10^{-7}$ & \Centering 0 \\
\bottomrule
\end{tabularx}\label{tab:pumpy}
\end{center} 
\end{table}

\section{Collection efficiency and state encoding for Poisson-distributed sources}\label{sec:PDS}

\subsection{With number coherence}

\noindent Coherent states may be expressed as a Poisson-distributed superposition of photon number states:

\begin{equation}
    \ket{\alpha} = \sum_{n=0}^{\infty} e^{-\frac{|\alpha|^2}{2}}\frac{\alpha^n}{\sqrt{n!}}\ket{n} =\sum_{n=0}^{\infty} C_\alpha\left(n\right)\ket{n},
\end{equation}
where $\{\ket{n}\}$ denote the photon number states and $\alpha$ is the coherent state amplitude. Using either polarization, time-bin or path encoding, the two-mode coherent states in all protocols may be expressed as:
\begin{equation}
\ket{\alpha_k} = \Ket{e^{i\theta}\frac{\alpha}{\sqrt{2}}}\otimes\Ket{e^{i(\theta+\phi_k)}\frac{\alpha}{\sqrt{2}}},
\label{eq:globalphase}
\end{equation}
where $\theta=0$ is a global phase and $\phi_k\in\{0,\frac{\pi}{2},2\pi,\frac{3\pi}{2}\}$ is the relative phase between the two modes, which can take one of four values depending on $k\in\{0,1,2,3\}$. In quantum cryptography, a potential eavesdropper or adversary must access $\phi_k$ to unveil the information encoded in the states. 

In a similar manner to \cite{DJL:pra00,BDG:PRA19}, we may only focus on the second mode which contains the relative phase $\phi_k$, and thus rewrite these four encoded states as $\ket{\widetilde{\alpha_k}}= \ket{e^{i\phi_k} \frac{\alpha}{\sqrt{2}}}$, with $k\in\{0,1,2,3\}$. To avoid truncating the infinite-dimensional Fock space in our state expressions, we notice that these four specific states may be expressed in a four-dimensional orthonormal basis $\{\ket{b_i}\}$ as:
\begin{equation}
\begin{aligned}
    \ket{\widetilde{\alpha_0}} &=  B_{0}\ket{b_0}+B_{1}\ket{b_1}+B_{2}\ket{b_2}+B_{3}\ket{b_3}  \\
    \ket{\widetilde{\alpha_1}} &= B_{0}\ket{b_0}+iB_{1}\ket{b_1}-B_{2}\ket{b_2}-iB_{3}\ket{b_3}\\%\displaybreak[1]
    \ket{\widetilde{\alpha_2}} &= B_{0}\ket{b_0}-B_{1}\ket{b_1}+B_{2}\ket{b_2}-B_{3}\ket{b_3}\\
    \ket{\widetilde{\alpha_3}} &= B_{0}\ket{b_0}-iB_{1}\ket{b_1}-B_{2}\ket{b_2}+iB_{3}\ket{b_3}
\end{aligned}\label{eq:fixedphase}
\end{equation}
where
\begin{align*}
    B_0 &= \frac{e^{-\frac{|\alpha|^2}{4}}}{\sqrt{2}}\sqrt{\cosh{\frac{\alpha^2}{2}}+\cos{\frac{\alpha^2}{2}}}\\
    B_1 &= \frac{e^{-\frac{|\alpha|^2}{4}}}{\sqrt{2}}\sqrt{\sinh{\frac{\alpha^2}{2}}+\sin{\frac{\alpha^2}{2}}}\\
    B_2 &= \frac{e^{-\frac{|\alpha|^2}{4}}}{\sqrt{2}}\sqrt{\cosh{\frac{\alpha^2}{2}}-\cos{\frac{\alpha^2}{2}}}\\
    B_3 &= \frac{e^{-\frac{|\alpha|^2}{4}}}{\sqrt{2}}\sqrt{\sinh{\frac{\alpha^2}{2}}-\sin{\frac{\alpha^2}{2}}}
\end{align*}

\subsection{Without number coherence}
 \noindent Phase randomization scrambles the global phase reference from Eq. (\ref{eq:globalphase}) by allowing $\theta$ to take values from $[0,2\pi]$ uniformly at random instead of a single value. By considering the state $\ket{e^{i\theta}\alpha}$ and integrating over all possible values of $\theta$, the adversary sees a classical mixture of Fock states given by \cite{LP:calt05}:
\begin{equation}
    \frac{1}{2\pi}\int_{0}^{2\pi} \ket{\sqrt{\mu} e^{i\theta}}\bra{\sqrt{\mu} e^{i\theta}}d\theta = e^{-\mu}\sum_{n=0}^{\infty}\frac{\mu^n}{n!}\ket{n}\bra{n},
%    \label{phaserand}
\end{equation}
where $\mu=|\alpha|^2$ is the average photon number, and $\ket{n}$ are the photon number states. As the coherent superpositions of number states vanish, the quantum-cryptographic security proofs may simply proceed according to the result of quantum non-demolition (QND) photon number measurements. If there is no photon in the state, then there is no information content. If there is $1$ photon, then the qubit security proof may be applied.
If there are more than $2$ photons in the pulse, it is assumed that perfect cheating is possible.

One can therefore express the phase randomized states $\rho_k$ in a $7$-dimensional orthonormal basis $\{\ket{v},\ket{q_0},\ket{q_1},\ket{m_0},\ket{m_1},\ket{m_2},\ket{m_3}\}$, where $\ket{v}$ is the vacuum state, $\ket{q_0}$ and $\ket{q_1}$ span a qubit space, and $\ket{m_i}$ constitute the four orthogonal outcomes which materialize the four perfectly distinguishable states in the multiphoton subspace. Our four phase-randomized coherent states may then be written as the following density matrices :
\begin{equation}
\begin{aligned}
    \rho_0 &= P_\mu(0)\:\ketbra{v}{v}+P_\mu(1)\:\ketbra{+}{+}+P_\mu(\geqslant 2)\:\ketbra{m_0}{m_0}\\
    \rho_1 &= P_\mu(0)\:\ketbra{v}{v}+P_\mu(1)\:\ketbra{+i}{+i}+P_\mu(\geqslant 2)\:\ketbra{m_1}{m_1}\\
    \rho_2 &= P_\mu(0)\:\ketbra{v}{v}+P_\mu(1)\:\ketbra{-}{-}+P_\mu(\geqslant 2)\:\ketbra{m_2}{m_2}\\
    \rho_3 &= P_\mu(0)\:\ketbra{v}{v}+P_\mu(1)\:\ketbra{-i}{-i}+P_\mu(\geqslant 2)\:\ketbra{m_3}{m_3},
\end{aligned}\label{eq:randomphase}
\end{equation}
where $P_\mu(n) =\left|C_\mu(n)\right|^2$, $\{\ket{+},\ket{+i},\ket{-},\ket{-i}\}$ are the usual $\sigma_x$ and $\sigma_y$ eigenstates in the qubit space spanned by $\ket{q_i}$, and the Poisson distribution coefficients are given by
\begin{equation}
\begin{aligned}
    P_\mu(0)&= e^{-\mu},&%\\
    P_\mu(1) &= \mu e^{-\mu},&%\\
    P_\mu(\geqslant 2) &= 1-(1+\mu)e^{-\mu}.
\end{aligned}\label{eq:PDScoeffs}
\end{equation}

\section{Collection efficiency and state encoding for quantum dots}\label{sec:QDS}

\subsection{Preliminary definitions}

\noindent Single photons are obtained by the action of the creation operator onto the vacuum. Beam splitters act linearly on creation operators, and leave the vacuum invariant. More precisely, a beam splitter of reflectivity $r$ acting on input modes $\left(k,l\right)$ maps the creation operators $\hat a_k^\dag,\hat a_l^\dag$ onto $\hat b_k^\dag,\hat b_l^\dag$ as:
\be
\begin{pmatrix}\hat b_k^\dag\\\hat b_l^\dag\end{pmatrix}=H_{kl}^{(r)}\begin{pmatrix}\hat a_k^\dag\\\hat a_l^\dag\end{pmatrix},
\ee
where
\be
H_{kl}^{(r)}=\begin{pmatrix}\sqrt r&\sqrt{1-r}\\\sqrt{1-r} & -\sqrt r\end{pmatrix}.
\ee
We similarly define the phase shift operation $P_{kl}^{(\phi)}$, acting on input modes $\left(k,l\right)$, as:
\be
P_{kl}^{(\phi)}=\begin{pmatrix} 1&0\\0 & e^{i\phi}\end{pmatrix}.
\ee

\subsection{Collection and encoding with number coherence}
 
\noindent Following Fig. \ref{fig:encoding}, we model the collection efficiency of the quantum dot as a beamsplitter of reflectivity $\eta$, and label the three input spatial modes as $0$, $1$ and $2$. The standard four protocol states are then encoded with a Mach-Zehnder interferometer, consisting of two beamsplitters described by $H_{01}^{(r)}$,  with tunable phase $\phi\in\{0,\pi\}$, corresponding to the Pauli Z eigenstates in a two-dimensional Hilbert space, and $\phi\in\{\frac{\pi}{2},\frac{3\pi}{2}\}$, corresponding to the Pauli X eigenstates in a two-dimensional Hilbert space. A general pure photonic state, with photon number distribution given by $\{p_n\}$ and input into spatial mode $0$, then evolves as:

\begin{equation}
\begin{aligned}
\sum_{n=0}^{\infty}\sqrt{\frac{p_n}{n!}}\left(a^\dagger_0\right)^n\ket{000}_{012}
          &\xrightarrow[]{H_{01}^{(\eta)}} \sum_{n=0}^{\infty}\sqrt{\frac{p_n}{n!}}\left(\sqrt{\eta}a^\dagger_0+\sqrt{1-\eta}a^\dagger_2\right)^n\ket{000}_{012}\\
         &\xrightarrow[]{H_{01}^{(1/2)}} \sum_{n=0}^{\infty}\sqrt{\frac{p_n}{n!}}\left(\sqrt{\frac{\eta}{2}}a^\dagger_0+\sqrt{\frac{\eta}{2}}a^\dagger_1+\sqrt{1-\eta}a^\dagger_2\right)^n\ket{000}_{012}\\
          &\xrightarrow[]{P_{01}^{(\phi)}} \sum_{n=0}^{\infty}\sqrt{\frac{p_n}{n!}}\left(\sqrt{\frac{\eta}{2}}a^\dagger_0+\sqrt{\frac{\eta}{2}}e^{i\phi}a^\dagger_1+\sqrt{1-\eta}a^\dagger_2\right)^n\ket{000}_{012}\\
          &\xrightarrow[]{H_{01}^{(1/2)}} \sum_{n=0}^{\infty}\sqrt{\frac{p_n}{n!}}\left(\sqrt{\frac{\eta}{4}}\left(1+e^{i\phi}\right)a^\dagger_0+\sqrt{\frac{\eta}{4}}\left(1-e^{i\phi}\right)a^\dagger_1+\sqrt{1-\eta}a^\dagger_2\right)^n\ket{000}_{012}\\
          &= \sum_{n=0}^{\infty}\sum_{k=0}^{n}\sum_{l=0}^{k}\sqrt{\frac{p_n}{n!}}\binom{n}{k}\binom{k}{l}\left(\sqrt{\frac{\eta}{4}}\left(1+e^{i\phi}\right)a^\dagger_0\right)^{k-l}\left(\sqrt{\frac{\eta}{4}}\left(1-e^{i\phi}\right)a^\dagger_1\right)^l\left(\sqrt{1-\eta}a^\dagger_2\right)^{n-k}\ket{000}_{012}\\
          &= \sum_{n=0}^{\infty}\sum_{k=0}^{n}\sum_{l=0}^{k}\sqrt{\frac{p_n}{n!}}\binom{n}{k}\binom{k}{l}\left(\frac{\eta}{4}\right)^\frac{k}{2}\left(1-\eta\right)^\frac{n-k}{2}\left(1+e^{i\phi}\right)^{k-l}\left(1-e^{i\phi}\right)^{l}\left(a^\dagger_0\right)^{k-l} \left(a^\dagger_1\right)^l\left(a^\dagger_2\right)^{n-k}\ket{000}_{012}
\end{aligned}\label{eq:encoding}
\end{equation}
The encoded state then reads:

\begin{equation}
   \ket{\psi_{\eta,\phi}}_{012} = \sum_{n=0}^{\infty}\sum_{k=0}^{n}\sum_{l=0}^{k} c_{nkl}(\eta,\phi)\left(a^\dagger_0\right)^{k-l} \left(a^\dagger_1\right)^l\left(a^\dagger_2\right)^{n-k}\ket{000}_{012},
\end{equation}
where

\begin{equation}
    c_{nkl}(\eta,\phi) = \sqrt{\frac{p_n}{n!}}\binom{n}{k}\binom{k}{l}\left(\frac{\eta}{4}\right)^\frac{k}{2}\left(1-\eta\right)^\frac{n-k}{2}\left(1+e^{i\phi}\right)^{k-l}\left(1-e^{i\phi}\right)^{l}.
\end{equation}
and $\{p_n\}$ take the values of Eq. (\ref{eq:probabilities}) in our work. 

To obtain the actual collected state $\rho^{(c)}_{\eta,\phi}$ used in our security analyses (where the $(c)$ superscript denotes coherence in Fock basis for further convenience), we simply trace out over spatial mode $2$:

\begin{figure*}
	\begin{center}
		\includegraphics[width=160mm]{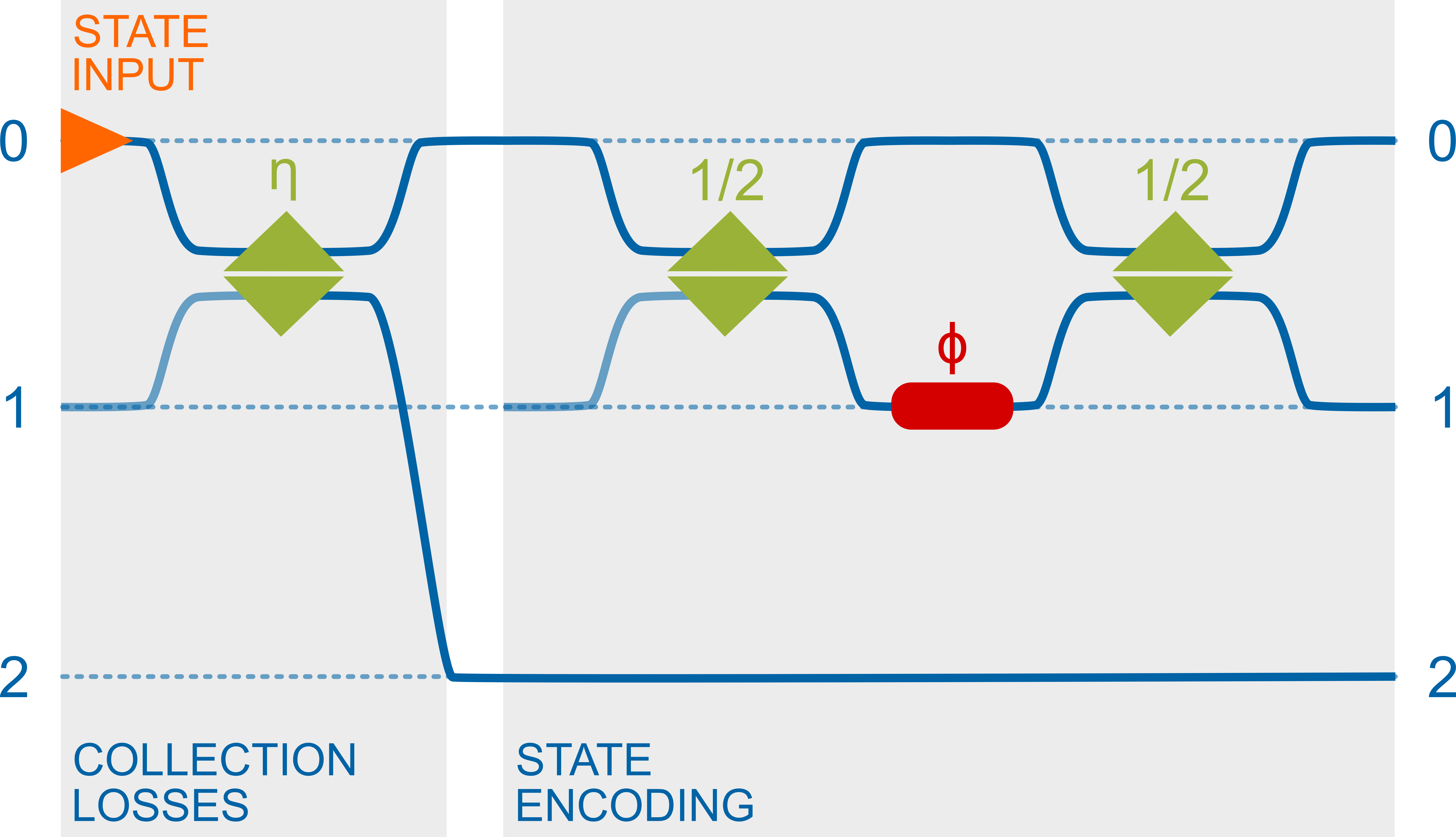}
		\caption{\textbf{Collection efficiency modeling and state encoding.} The collection efficiency of the quantum dot is modelled as a beamsplitter of reflection $\eta$, with two input spatial modes as $0$ and $1$. The quantum information is then encoded with a Mach-Zehnder interferometer, with tunable phase $\phi\in\{0,\pi\}$ for Z eigenstates and $\phi\in\{\frac{\pi}{2},\frac{3\pi}{2}\}$ for X eigenstates. }
		\label{fig:encoding}
	\end{center}
\end{figure*}

\begin{equation}
\begin{aligned}
 \rho^{(c)}_{\eta,\phi}&= \Tr_2\left( \ket{\psi_{\eta,\phi}}_{012} \bra{\psi_{\eta,\phi}}_{012} \right)\\
 &= \Tr_2\left( \sum_{n,m=0}^{\infty}\sum_{k=0}^{n}\sum_{l=0}^{k}\sum_{p=0}^{m}\sum_{q=0}^{p}c_{nkl}(\eta,\phi)c_{mpq}^{*}(\eta,\phi)\left(a^\dagger_0\right)^{k-l} \left(a^\dagger_1\right)^l\left(a^\dagger_2\right)^{n-k}\ket{000}_{012}\bra{000}_{012}a_0^{p-q} a_1^q a_2^{m-p}\right). 
 % &= \sum_{n,m=0}^{\infty}\sum_{k=0}^{n}\sum_{l=0}^{k}\sum_{s=0}^{m}\sum_{q=0}^{m-s}c_{nkl}(\eta,\phi)c_{m(m-s)q}^{*}(\eta,\phi) \:s!\:\left(a^\dagger_0\right)^{k-l} \left(a^\dagger_1\right)^l\ket{00}_{01}\bra{00}_{01}a_0^{m-s-q} a_1^q.
\end{aligned}\label{eq:numbercoherence}
\end{equation}

\subsection{Collection and encoding without number coherence}

 \noindent Similar calculations lead to the expression for the collected state $\rho^{(nc)}_{\eta,\phi}$, where the superscript $(nc)$ this time accounts for "no coherence" in Fock basis. In essence, we set $n=m$ in Eq. (\ref{eq:numbercoherence}) and obtain the following state:

\begin{equation}
\begin{aligned}
 \rho^{(nc)}_{\eta,\phi}= \Tr_2\left( \sum_{n=0}^{\infty}\sum_{k=0}^{n}\sum_{l=0}^{k}\sum_{p=0}^{n}\sum_{q=0}^{p}c_{nkl}(\eta,\phi)c_{npq}^{*}(\eta,\phi)\left(a^\dagger_0\right)^{k-l} \left(a^\dagger_1\right)^l\left(a^\dagger_2\right)^{n-k}\ket{000}_{012}\bra{000}_{012}a_0^{p-q} a_1^q a_2^{n-p}\right).
\end{aligned}\label{eq:diagonal}
\end{equation}

\section{QDS performance for quantum primitives}\label{sec:resultsanddescriptions}

\subsection{BB84 quantum key distribution}\label{sec:decoydescription}

\subsubsection{Brief introduction}

Quantum key distribution (QKD) is one of the most mature quantum-cryptographic primitives implemented so far. In its simplest form, it allows two parties, Alice and Bob, to establish a secret key over a public quantum channel, provided that a public, authenticated classical channel is available. The parties must ensure that the unwanted presence of an eavesdropper, Eve, on the channel is detected with arbitrarily high probability. Once a secret key has successfully been established, Alice and Bob may use it to encrypt a secret message through one-time padding \cite{M:CM82}. This encryption technique requires a secret key whose length is at least equal to the message length.   

In a standard protocol, Alice encodes $N$ bits of the secret key she wishes to share into $N$ qubit states. She randomly picks the encoding basis for each qubit ($\sigma_z$ or $\sigma_x$), and stores this classical information. Bit $0$ is therefore randomly encoded in either $\ket{0}$ or $\ket{+}$, while bit $1$ is randomly encoded in $\ket{1}$ or $\ket{-}$. The states are sent over a quantum channel to distant Bob, who does not know the encoding basis, and thus randomly measures each qubit in either $\sigma_z$ or $\sigma_x$. He records each qubit's measurement basis, along with the associated measurement outcomes. Once the quantum communication stage is over, Alice and Bob proceed to a classical reconciliation stage: Bob communicates his sequence of measurement bases (without the measurement outcomes) to Alice. After comparing it with her stored sequence, Alice reports to Bob the elements for which her preparation basis does not match Bob's measurement basis. They both agree to dismiss all bits which correspond to a basis mismatch from the final key. After basis reconciliation, Alice and Bob then compare a pre-agreed random subset of their corrected key to ensure that all bits match. If any of the bits disagree, they may conclude on the presence of Eve and abort the protocol. If all bits agree, the key has then successfully been established, and they may use it to encrypt a secret message. Note that an additional stage, known as privacy amplification, is required in the noisy setting, in order to decrease the amount of information that Eve acquires from Alice and Bob's classical error correction routine \cite{DEJ:PRL96}.

\subsubsection{Results}

\begin{figure*}
	\begin{center}
		\includegraphics[width=180mm]{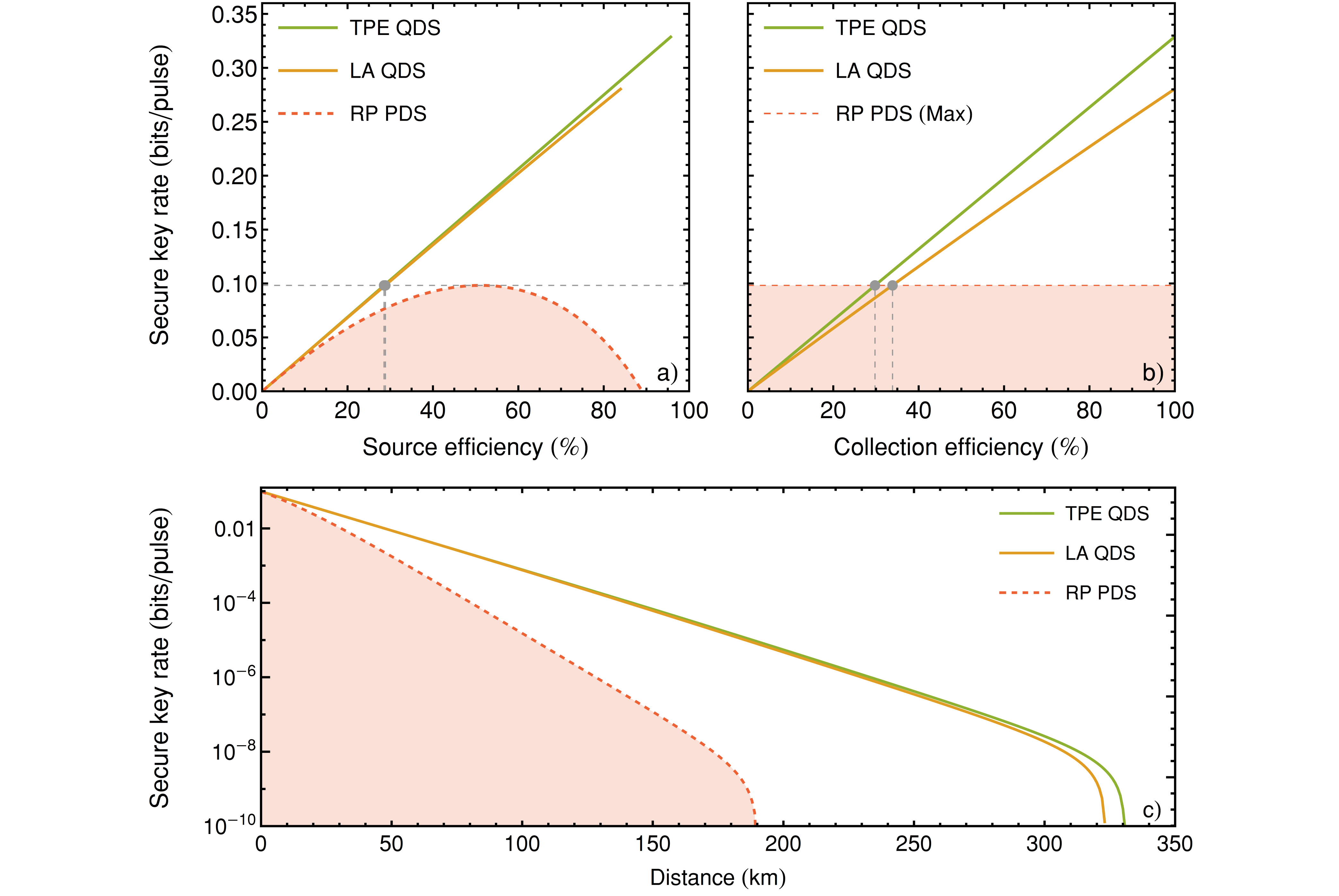}
		\caption{\textbf{Source comparison for BB84 QKD without decoy states.} (a) Simulated secret key rates from Eq. (\ref{eq:skrbb84}) as a function of source efficiency for LA and TPE QDS, along with randomized-phase (RP) PDS. Source efficiency is defined as $1-e^{-\mu}$ for PDS and $1-\sum_{n=0}^\infty p_n(1-\eta)^n$ for QDS, where $\eta$ is the QDS collection efficiency. Chosen pulse lengths, pulse areas, and photon number populations $\{p_n\}$ are displayed in Table \ref{tab:pumpy}. (b) Simulated secret key rates as a function of QDS collection efficiency, compared to the best performance of RP PDS sources (dashed line). (c) Simulated secret key rates as a function of distance, assuming single mode telecom fiber losses of $0.21$ dB/km. The QDS collection efficiencies were chosen as the intersection points from Fig (b). Parameters for all plots are: alignment error rate $e_d=2\%$, dark count probability $Y_0 =10^{-9}$, detection efficiency $\eta_d=100\%$, and error-correcting code inefficiency $f=1.2$.  }
		\label{fig:bb84}
	\end{center}
\end{figure*}

\begin{figure*}
	\begin{center}
		\includegraphics[width=180mm]{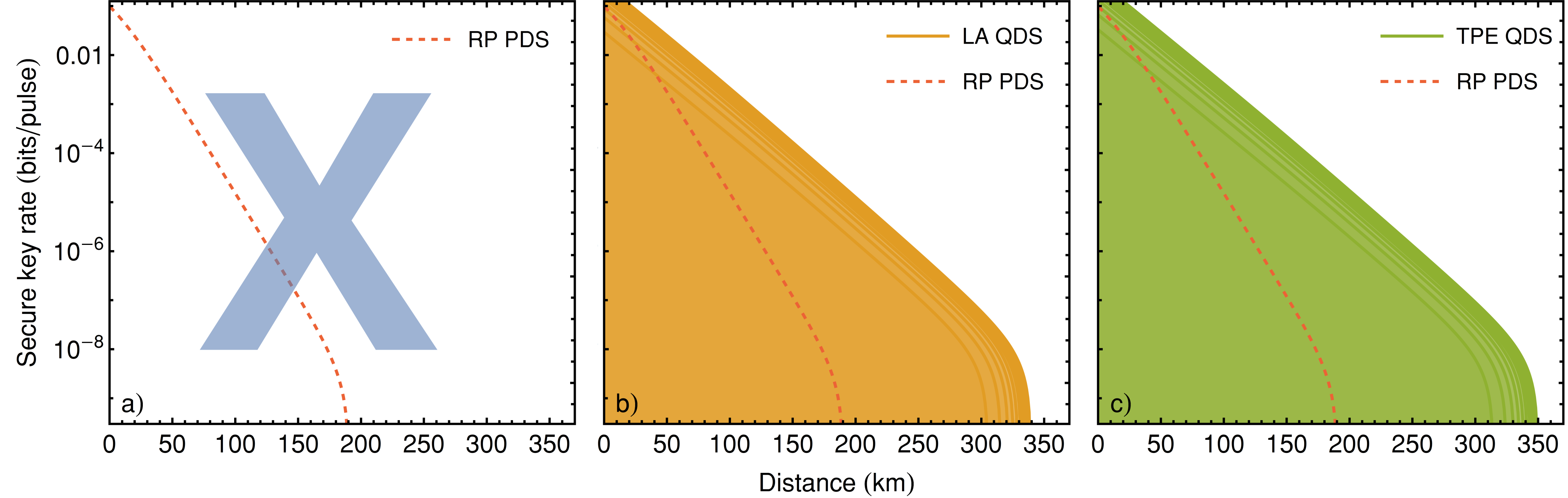}
		\caption{\textbf{Collection efficiency comparison for BB84 QKD without decoy states.} Simulated secret key rates from Eq. (\ref{eq:skrbb84}) as a function of distance for (a) RE QDS (b) LA QDS (c) TPE QDS, with collection efficiencies ranging from $\eta=1\%$ (bottom curves) to $\eta=100\%$ (top curves), in steps of $10\%$. The optimal performance of randomized-phase (RP) PDS is also plotted in dashed lines, in order to identify which QDS collection efficiencies are required to overcome PDS for each pumping. Single mode telecom fiber losses of $0.21$ dB/km are assumed. Parameters for all plots are: alignment error rate $e_d=2\%$, dark count probability $Y_0 =10^{-9}$, detection efficiency $\eta_d=100\%$, and error-correcting code inefficiency $f=1.2$. }
		\label{fig:bb84_collection}
	\end{center}
\end{figure*}

\begin{figure*}
	\begin{center}
		\includegraphics[width=180mm]{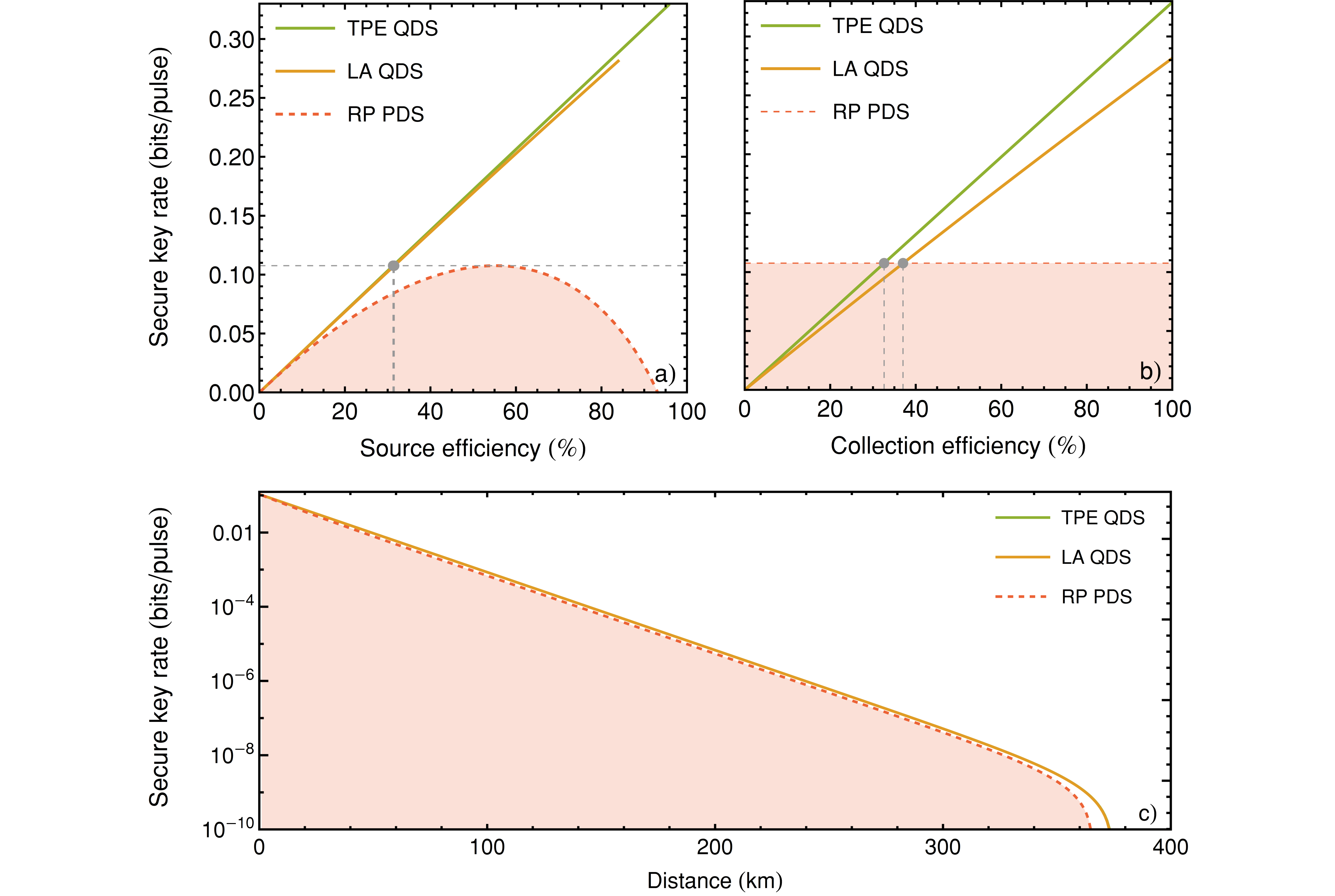}
		\caption{\textbf{Source comparison for BB84 QKD with decoy states.} (a) Simulated secret key rates from Eq. (\ref{eq:skrbb84}) as a function of source efficiency for LA and TPE QDS, along with randomized-phase (RP) PDS. Source efficiency is defined as $1-e^{-\mu}$ for PDS and $1-\sum_{n=0}^\infty p_n(1-\eta)^n$ for QDS, where $\eta$ is the QDS collection efficiency. Chosen pulse lengths, pulse areas, and photon number populations $\{p_n\}$ are displayed in Table \ref{tab:pumpy}. (b) Simulated secret key rates as a function of QDS collection efficiency, compared to the best performance of RP PDS sources (dashed line). (c) Simulated secret key rates as a function of distance, assuming single mode telecom fiber losses of $0.21$ dB/km. The QDS collection efficiencies were chosen as the intersection points from Fig (b). Parameters for all plots are: alignment error rate $e_d=2\%$, dark count probability $Y_0 =10^{-9}$, detection efficiency $\eta_d=100\%$, and error-correcting code inefficiency $f=1.2$.  }
		\label{fig:bb84_decoy}
	\end{center}
\end{figure*}

\begin{figure*}
	\begin{center}
		\includegraphics[width=180mm]{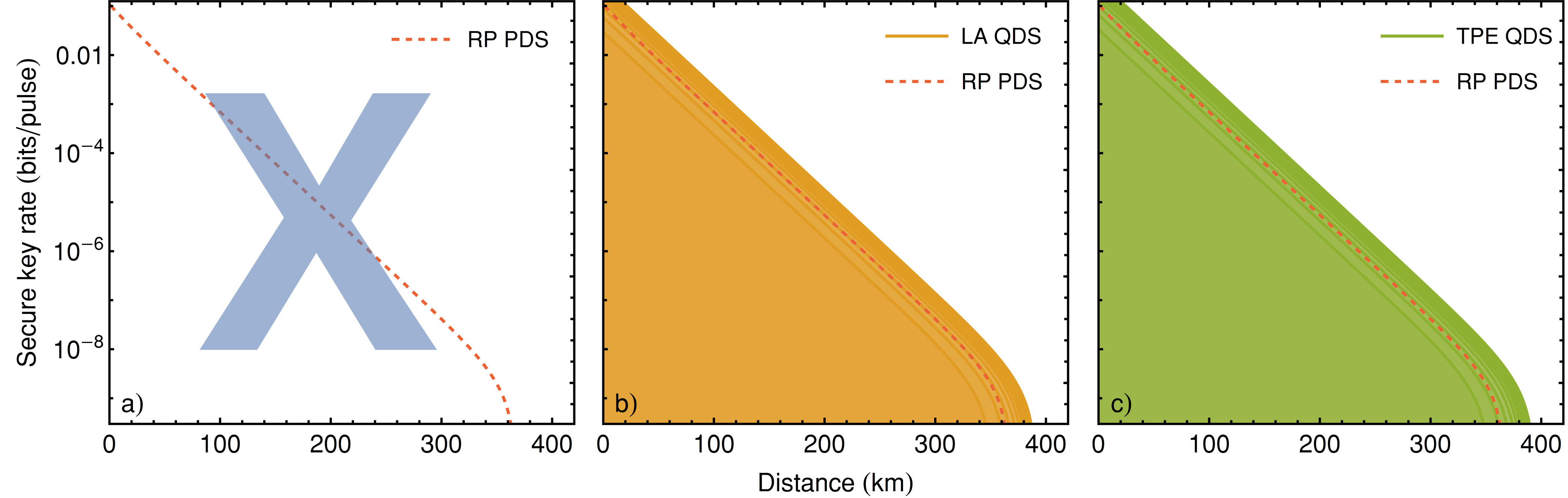}
		\caption{\textbf{Collection efficiency comparison for BB84 QKD with decoy states.} Simulated secret key rates from Eq. (\ref{eq:skrbb84}) as a function of distance for (a) RE QDS (b) LA QDS (c) TPE QDS, with collection efficiencies ranging from $\eta=1\%$ (bottom curves) to $\eta=100\%$ (top curves), in steps of $10\%$. The optimal performance of randomized-phase (RP) PDS is also plotted in dashed lines, in order to identify which QDS collection efficiencies are required to overcome PDS for each pumping. Single mode telecom fiber losses of $0.21$ dB/km are assumed. Parameters for all plots are: alignment error rate $e_d=2\%$, dark count probability $Y_0 =10^{-9}$, detection efficiency $\eta_d=100\%$, and error-correcting code inefficiency $f=1.2$. }
		\label{fig:bb84_decoy_collection}
	\end{center}
\end{figure*}

In order to compare the performance of all sources for BB84 QKD, we study the protocol with and without the decoy state countermeasure (see Section \ref{sec:decoy} for details), assuming one-way classical post-processing \cite{KGR:PRL05}. Without decoy states, we plot the secure key rate per pulse as a function of source efficiency, collection efficiency, and distance in Fig. \ref{fig:bb84}. We then display the performance of QDS pumping schemes for collection efficiencies ranging from $1\%$ to $100\%$ in Fig. \ref{fig:bb84_collection}, and compare these to the best performance of randomized-phase PDS. We then proceed to similar plots for BB84 QKD with decoy states, in Figs. \ref{fig:bb84_decoy} and \ref{fig:bb84_decoy_collection}. In general, the secure key rate as a function of source efficiency reaches a maximum for PDS, after which the multiphoton component becomes too significant. Regarding QDS, the key rate evolves almost linearly with source efficiency, since the vacuum and single photon components dominate at all source efficiencies.

\subsection{Twin-field quantum key distribution}

\subsubsection{Brief introduction}

 \noindent Twin-field QKD (TF-QKD) was proposed as a new protocol configuration to overcome the repeaterless rate-distance limit of standard QKD \cite{LYD:Nature18}. By delegating the measurement setup to a third untrusted party situated halfway between Alice and Bob, the optical fields sent by each party travel only half the communication distance of standard QKD. Since the key bits are extracted from the resulting single-photon interference at Charlie's station, the secure key rate scales with the square root of the channel transmittance instead of scaling linearly.
 
 \subsubsection{Results}
 
\begin{figure*}
	\begin{center}
		\includegraphics[width=180mm]{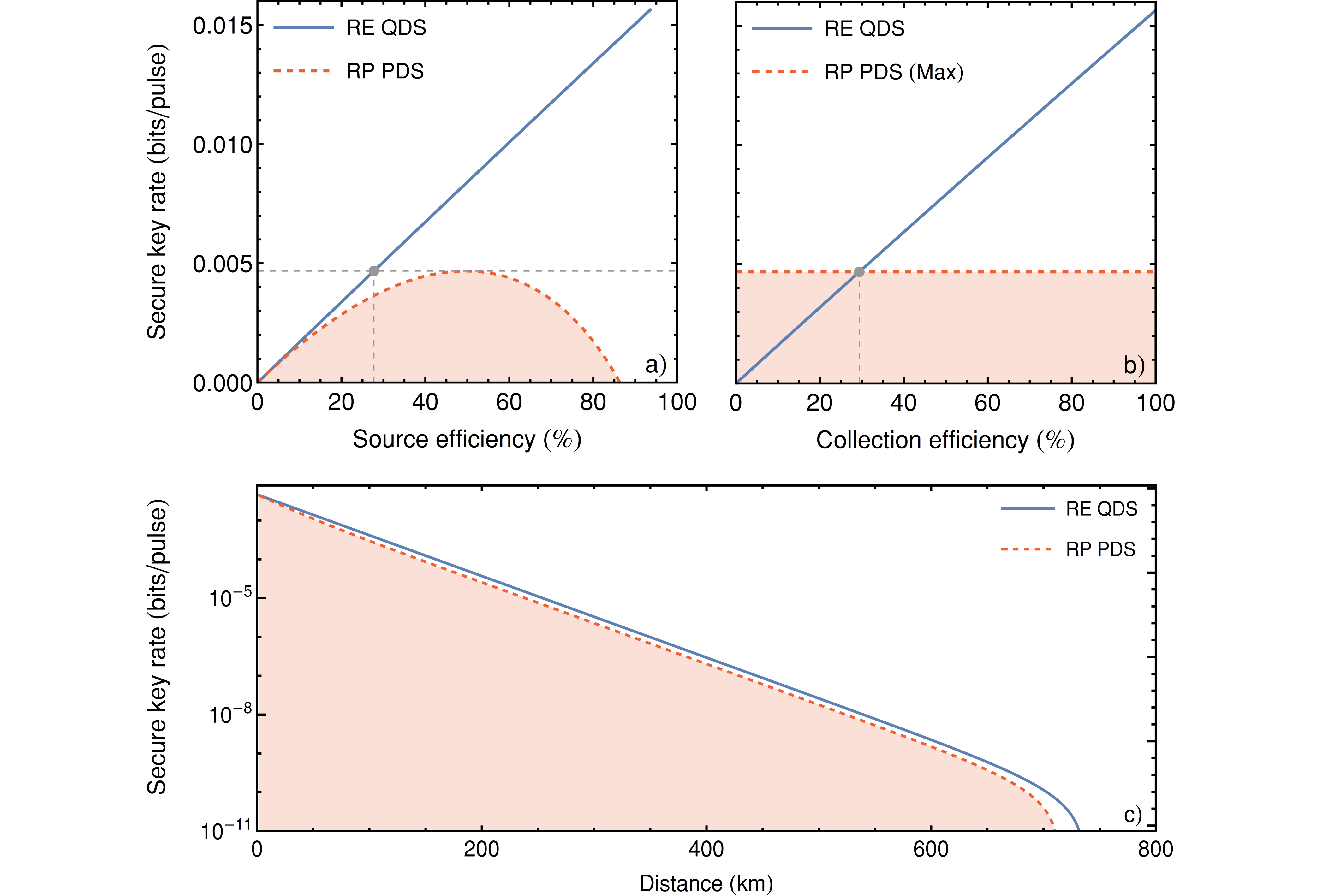}
		\caption{\textbf{Source comparison for twin-field QKD.} (a) Simulated secret key rates from Eq. (\ref{eq:skrtf}) as a function of source efficiency for RE QDS, along with randomized-phase (RP) PDS. Source efficiency is defined as $1-e^{-\mu}$ for PDS and $1-\sum_{n=0}^\infty p_n(1-\eta)^n$ for QDS, where $\eta$ is the QDS collection efficiency. Chosen pulse lengths, pulse areas, and photon number populations $\{p_n\}$ are displayed in Table \ref{tab:pumpy}. (b) Simulated secret key rates as a function of QDS collection efficiency, compared to the best performance of RP PDS sources (dashed line). (c) Simulated secret key rates as a function of distance, assuming single mode telecom fiber losses of $0.21$ dB/km. The QDS collection efficiencies were chosen as the intersection points from Fig (b). Parameters for all plots are: alignment error rate $e_d=2\%$, dark count probability $Y_0 =10^{-9}$, detection efficiency $\eta_d=100\%$, and error-correcting code inefficiency $f=1.2$.   }
		\label{fig:twinfield}
	\end{center}
\end{figure*}

\begin{figure*}
	\begin{center}
		\includegraphics[width=180mm]{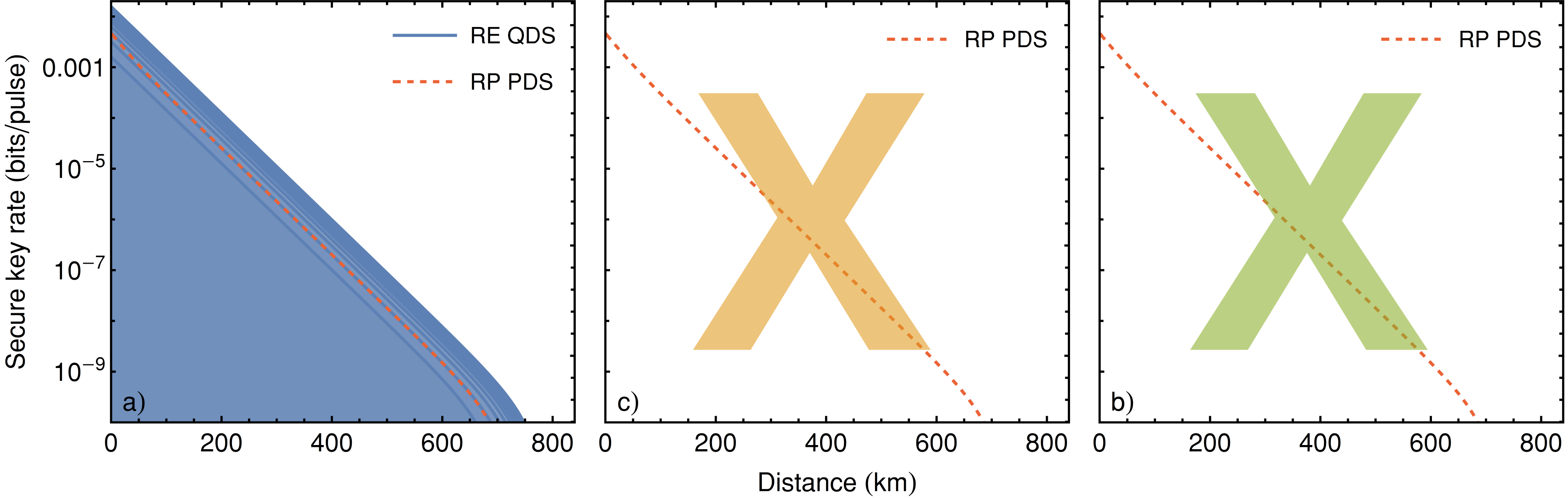}
		\caption{\textbf{Collection efficiency comparison for twin-field QKD.} Simulated secret key rates from Eq. (\ref{eq:skrtf}) as a function of distance for (a) RE QDS (b) LA QDS (c) TPE QDS, with collection efficiencies ranging from $\eta=1\%$ (bottom curves) to $\eta=100\%$ (top curves), in steps of $10\%$. The optimal performance of randomized-phase (RP) PDS is also plotted in dashed lines, in order to identify which QDS collection efficiencies are required to overcome PDS for each pumping. Single mode telecom fiber losses of $0.21$ dB/km are assumed. Parameters for all plots are: alignment error rate $e_d=2\%$, dark count probability $Y_0 =10^{-9}$, detection efficiency $\eta_d=100\%$, and error-correcting code inefficiency $f=1.2$. }
		\label{fig:twinfield_collection}
	\end{center}
\end{figure*}

 In order to compare the performance of all sources for twin-field QKD, we plot the secure key rate from Eq. (\ref{eq:skrtf}) as a function of source efficiency, collection efficiency, and distance in Fig. \ref{fig:twinfield}. We then display the performance of QDS pumping schemes for collection efficiencies ranging from $1\%$ to $100\%$ in Fig. \ref{fig:twinfield_collection}, and compare these to the best performance of randomized-phase PDS.
 
\subsection{Unforgeable quantum tokens}\label{sec:descriptionqmoney}

\subsubsection{Brief introduction}

This primitive, in its private-key form, allows a central authority to issue tokens comprising of quantum states, whose unforgeability is intrinsically guaranteed by the no-cloning theorem. One famous application is quantum money, which can prevent banknote forgery \cite{Wie:acm83}, double-spending with credit cards \cite{BDG:PRA19,BOV:npj18}, and also guarantee features such as user privacy \cite{Kent:npjQI22}. 

In a private-key scheme, the quantum state is encoded according to a secret pre-shared classical key, which is known by the central authority and the verifier(s) only. The key contains a sequence of secret information bits, as well as a sequence of secret basis bits, which
indicate the random preparation basis of each information bit (the states used are the standard BB84 states from QKD). This ensures that a dishonest client willing to duplicate the money state will introduce errors in at least one
of two states, due to no-cloning. Upon verification, these errors will be detected by the verifier(s), who measure(s) each sub-system of the money state in the correct
basis and compares the measurement outcomes with the secret key.

There exist different forms of quantum token protocols, from private-key to public-key, with quantum verification \cite{Wie:acm83} or classical verification \cite{BDG:PRA19}. Some schemes require quantum storage \cite{BOV:npj18,GAA:pra18}, while others replace this requirement with no-signalling constraints \cite{Kent:PRA20}. Here, we focus on private-key quantum token schemes with quantum verification that assume quantum storage, which can provide information-theoretic security for unforgeability. 

\subsubsection{Results}

The exact protocol considered here is described in \cite{BDG:PRA19}, and we extend its security analysis to the quantum dot framework in Section \ref{sec:qtokens}.

\begin{figure*}
	\begin{center}
		\includegraphics[width=180mm]{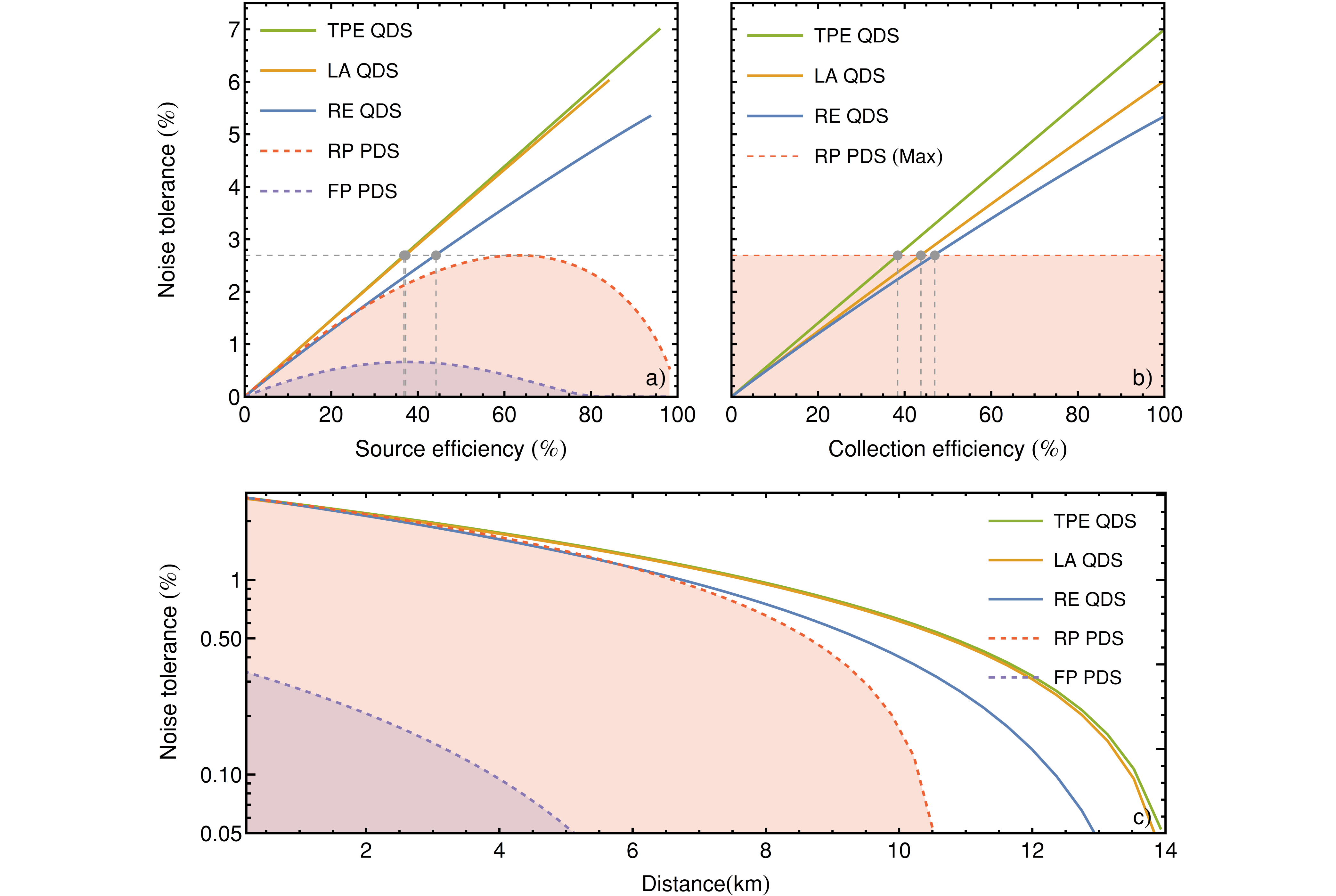}
		\caption{\textbf{Source comparison for unforgeable quantum tokens.} (a) Numerical optimal noise tolerance from Eq. (\ref{eq:noisetol}) as a function of source efficiency for RE, LA and TPE QDS, along with fixed-phase (FP) and randomized-phase (RP) PDS. Source efficiency is defined as $1-e^{-\mu}$ for PDS and $1-\sum_{n=0}^\infty p_n(1-\eta)^n$ for QDS, where $\eta$ is the QDS collection efficiency. Chosen pulse lengths, pulse areas, and photon number populations $\{p_n\}$ are displayed in Table \ref{tab:pumpy}. RE photonic states were assumed to be maximally pure in number basis, expressed as $\sum_{n=0}^\infty \sqrt{p_n}\ket{n}$, while LA states were expressed as diagonal states $\sum_{n=0}^\infty p_n\ket{n}\bra{n}$. (b) Numerical noise tolerance as a function of QDS collection efficiency, compared to the best performance of PDS  sources (dashed line). (c) Numerical noise tolerance plotted as a function of distance, assuming single mode telecom fiber losses of $0.21$ dB/km. The QDS collection efficiencies were chosen as the intersection points from Fig (b). }
		\label{fig:qtoken}
	\end{center}
\end{figure*}

\begin{figure*}
	\begin{center}
		\includegraphics[width=180mm]{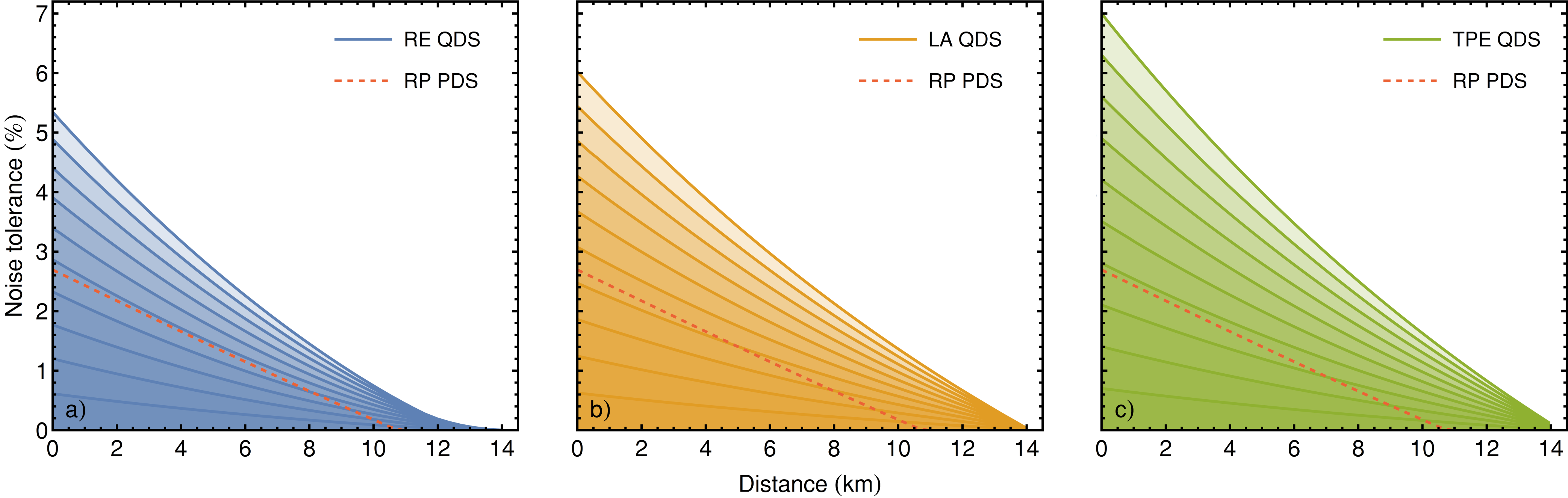}
		\caption{\textbf{Collection efficiency comparison for unforgeable quantum tokens.} Numerical optimal noise tolerance from Eq. (\ref{eq:noisetol}) as a function of distance for (a) RE QDS (b) LA QDS (c) TPE QDS, with collection efficiencies ranging from $\eta=10\%$ (bottom curves) to $\eta=100\%$ (top curves), in steps of $10\%$. The optimal performance of randomized-phase (RP) PDS is also plotted in dashed lines, in order to identify which QDS collection efficiencies are required to overcome PDS for each pumping. Single mode telecom fiber losses of $0.21$ dB/km are assumed.}
		\label{fig:qtoken_collection}
	\end{center}
\end{figure*}

In order to compare the performance of all sources for unforgeable quantum tokens, we solve problem (\ref{eq:noisetol}) from Section \ref{sec:qtokens} numerically using the MATLAB cvx package with solver SDPT3. In this way, we plot the evolution of the noise tolerance as a function of source efficiency, collection efficiency, and distance in Fig. \ref{fig:qtoken}. We then display the performance of all three QDS pumping schemes for collection efficiencies ranging from $1\%$ to $100\%$ in Fig. \ref{fig:qtoken_collection}, and compare these to the best performance of randomized-phase PDS.

Naturally, PDS reach a maximal noise tolerance for source efficiencies around $63\%$, corresponding to $\mu\approx 1$, before dropping again when the multiphoton contribution becomes too significant. For QDS, we notice a striking difference between schemes with coherence (RE) and those without (LA and TPE): the latters give an overhead of almost $2\%$ on the noise tolerance with respect to RE at high source efficiencies. This difference is crucial in making implementations feasible, since boosting the fidelity of quantum state preparation and quantum storage by a few percent can be extremely challenging. These differences are also reflected in Fig. \ref{fig:qtoken}.b., which identifies the collection efficiencies at which QDS can outperform the best PDS performance: while LA and TPE require $44\%$ and $38\%$, respectively, RE must be pushed to $47\%$ to beat PDS. For information purposes, we also select three state-of-the art experimental QDS, and show how they would perform in such a beyond-QKD protocol with their reported values of brightness and single-photon purity. 

Fig. \ref{fig:qtoken}.c. finally compares the performance of each source as a function of distance. Once again, the difference between LA/TPE and RE is significant due to the coherence feature. We notice here that the maximal distance for all sources is much shorter than in QKD schemes, since our selected quantum token scheme bears a maximal loss tolerance of $50\%$: above this limit, an adversary can clone the quantum token without introducing any errors \cite{BDG:PRA19}. 

\subsection{Quantum strong coin flipping}

\subsubsection{Brief introduction}

Strong coin flipping (SCF) allows two distant parties, Alice and Bob, to generate and agree on a random bit. They do not trust each other and wish to ensure that the bit is truly random.  We call the coin flip \textit{fair} when two honest parties each win with probability $1/2$. On the other hand, security for this task must guarantee that none of the two parties can force the other to declare outcome $i\in\{0,1\}$ with probability higher than $P=\frac{1}{2}+\epsilon^{(i)}$, where $\epsilon^{(i)}$ is the protocol \textit{bias}. In its most general form, SCF does not necessarily involve equal cheating probabilities for both parties, but when it does, the protocol is labelled \textit{balanced}. We define the following upper bounds on Alice and Bob's probabilities of forcing their opponent to declare outcome $i$:

\begin{equation}
    \begin{aligned}
    &P_{A}^{(i)}\leqslant\frac{1}{2}+ \epsilon_A^{(i)} \:\:\:\:\:\:\:\:\:\: \text{Alice forces Bob to declare $i$}\\
    &P_{B}^{(i)}\leqslant\frac{1}{2}+ \epsilon_B^{(i)} \:\:\:\:\:\:\:\:\:\: \text{Bob forces Alice to declare $i$}\\
    \end{aligned}
\end{equation}
The bias $\epsilon$ of a given SCF protocol is then defined as the highest of all four biases:

\begin{equation}
   \epsilon= \max\Bigg\{\epsilon_A^{(0)},\epsilon_A^{(1)},\epsilon_B^{(0)},\epsilon_B^{(1)}\Bigg\}.
\end{equation}

 \noindent Information-theoretic strong coin flipping with arbitrarily small bias cannot be reached with quantum mechanics alone \cite{K:PC03,IEEE:CK09}, but requires additional space-time constraints \cite{BBB:PRA09}, which can be experimentally challenging. It was shown, however, that even without such constraints, quantum mechanics can provide strong coin flipping protocols that perform better than any classical coin flipping protocol with information-theoretic security (in terms of bias) \cite{HW:TCC11}. This led to experimental demonstrations of quantum strong coin flipping, namely \cite{PJ+:natcomm14,BBB:NC11,MTV:PRL05}.

In quantum coin flipping, the states generated by Alice are usually not encoded in the standard way of Fig. \ref{fig:encoding} (i.e. they are not an extension of the BB84 states $\{\ket{+},\ket{-},\ket{+i},\ket{-i}\}$ to infinite Hilbert spaces). Instead, the four coin flipping states must allow for an extra free parameter $y$, which will be varied to guarantee a fair (resp. balanced) coin flip, i.e. a coin flip in which Alice and Bob have equal honest (resp. dishonest) winning probabilities. 

The required states in a two-dimensional qubit space spanned by $\{\ket{v_0},\ket{v_1}\}$ are the following:

\begin{equation}
    \begin{aligned}
    \ket{\Phi^{\left(y\right)}_{\alpha, 0}} = \sqrt{y}\ket{v_0}+(-1)^{\alpha}\sqrt{1-y}\ket{v_1},\\
    \ket{\Phi^{\left(y\right)}_{\alpha, 1}} = \sqrt{1-y}\ket{v_0}-(-1)^{\alpha}\sqrt{y}\ket{v_1},
    \end{aligned}
\end{equation}
where $\alpha\in\{0,1\}$ denotes the encoding basis. In order to extend them to the full Hilbert space required in photonic setups, we simply change the two $H_{01}^{(1/2)}$ beamsplitter transformations from Fig. \ref{fig:encoding} to  $H_{01}^{(y)}$. We then reproduce the workings from Eqs. (\ref{eq:encoding}), (\ref{eq:numbercoherence}) and (\ref{eq:diagonal}) with these new coefficients. The resulting coin flipping states are labelled as $\{\sigma_{\alpha,0}^{\left(y,\eta,\phi\right)},\sigma_{\alpha,1}^{\left(y,\eta,\phi\right)}\}$, to account for PDS average photon number or QDS collection efficiency $\eta$, and phase encoding $\phi$. 

\subsubsection{Results}

 We focus here on the quantum protocol from \cite{PJ+:natcomm14}, and additionally study the effect of photon number coherence on the protocol bias in the security proof (see Section \ref{sec:coinflipping} for details).
 
\begin{figure*}
	\begin{center}
		\includegraphics[width=180mm]{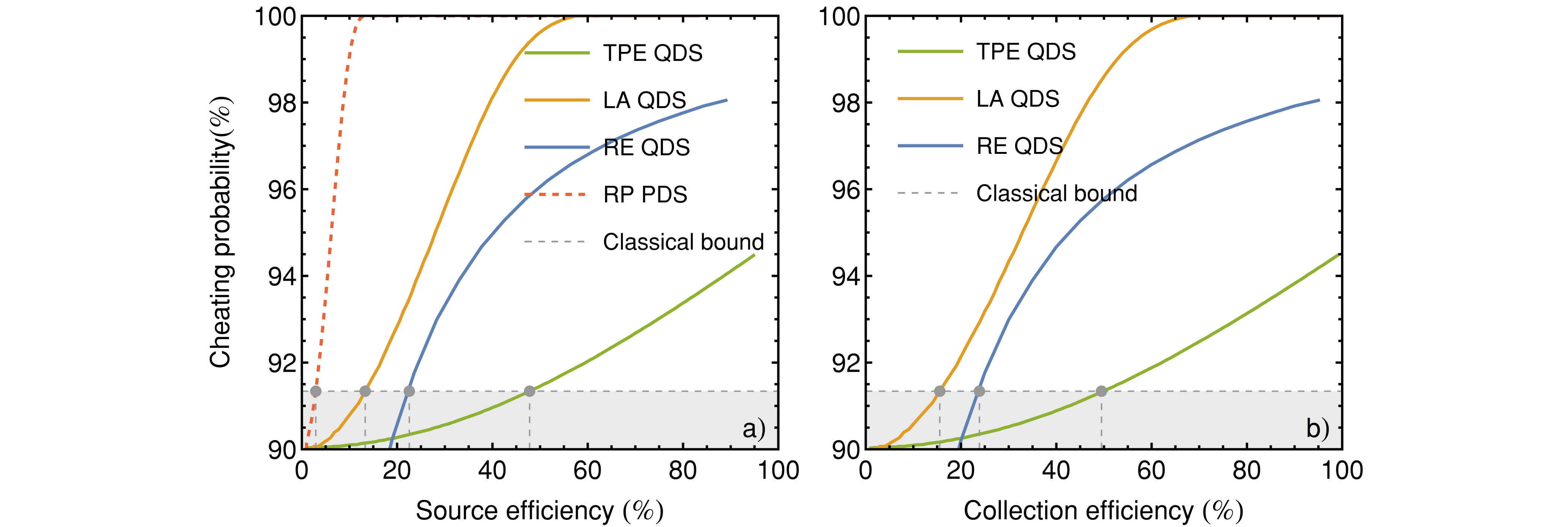}
		\caption{\textbf{Source comparison for quantum strong coin flipping.} Cheating probability as a function of (a) source efficiency and (b) collection efficiency for a balanced protocol with number of states $N=1000$, using RE, LA, TPE QDS, and randomized-phase (RP) PDS. The best achievable classical bound from Eq. (\ref{eq:quantumadvantage}) is plotted in gray dashed lines, for an error rate of $e=1.5\%$.  Source efficiency is defined as $1-e^{-\mu}$ for PDS and $1-\sum_{n=0}^\infty p_n(1-\eta)^n$ for QDS, where $\eta$ is the QDS collection efficiency. Chosen pulse lengths, pulse areas, and photon number populations $\{p_n\}$ are displayed in Table \ref{tab:pumpy}. RE photonic states were assumed to be maximally pure in number basis, expressed as $\sum_{n=0}^\infty \sqrt{p_n}\ket{n}$, while LA states were expressed as diagonal states $\sum_{n=0}^\infty p_n\ket{n}\bra{n}$. As discussed in Section \ref{sec:coinflipping}, note that the upper bounds on the cheating probability for LA and XX are general and may be over-estimated, while the upper bound for RE considers a specific attack only (i.e., the cheating probability may in fact be higher).  }
		\label{fig:coin}
	\end{center}
\end{figure*}

 We show in Fig. \ref{fig:coin} how the cheating probability evolves as a function of source and collection efficiencies in a balanced protocol (i.e., a protocol in which Alice and Bob have equal cheating probabilities). For the purpose of this example, we fix the number of states to $N=1000$, and calculate the subsequent honest abort probability:

\begin{equation}
    \mathcal{P}_{ab} = Z + \left(1-Z\right)\frac{e}{2},
\end{equation}
where $Z$ is the probability that Honest Bob does not register any click after the $N$ states have been sent, and $e$ is the quantum error rate. Using the results derived in \cite{BMH:NatComms14,HW:TCC11}, we deduce the best achievable classical bound, thus identifying where QDS and PDS allow for quantum advantage in terms of cheating probability. For balanced, strong coin flipping protocols, the optimal classical bound reads:

\begin{equation}
    \mathcal{P}_{c} = 1-\sqrt{\frac{\mathcal{P}_{ab}}{2}}.
    \label{eq:quantumadvantage}
\end{equation}

 As can be seen from Fig. \ref{fig:coin}, the cheating probability for RE is under-estimated, since we have only considered a specific discrimination attack in our security proof (Section \ref{sec:coinflipping}). The bounds for no coherence (LA/TPE), on the other hand, are over-estimated, since we have used the framework from \cite{PJ+:natcomm14}, which does not provide tight bounds on the cheating probability. The combination of both features does not allow us to directly conclude that LA performs better than RE. However, for comparable brightness and purity, we know that it should perform better than RE, since the optimal attacks in quantum strong coin flipping protocols involve some form of discrimination, which is always more successful in the presence of photon-number coherence. 
 
Note that we have not plotted the curves for fixed-phase PDS. For LA, TPE, and RE QDs, the dimension of the Hilbert space is upper-bounded (since $p_n = 0$ for $n>=3$). For randomized-phase PDS, the required dimension is also upper-bounded, since perfect cheating is assumed for higher photon number terms. On the other hand, deriving the bounds for fixed-phase PDS would imply solving the semidefinite program from Eq. (\ref{eq:maxBob}) in an infinite-dimensional Hilbert space.

\begin{figure*}
	\begin{center}
		\includegraphics[width=180mm]{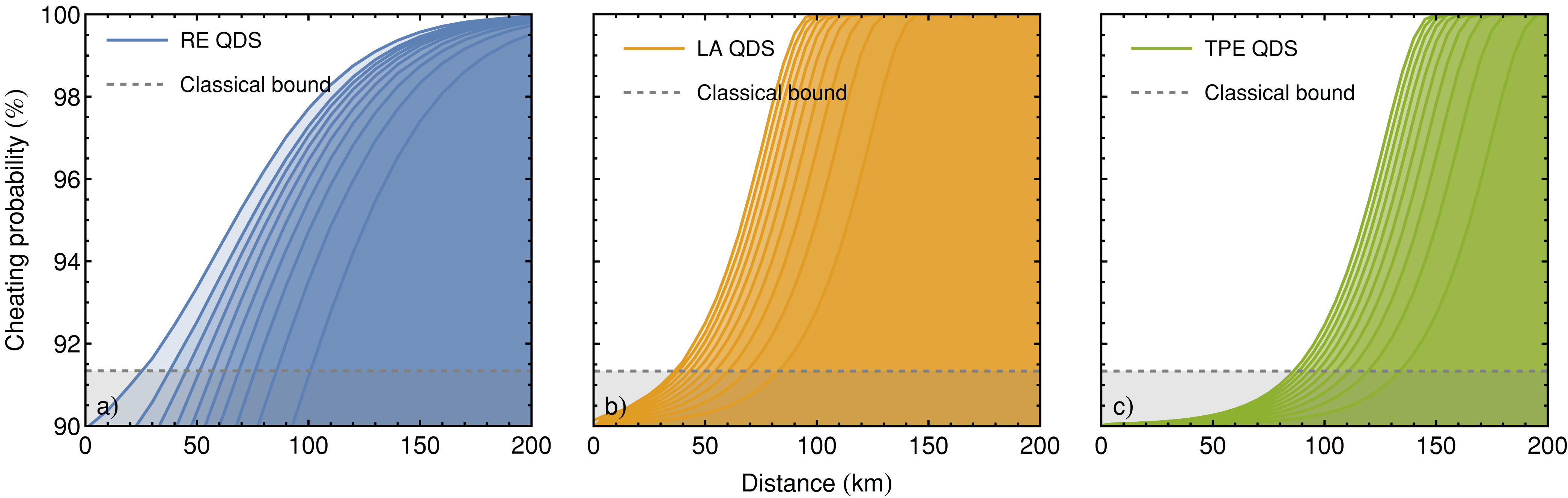}
		\caption{\textbf{Collection efficiency comparison for quantum strong coin flipping.} 
		Cheating probability as a function of distance for a balanced protocol using (a) RE, (b) LA, and (c) TPE QDS. The honest abort probability $\mathcal{P}_{ab}=2.5\%$ (of which $1.5\%$ come from the error rate), and the number of states $N$ are varied for each point to achieve $\mathcal{P}_{ab}$. The best achievable classical bound from Eq. (\ref{eq:quantumadvantage}) is plotted in gray dashed lines. Collection efficiencies ranging from $\eta= 10\%$ to $\eta= 100\%$ are plotted from right to left, in steps of $10\%$. Single mode telecom fiber losses of $0.21$dB/km are assumed. }
		\label{fig:coinlosses}
	\end{center}
\end{figure*}

In Fig. \ref{fig:coinlosses}, we show how the cheating probability evolves as a function of distance for various collection efficiencies and a fixed honest abort probability of $\mathcal{P}_{ab} =2.5\%$. Since, for PDS, a given abort probability can be achieved by varying either the number of states $N$ or the average photon number per pulse $\mu$, we choose $N$ to be equal to the number of states required for the best performing quantum dot scheme (TPE). Essentially, the number of states $N$ dictates the time duration of the protocol (for a fixed repetition rate), so the plots compare QDS to PDS performance for a fixed protocol duration, and fixed abort probability.

Here, we note the striking feature that QDS perform better at lower collection efficiencies, due to lower multiphoton component. This is in contrast with QKD, in which the main figure of merit is key rate, and the trade-off between brightness and single-photon purity is more crucial. 

Interestingly, we have not considered the weak coin flipping protocol from \cite{BCK:PRA20} in our work, since the information there is not encoded onto the phase between various photon number terms (as is the case in our polarization, time-bin or path-encoded primitives), but rather onto the photon number itself. This means that having a phase reference between the photon number terms should not leak any information to an adversary, since it does not reveal whether the state was a vacuum state or a 1-photon Fock state. It therefore should not matter whether one uses RE, LA or TPE pumping schemes.

\subsection{Quantum bit commitment}

\subsubsection{Brief introduction}

A bit commitment protocol consists of two phases: the \textit{commit} phase and the \textit{open} phase. In the commit phase, Honest Alice chooses a bit $b\in\{0,1\}$ and provides Honest Bob with some form of evidence that she has committed to this choice. In the open phase, which happens some time after the commit phase, Honest Alice reveals $b$ to Honest Bob. The desired security features are the following:

\begin{itemize}
    \item Dishonest Alice cannot change $b$ after the commit phase,
    \item Dishonest Bob cannot access $b$ before the open phase.
\end{itemize}

Just like its strong coin flipping counterpart, the quantum version of bit commitment cannot provide perfect information-theoretic security for both parties without additional spacetime constraints \cite{M:PRL97,LC:PRL97}. However, it is possible to circumvent this no-go theorem for two-party computations \cite{M:PRL97,LC:PRL97} by placing restrictions on Dishonest Bob's storage capabilities. 

We focus on the quantum protocol from \cite{NJC:NatComms12}, secure under a bounded storage assumption. During the commit phase, Honest Alice generates $N$ BB84 states similarly to the QKD and quantum token protocols from Sections \ref{sec:decoydescription} and \ref{sec:descriptionqmoney}, and Honest Bob performs random X or Z measurements on each state. Both parties wait a pre-agreed amount of time $\Delta t$, during which it is assumed that a Dishonest Bob may only store and retrieve $S$ of the $N$ quantum states sent by Alice. After waiting $\Delta t$, Honest Alice sends her preparation basis for each of the $N$ states to Honest Bob, who compares them with his measurement basis choices. An error correction and privacy amplification subroutine is then performed between the two parties. During the open phase, Honest Alice reveals the encoding of each of the $N$ states, along with her committed bit $b$. Honest Bob performs consistency checks and accepts or rejects the commitment depending on the outcome. 

\subsubsection{Results}

\begin{figure*}
	\begin{center}
		\includegraphics[width=180mm]{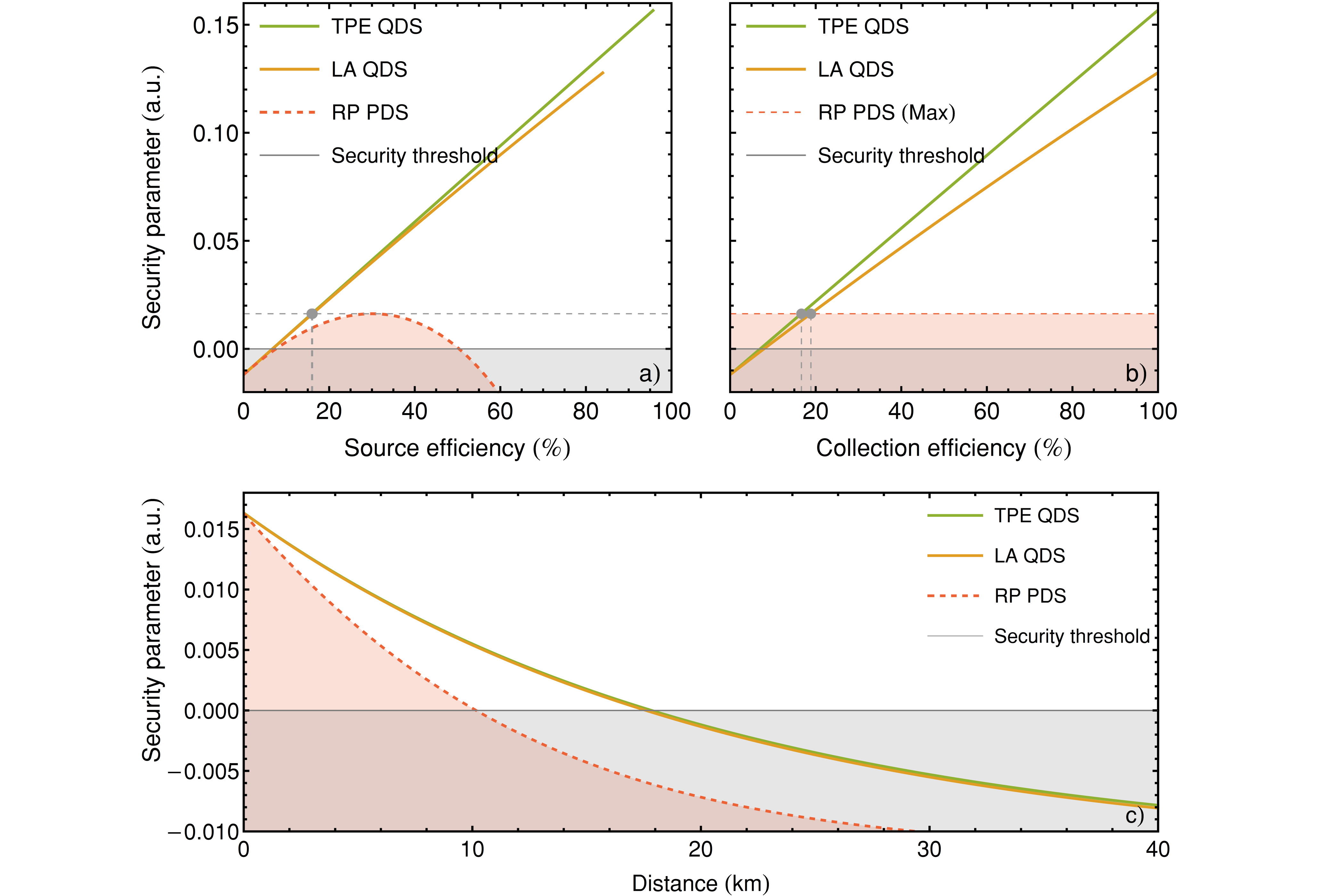}
		\caption{\textbf{Source comparison for quantum bit commitment under the bounded storage assumption.} (a) Security condition from Eq. (\ref{eq:bitcommcondition}) as a function of source efficiency for LA and TPE QDS, along with randomized-phase (RP) PDS. Source efficiency is defined as $1-e^{-\mu}$ for PDS and $1-\sum_{n=0}^\infty p_n(1-\eta)^n$ for QDS, where $\eta$ is the QDS collection efficiency. Chosen pulse lengths, pulse areas, and photon number populations $\{p_n\}$ are displayed in Table \ref{tab:pumpy}. (b) Security condition from Eq. (\ref{eq:bitcommcondition}) as a function of QDS collection efficiency, compared to the best performance of RP PDS sources (dashed line). (c)Security condition from Eq. (\ref{eq:bitcommcondition}) as a function of distance, assuming single mode telecom fiber losses of $0.21$ dB/km. The QDS collection efficiencies were chosen as the intersection points from Fig (b). The error rate is $e=2\%$, detection efficiency $\eta_d=100\%$, and the chosen parameters for Eq. (\ref{eq:bitcommcondition}) are $\epsilon = 2\times 10^{-5}$, $\beta = 0.007$ and $\gamma = 0.008$.}
		\label{fig:bitcomm}
	\end{center}
\end{figure*}

\begin{figure*}
	\begin{center}
		\includegraphics[width=180mm]{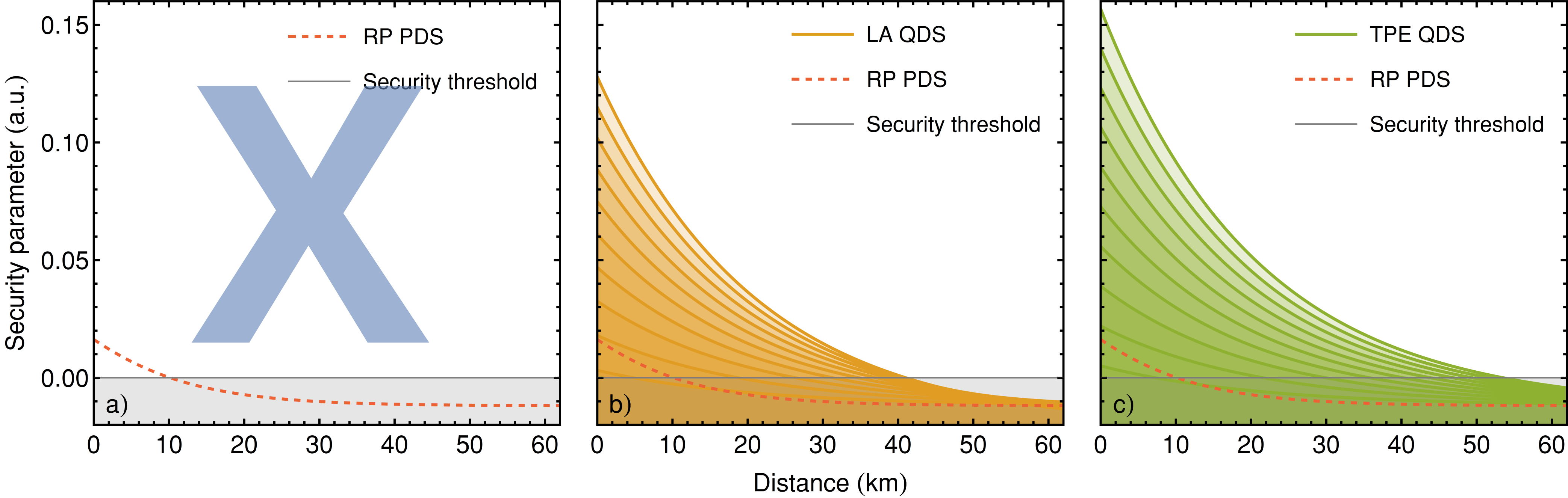}
		\caption{\textbf{Collection efficiency comparison for quantum bit commitment under the bounded storage assumption.} Security condition from Eq. (\ref{eq:bitcommcondition}) as a function of distance for (a) RE QDS (b) LA QDS (c) TPE QDS, with collection efficiencies ranging from $\eta=1\%$ (bottom curves) to $\eta=100\%$ (top curves), in steps of $10\%$. The optimal performance of randomized-phase (RP) PDS is also plotted in dashed lines, in order to identify which QDS collection efficiencies are required to overcome PDS for each pumping. Single mode telecom fiber losses of $0.21$ dB/km are assumed.  The error rate is $e=2\%$, detection efficiency $\eta_d=100\%$, and the chosen parameters for Eq. (\ref{eq:bitcommcondition}) are $\epsilon = 2\times 10^{-5}$, $\beta = 0.007$ and $\gamma = 0.008$. }
		\label{fig:bitcomm_collection}
	\end{center}
\end{figure*}

In Figs \ref{fig:bitcomm} and \ref{fig:bitcomm_collection}, we plot the security condition from Eq. (\ref{eq:bitcommcondition}) presented in Section \ref{sec:bitcommitment}. Similarly to \cite{NJC:NatComms12}, we assume that Alice sends $N = 10^8$ states, and that Dishonest Bob's storage size is $S = 972$. Once again, the security condition as a function of source efficiency reaches a maximum for PDS, after which the multiphoton component becomes too significant. Regarding QDS, the condition evolves almost linearly with source efficiency, since the vacuum and single photon components dominate at all source efficiencies.

\section{Mathematical tools for quantum cryptography}\label{sec:tools}

\subsection{Semidefinite programming}

 \noindent Quantum theory relies on linear algebra. In quantum cryptography, security analyses often involve optimizing over semidefinite positive objects to find the adversary's optimal cheating strategy. Most of the time, these objects are density matrices, measurement operators, or more general completely positive trace-preserving (CPTP) maps. Semidefinite programming provides a suitable framework for this, as it allows to optimize over semidefinite positive variables, given linear constraints.

A semidefinite program may be defined as a triple $\left(\Lambda,F,C\right)$ where $\Lambda$ is a Hermitian-preserving CPTP map, and $F$ and $C$ are Hermitian operators living in complex Hilbert spaces $\mathcal{H}_F$ and $\mathcal{H}_C$, respectively.

We start by defining a maximization problem, which will serve as our \textit{primal problem}. The primal problem maximizes a \textit{primal objective function}, $\Tr\left(F^{\dagger}X\right)$,  over all positive semidefinite variables $X$, given a set of linear constraints expressed as a function of $C$:

\begin{equation}
\begin{aligned}
\text{maximize} &&& \Tr\left(F^{\dagger}X\right)\\
\text{s.t.}  &&& \Lambda(X) = C\\
&&& X \geqslant 0.
\end{aligned}\label{primal}
\end{equation}
If it exists, the operator $X$ which maximizes $\Tr\left(F^{\dagger}X\right)$ given these constraints is the \textit{primal optimal solution}, and the corresponding value of $\Tr\left(F^{\dagger}X\right)$ is the  \textit{primal optimal value}.

Semidefinite programs present an elegant dual structure, which associates a dual minimization problem to each primal maximization problem. Effectively, the new variable(s) of the dual problem may be understood as the Lagrange multipliers associated with the constraints of the primal problem (one for each constraint). 

The dual problem associated with (\ref{primal}) reads:

\begin{equation}
\begin{aligned}
\text{minimize} &&& \Tr\left(C^{\dagger}Y\right)\\
\text{s.t.}  &&& \Lambda^*(Y) - F \geqslant 0\\
&&& Y = Y^{\dagger}.
\end{aligned}\label{duall}
\end{equation}
 Similarly to the primal problem, the operator $Y$ which minimizes $\Tr\left(C^{\dagger}Y\right)$ given these constraints, if it exists, is the \textit{dual optimal solution}, and the corresponding value of $\Tr\left(C^{\dagger}Y\right)$ is the  \textit{dual optimal value}.

The Lagrange multiplier method allows to find the local extremum of a constrained function. The optimal value $s_p$ of the primal problem therefore upper bounds the optimal value $s_d$ of the dual problem, while the optimal value of the dual lower bounds that of the primal. This property is known as \textit{weak duality}, and may be simply expressed as:

\begin{equation}
    s_p\leqslant s_d.
\end{equation}
In many quantum-cryptographic applications however, we wish to ensure that the upper bound derived in the primal problem is \textit{tight}, i.e. that the local maximum is in fact a global maximum for the objective function. The dual problem will help to prove this when there exists \textit{strong duality}:

\begin{equation}
    s_p= s_d.
\end{equation}

\subsection{Choi's theorem on completely positive maps}

 \noindent Let us consider a tensor product of two $d$-dimensional Hilbert spaces $\mathcal{H}=\mathcal{H}_1^d\otimes\mathcal{H}_2^d$, and then define the maximally entangled state $\ket{\Phi^{+}}\bra{\Phi^{+}}$ on $\mathcal{H}$ as
\begin{equation}
\ket{\Phi^{+}}\bra{\Phi^{+}} = \frac{1}{d}\sum_{i,j=1}^{d} \ket{i}\bra{j}\otimes\ket{i}\bra{j}
\end{equation}
We introduce a completely positive linear map $\Lambda : \mathcal{H}_1^d \rightarrow \mathcal{H}_3^{d'}$, and define the Choi-Jamiolkowski operator $J(\Lambda) : \mathcal{H}_1^d\otimes\mathcal{H}_2^d \rightarrow \mathcal{H}_3^{d'}\otimes\mathcal{H}_2^d$ as the operator which applies $\Lambda$ to the first half of the maximally entangled state $\ket{\Phi^{+}}\bra{\Phi^{+}}$:
\begin{equation}
J(\Lambda) = \frac{1}{d}\sum_{i,j=1}^{d} \Lambda(\ket{i}\bra{j})\otimes\ket{i}\bra{j}.
\label{eq:choi}
\end{equation}
Choi's theorem then states that $\Lambda$ is completely positive if and only if
$J(\Lambda)$ is positive semidefinite. We also have that $\Lambda$ is a trace-preserving map if and only if $\Tr_{\mathcal{H}_3^{d'}}(J(\Lambda)) = \mathbb{1}_{\mathcal{H}_2^d}$ \cite{MVW:tqc12,W:LN11,VB:SIAM96}. These properties are implemented as constraints in the optimization problem from Eq. (\ref{eq:noisetol}).
%TODO find ref !

\section{Security of BB84 quantum key distribution}\label{sec:decoy}

\subsection{With Poisson-distributed sources}

 \noindent In the majority of QKD implementations so far, highly attenuated laser states are used instead of single photons \cite{Pan:RevMod20}. New techniques, such as the insertion of decoy states into the protocol, have been developed to provide significant secure key rates despite the presence of multiphoton noise \cite{GLLP04,LMC:PRL05}. We note that the derivation of the secure key rate in this setting assumes that the states sent by Alice bear no coherence in Fock basis, in order to satisfy the photon number channel assumption \cite{LMC:PRL05,Ma:Thesis08}. This implies that the global phase of each state must be actively randomized, where the random phases are chosen from a given set of $m$ phases $\mathcal{S}_m\in\left[0,2\pi\right)$. In this work, we assume $m\rightarrow\infty$, but note that security proofs with discrete phase randomization also exist \cite{CZL:NJP15}.  

We briefly recall the workings from \cite{GLLP04,LMC:PRL05,Ma:Thesis08} for practical BB84 QKD without and with infinite decoy states, respectively. Let us define the yield $Y_k$ of a $k$-photon state, which gives the conditional probability of a detection on Bob's detector given that Alice generates a $k$-photon state:
\begin{equation}
    Y_k = Y_0 + \left(1-Y_0\right) \left[1-\left(1-\eta_d\eta_t\right)^k\right],
    \label{eq:yield}
\end{equation}
where $\eta_d$ is Bob's detection efficiency, $\eta_t$ is the channel transmission, and $Y_0$ is the dark count probability. We may then define the gain $Q_k$ of a $k$-photon state as the probability that Alice sends a $k$-photon state \textit{and} Bob gets a detection:
\begin{equation}
    Q_k = Y_k P_\mu(k),
    \label{eq:gains}
\end{equation}
where $P_\mu(k)$ are the Poisson distributed coefficients from Eq. (\ref{eq:PDScoeffs}) with average photon number $\mu$. From \cite{Ma:Thesis08}, the secure key rate $R$ after privacy amplification and error correction may be lower-bounded as:
\begin{equation}
R \geqslant \frac{1}{2}\left[Q_1\left(1-H_2(e_1)\right)-fQ_\mu H_2\left(E_\mu\right)\right],
\label{eq:skrbb84}
\end{equation}
where $Q_\mu$ and $E_\mu$ are the gain and quantum bit error rate (QBER) of the signal state, respectively, $e_1$ is the QBER generated by single-photon states only, $f$ is the error correcting code inefficiency assuming one-way classical post-processing, and $H_2(x)=-x\log_2(x)-(1-x)\log_2(1-x)$ is the binary entropy function for $0<x\leqslant 1$. The first term states that only single-photon states contribute positively to the secure key rate, since multi-photon pulses leak information on lossy channels \cite{BLM:PRL00}. The second term materializes the cost of error correction.

In practical BB84 QKD without decoy states, Alice and Bob cannot estimate $Q_1$ and $e_1$ from their eavesdropped channel, as an eavesdropper may influence the photon number statistics observed by Alice and Bob. In order to nevertheless estimate the secure key rate from Eq. (\ref{eq:skrbb84}), Alice and Bob must make a pessimistic (yet secure) assumption, namely that \textit{all} losses and error come from single-photon states \cite{Ma:Thesis08,LMC:PRL05}:
\begin{equation}
    \left\{
    \begin{array}{ll}
        &Y_k = 1\\
        &e_k = 0 
    \end{array}
\right. \:\:\:\: \:\:\:\:\text{for}\:\:\: k\geqslant 2
\end{equation}
We may then estimate the required single-photon parameters as:
\begin{equation}
    \left\{
    \begin{array}{ll}
        &Q_1\geqslant Q_\mu - \sum\limits_{k=2}^{\infty}P_\mu(k)\\
        &e_1 \leqslant \frac{E_\mu Q_\mu}{Q_1} 
    \end{array}
\right.\label{eq:single}
\end{equation}
where $Q_\mu=\sum\limits_{k=0}^{\infty}Q_k$ and  $E_\mu=\frac{1}{Q_\mu}\sum\limits_{k=0}^{\infty}e_k Q_k$ and $e_k= \frac{e_0 Y_0 +e_d \left[1-\left(1-\eta_d\eta_t\right)^k\right]}{Y_k}$. The parameter $e_d$ characterizes the detection error probability, which depends on the optical alignment of the entire system.

In practice, this pessimistic assumption places a limit on the secure communication distance between Alice and Bob: since single-photon states are the only signals that contribute positively to the secret key rate from (\ref{eq:skrbb84}), assuming that all channel losses come from these effectively reduces the key rate. 

Decoy state QKD \cite{LMC:PRL05}, on the other hand, fixes this issue by proposing that Alice varies the average photon number $\mu$ of her signal. Some pulses are then used as decoy states to estimate the channel statistics, while the others are used as true signal states for the protocol. In the limit of infinite decoy states, since the eavesdropper cannot differentiate between the two, the estimation of $Q_1$ and $e_1$ can be performed much more accurately, as:
\begin{equation}
\left\{
    \begin{array}{ll}
        Q_1 = Y_1 P_\mu(1)\\
        e_1 = \frac{e_0 Y_0 +e_d \eta_d\eta_t}{Y_1}
    \end{array}
\right.\label{eq:one}
\end{equation}

\subsection{With quantum dot sources}

For QDS with collection efficiency $\eta$, we simply replace the $P_\mu(k)$ photon number coefficients in Eqs. (\ref{eq:gains}), (\ref{eq:single}) and (\ref{eq:one}) by the following $P_\eta(k)$ coefficients:

\begin{equation}
\begin{aligned}
    &P_\eta(0) = p_0 + p_1\left(1-\eta\right) +  p_2\left(1-\eta\right)^2 \\
    &P_\eta(1) = p_1\eta + p_2 \left[1-\eta^2-\left(1-\eta\right)^2\right] \\
    &P_\eta(k\geqslant 2) = 1-P_\eta(0)-P_\eta(1),
\end{aligned}\label{eq:dotss}
\end{equation}
where $\{p_n\}$ are the photon number populations from Eq. (\ref{eq:probabilities}).

\section{Security of twin-field quantum key distribution}\label{sec:twinfield}

\subsection{With Poisson-distributed sources}

The decoy method presented in Section \ref{sec:decoy} can also be applied in TF-QKD, which implies that Alice and Bob must both randomize their pulses' global phase. However, a global phase reference must be shared between the two parties at some stage of the protocol, without leaking any information to Charlie. In \cite{LYD:Nature18}, the proposed method is for Alice and Bob to agree on a fixed number of global phase slices $m$, equally splitting the interval $\left[0,2\pi\right)$, from which they uniformly sample a global phase for each state. After Charlie's announcement of the measurement outcomes, they reveal which phase slice was chosen, and sift the raw key in such a way that they keep only the elements for which their chosen phase slices match. This method can potentially leak information to eavesdroppers, and variants that offer alternative methods have been proposed \cite{Cui_2019,CAL:npjQI19}.

In this setting, the secret key rate for TF-QKD resembles that derived in Eq. (\ref{eq:skrbb84}) for standard decoy QKD, with a few amendments. First, the channel transmittance $\eta_t$ presented in Eq. (\ref{eq:yield}) must be corrected to $\sqrt{\eta_t}$, since each optical pulse travels only half the distance between Alice and Bob. This gives the following expression for the yield:

\begin{equation}
    Y_k^{(TF)} = Y_0 +\left(1-Y_0\right) \left[1-\left(1-\eta_d\sqrt{\eta_t}\right)^k\right].
    \label{eq:yieldTF}
\end{equation}
Then, an extra factor, dependent on the number $m$ of chosen phase slices and the duty cycle $d$ of quantum vs. classical signals, must be added to correct the key rate. This factor reads $d/m$. Finally, the intrinsic error rate $e_s$ generated by the finite phase slicing must be taken into account on top of the overall setup alignment error $e_d$. For illustration purposes, we here assume that the setup alignment error $e_d=2\%$, and, for $d=1$ and $m=16$, that the error due to phase slicing is $e_s=1.275\%$ \cite{LYD:Nature18}. Following the workings from \cite{LYD:Nature18} with optimal $\mu=0.765$, this provides the following modified expression for the TF key rate :

\begin{equation}
R^{(TF)} \geqslant \frac{d}{2m}\left[Q_1^{(TF)}\left(1-H_2\left(e_1^{(TF)}\right)\right)-fQ_\mu^{(TF)} H_2\left(E^{(TF)}_\mu\right)\right],
\label{eq:skrtf}
\end{equation}
where

\begin{equation}
    \left\{
    \begin{array}{ll}
        &Q_k^{(TF)} = Y_k^{(TF)} P_\mu(k),\\
        &Q_\mu^{(TF)} = \sum\limits_{k=0}^{\infty}Q_k^{(TF)}\\
        &e_k^{(TF)}= \frac{e_0 Y_0 +(e_d+e_s) \left[1-\left(1-\eta_d\sqrt{\eta_t}\right)^k\right]}{Y_k^{(TF)}}\\
        &E_\mu^{(TF)}=\frac{1}{Q_\mu^{(TF)}}\sum\limits_{k=0}^{\infty}e_k^{(TF)} Q_k^{(TF)}
    \end{array}
\right.\label{eq:tfsystem}
\end{equation}

\subsection{With quantum dot sources}

For QDS with collection efficiency $\eta$, we simply replace the $P_\mu(k)$ photon number coefficients in Eq. (\ref{eq:tfsystem}) by the following $P_\eta(k)$ coefficients:

\begin{equation}
\begin{aligned}
    &P_\eta(0) = p_0 + p_1\left(1-\eta\right) +  p_2\left(1-\eta\right)^2 \\
    &P_\eta(1) = p_1\eta + p_2 \left[1-\eta^2-\left(1-\eta\right)^2\right] \\
    &P_\eta(k\geqslant 2) = 1-P_\eta(0)-P_\eta(1),
\end{aligned}
\end{equation}
where $\{p_n\}$ are the photon number populations from Eq. (\ref{eq:probabilities}).

\section{Security of unforgeable quantum tokens (or money)}\label{sec:qtokens}

We start with a brief introduction to the semidefinite programming techniques required for this security analysis, followed by Choi's theorem on completely positive maps, before deriving the unforgeability regions of the protocol.

\subsection{Unforgeability analysis for quantum tokens}

 \noindent The exact protocol considered here is described in \cite{BDG:PRA19}, and we extend its security analysis to the quantum dot framework. A successful forging attack is one in which two copies of the quantum token state are simultaneously accepted at two spatially separated verification points. Let $\Lambda$ be the optimal adversarial map which produces two copies (living in $\mathcal{H}_1\otimes\mathcal{H}_2$) of the following original quantum token state living in $\mathcal{H}_{\text{ini}}$:

\begin{equation}
\rho_{\text{ini}} = \frac{1}{4}\sum_{k=0}^3 \sigma^{\left(\eta,\phi\right)}_k.
\end{equation}
Here, the set $\{\sigma^{\left(\eta,\phi\right)}_k\}$ contains either the fixed-phase coherent states from Eq. (\ref{eq:fixedphase}), the randomized phase coherent states from Eq. (\ref{eq:randomphase}), the quantum dot states exhibiting number coherence from Eq. (\ref{eq:numbercoherence}), or the quantum dot states without number coherence from Eq. (\ref{eq:diagonal}). The superscripts $\left(\eta,\phi\right)$ serve as a reminder that these states depend on the PDS average photon number or QDS collection efficiency $\eta$, and encoding phase $\phi$.

This proof makes use of the existence of a squashing model for the measurement setup \cite{BML:prl08}. Essentially, this model allows to express the infinite-dimensional measurement operators in a 3-dimensional space spanned by $\{\ket{0},\ket{1},\ket{\varnothing}\}$, by imposing a condition on the terminal's postprocessing, consisting of assigning a random measurement outcome $\ket{0}$ or $\ket{1}$ to any double click, and declaring a $\ket{\varnothing}$ flag when no detection is registered. The probability that a verifier declares an incorrect measurement outcome for token $1$ is given by:

\begin{equation}
V_1 = \Tr\sum_{k=0}^3\left(\frac{1}{2}\ket{\beta_{k}^\perp}\bra{\beta_{k}^\perp}\otimes \mathbb{1}\right) \Lambda\left(\frac{1}{4}\sigma^{\left(\eta,\phi\right)}_k\right),
\end{equation} 
while that for token $2$ reads:

\begin{equation}
V_2=\Tr\sum_{k=0}^3\left(\mathbb{1}\otimes\frac{1}{2}\ket{\beta_{k}^\perp}\bra{\beta_{k}^\perp}\right) \Lambda\left(\frac{1}{4}\sigma^{\left(\eta,\phi\right)}_k\right),
\end{equation}
where $\ket{\beta_{k}}$ is the squashed qubit associated with the original state $\sigma^{\left(\eta,\phi\right)}_k$, i.e.\@ $\ket{\beta_0} = \ket{+}$, $\ket{\beta_1} = \ket{+i}$, $\ket{\beta_2} = \ket{-}$, $\ket{\beta_3} = \ket{-i}$, and $\ket{\beta_{k}^\perp}$ is its orthogonal qubit state. The factor $1/4$ indicates that each $\sigma_k$ is equally likely to occur, while $1/2$ accounts for the verifier's random measurement basis choice. Using Eq. (\ref{eq:choi}), we may rewrite these expressions as $V_1 =\Tr\left(E_1(\mu)J(\Lambda)\right)$ and $V_2 = \Tr\left(E_2(\mu)J(\Lambda)\right)$, where $E_1(\mu)$ and $E_2(\mu)$ are the \textit{error operators}:

\begin{equation}
\begin{aligned}
   E_1(\mu)=&\frac{1}{4}\sum_{k=0}^{3}\frac{1}{2} \ket{\beta_{k}^\perp}\bra{\beta_{k}^\perp}\otimes\mathbb{1}\otimes\overline{\sigma^{\left(\eta,\phi\right)}_k}, \\
   E_2(\mu)=&\frac{1}{4}\sum_{k=0}^{3}\mathbb{1}\otimes\frac{1}{2} \ket{\beta_{k}^\perp}\bra{\beta_{k}^\perp}\otimes \overline{\sigma^{\left(\eta,\phi\right)}_k} .%\nonumber\\
\end{aligned}
\end{equation}

Following a similar method, the probability that verifier $1$ (resp. 2) registers a no-detection event for token $1$ (resp. $2$) reads $\Tr\left(L_1(\mu)J(\Lambda)\right)$ (resp. $\Tr\left(L_2(\mu)J(\Lambda)\right)$), where $L_1(\mu)$ and $L_2(\mu)$ are the \textit{loss operators}, which contain the projection onto the state $\ket{\varnothing}$:
\begin{equation}
\begin{aligned}
L_1(\mu) = \frac{1}{4}\sum_{k=0}^3 \ket{\varnothing}\bra{\varnothing}\otimes\mathbb{1} \otimes\overline{\sigma^{\left(\eta,\phi\right)}_k}, \\
L_2(\mu) = \frac{1}{4}\sum_{k=0}^3
\mathbb{1}\otimes\ket{\varnothing}\bra{\varnothing} \otimes\overline{\sigma^{\left(\eta,\phi\right)}_k} .
\label{will}
\end{aligned}
\end{equation}
We now search for the optimal cloning map $\Lambda$ that minimizes the noise that the adversary must introduce for both tokens given a fixed combined channel and detection losses $l$. We cast this problem in the following SDP for a card with a single state,
\begin{equation}
\begin{aligned}
\min\quad& \Tr\left(E_1(\mu) J(\Lambda) \right) \\
\text{s.t. }&  \Tr_{\mathcal{H}_1\otimes\mathcal{H}_2}\left(J(\Lambda)\right) = \mathbb{1}_{\mathcal{H}_{\text{ini}}} \\
    &\Tr\left(E_1(\mu) J(\Lambda) \right) \geqslant \Tr\left(E_2(\mu) J(\Lambda) \right) \\
    &\Tr\left(L_1(\mu) J(\Lambda) \right) \leqslant l \\
    &\Tr\left(L_2(\mu) J(\Lambda) \right) \leqslant l \\
    &J(\Lambda) \geqslant 0
\end{aligned}\label{eq:noisetol}
\end{equation}
The first constraint imposes that $\Lambda$ is trace-preserving, the second imposes that the error rate measured for token $1$ is at least equal to the one measured for token $2$, the third and fourth impose that the losses measured for tokens $1$ and $2$ do not exceed the expected honest losses, and the fifth imposes that $\Lambda$ is completely positive. 

Using similar techniques to \cite{BDG:PRA19}, it can be shown that this lower bound is in fact optimal, and that the adversary does not succeed better by performing a general attack on the full tensor product of the $N$ states contained in the token.

\section{Security of quantum strong coin flipping}\label{sec:coinflipping}

We focus here on the quantum protocol from \cite{PJ+:natcomm14}, and additionally study the effect of photon number coherence on the protocol bias in the security proof.

\subsection{Dishonest Alice}

 \noindent Since Dishonest Alice can send any arbitrary quantum state to Honest Bob, the corresponding security proof does not depend on the emitted PDS or QDS state, but depends only on the parameter $y$ imposed by the protocol. Using the notations from \cite{PJ+:natcomm14}, we assume here that Dishonest Alice wishes to bias the outcome towards $x=0$, corresponding to $b=c$. Note that the security analysis proceeds similarly for outcome $x=1$ with $b\neq c$. 

From \cite{PJ+:natcomm14}, Dishonest Alice's optimal strategy consists in sending the state that maximizes the average probability
of revealing $(\alpha= 0, c = 0)$ and $(\alpha= 1, c = 1)$ or of revealing $(\alpha= 1, c = 0)$ and $(\alpha = 0, c = 1)$. Note that these four pairs yield a better cheating strategy than the other pairs, since the states in the pairs $\{\ket{\Phi^{\left(y\right)}_{00}},\ket{\Phi^{\left(y\right)}_{11}}\}$ and $\{\ket{\Phi^{\left(y\right)}_{10}},\ket{\Phi^{\left(y\right)}_{01}}\}$ have a larger overlap than the states in the pairs $\{\ket{\Phi^{\left(y\right)}_{00}},\ket{\Phi^{\left(y\right)}_{01}}\}$ and $\{\ket{\Phi^{\left(y\right)}_{10}},\ket{\Phi^{\left(y\right)}_{11}}\}$. 

Following the arguments from \cite{PJ+:natcomm14,BBB:PRA09}, Dishonest Alice's optimal cheating strategy is to create an entangled state, of which she sends one half to Bob, waits for his measurement and declaration of classical data, and finally performs a measurement on her part of the state to decide which outcome she reveals. This yields the following upper-bound on Alice's cheating probability:

\begin{equation}
    \mathcal{P}_A^{bias} \leqslant \frac{3}{4}  + \frac{1}{2}\sqrt{y(1-y)}.
\end{equation}

%Since Honest Bob's basis choice is uniformly random, Dishonest Alice must generate a state $\rho$ that maximizes the overlap with $\Pi_{acc}=\frac{1}{2}\ket{\Phi^{\left(y\right)}_{00}}\bra{\Phi^{\left(y\right)}_{00}}+\frac{1}{2}\ket{\Phi^{\left(y\right)}_{11}}\bra{\Phi^{\left(y\right)}_{11}}$ or $\Pi_{acc}=\frac{1}{2}\ket{\Phi^{\left(y\right)}_{10}}\bra{\Phi^{\left(y\right)}_{10}}+\frac{1}{2}\ket{\Phi^{\left(y\right)}_{01}}\bra{\Phi^{\left(y\right)}_{01}}$ . Once Honest Bob has declared a bit $b$, she then declares $\alpha=b$ and $c=b$. This maximization problem may be recast as the following SDP:

%\begin{equation}
%\begin{aligned}
%\max\quad& \Tr\left(\Pi_{acc} \rho \right) \displaybreak[1]\\
%\text{s.t. }&  \Tr\left(\rho\right) = 1 \\
%    &\rho \geqslant 0.
%\end{aligned}
%\label{sdp:SCFAlice}
%\end{equation}
%
%From the protocol description, we know that Honest Bob always accepts the coin when his measurement %basis is different to Alice's. Since Bob’s
%basis choice is uniformly random and independent of Alice, she can then finally bias the coin with %probability:

%\begin{equation}
%    P_A^{bias} = \frac{1}{2} P_A + \frac{1}{2}\times 1,
%\end{equation}
%
%where $P_A$ is the optimal cheating probability calculated from problem (\ref{sdp:SCFAlice}). From \cite{PJ+:natcomm14,BBB:PRA09}, we know that Dishonest Alice cannot cheat better by sending states living in a larger Hilbert space. The upper bound on Alice's cheating probability then finally reads: 

\subsection{Dishonest Bob}

\subsubsection{Without photon number coherence}

 \noindent In general, Dishonest Bob's optimal cheating strategy involves some form of discrimination problem, in which he tries to identify which state was sent by Alice from a known and pre-agreed set of states. The works from  \cite{PJ+:natcomm14,BBB:NC11,MTV:PRL05} provided a practical security analysis for PDS (either SPDC or attenuated laser states), with the crucial assumption that \textit{no coherence} is present in the photon number basis. Thus, we can use these security analyses for phase-randomized PDS, as well as for LA and TPE QDS. For RE QDS however, we must extend the security analysis to incorporate the presence of photon number coherence in the states generated by Alice, which is performed in the next section.

To upper bound Bob's cheating probability without photon number coherence, we use the expression derived in \cite{PJ+:natcomm14}, noting that it may not be a tight upper bound. However, this does not compromise the security of the protocol, as it only increases Dishonest Bob's power. Similarly to Dishonest Alice, we assume that Dishonest Bob wishes to bias the flip outcome towards $x=0$, corresponding to $b=c$. Note that the security analysis proceeds similarly for outcome $x=1$ with $b\neq c$. As a brief summary, the cases in which Bob \textit{cannot} perfectly cheat, considered in \cite{PJ+:natcomm14}, are the following:

\begin{itemize}
    \item $A_1$: Alice sends only vacuum states, with probability $\mathcal{P}\left(A_1\right)= P^N_x(0)$.
    \item  $A_2$: Alice sends at least one single-photon pulse, and vacuum states, with probability $\mathcal{P}\left(A_2\right)= \left(P_x(0)+P_x(1)\right)^N-P^N_x(0)$.
    \item  $A_3$:  Alice sends one two-photon pulse, and vacuum states, with probability $\mathcal{P}\left(A_3\right) = NP_x(k\geqslant 2)P^{N-1}_x(0)$.
    \item $A_4$: Alice sends one two-photon pulse, at least one single-photon pulse, and vacuum states, with probability $\mathcal{P}\left(A_4\right)= N P_x(k\geqslant 2) \left(\left(P_x(0)+P_x(1)\right)^{N-1}-P^{N-1}_x(0)\right)$
\end{itemize}
where $\{P_x(k)\}$ are the PDS coefficients from Eq. (\ref{eq:PDScoeffs}) for $x=\mu$ or the QDS coefficients from Eq. (\ref{eq:dotss}) for $x=\eta$.

Assuming no coherence in photon number basis, Bob's cheating probability can be upper-bounded individually for each of the four cases \cite{PJ+:natcomm14}. In case $A_1$, Bob's optimal strategy involves declaring a random bit $b'$, which will make him successfully bias the coin towards his desired outcome with probability $\mathcal{P}\left(b'|A_1\right)=1/2$. In case $A_2$, Bob's optimal strategy consists in performing a Helstrom measurement, which is successful with probability $\mathcal{P}\left(b'|A_2\right)=y$. In case $A_3$, Bob's optimal cheating strategy also reads $\mathcal{P}\left(b'|A_3\right)=y$. In case $A_4$, an optimization must be performed to find the best set of discrimination measurement operators within the full spectrum of conclusive and inconclusive measurements. This allows to upper bound his cheating probability as $\mathcal{P}\left(b'|A_4\right)\leqslant -2y^2+4y-1$ \cite{PJ+:natcomm14}.

Summing the contributions from all four cases, and assuming that Bob can cheat with probability $1$ in all other cases, provides the final upper bound:

\begin{equation}
    \mathcal{P}_B^{bias} \leqslant \sum_{i=1}^4 \mathcal{P}\left(A_i\right)\mathcal{P}\left(b'|A_i\right) + \left[1-\sum_{i=1}^4 \mathcal{P}\left(A_i\right)\right]\times 1 .
    \label{eq:bobbias}
\end{equation}

\subsubsection{With photon number coherence}

 \noindent Extending the security proof to account for photon number coherence in the RE-pumped QDS is challenging: unlike in the previous subsection, the security analysis cannot be decomposed into independent security analyses that are conditioned on the outcome of a photon number measurement. Dishonest Bob can indeed exploit the presence of coherence to perform more general discrimination attacks. 
 
 We note that Bob's aim is to discriminate between the following two states sent by Honest Alice with equal probabilities, in such a way that the outcome of the flip is optimally biased towards $x=0$:
 
 \begin{equation}
 \begin{aligned}
     \sigma_0^{\left(y,\eta,\phi\right)} &= \frac{1}{2}\left( \sigma_{00}^{\left(y,\eta,\phi\right)}+\sigma_{10}^{\left(y,\eta,\phi\right)}\right)\\
      \sigma_1^{\left(y,\eta,\phi\right)}& =\frac{1}{2}\left( \sigma_{01}^{\left(y,\eta,\phi\right)}+\sigma_{11}^{\left(y,\eta,\phi\right)}\right).
\end{aligned}
 \end{equation}

There exists a broad spectrum of discrimination measurements, which may be characterized by a set of parameters $\{p_{conc},p_{corr}\}$ \cite{BBB:PRA09}. The first parameter $p_{conc}$ gives the probability that implementing the POVM will yield a conclusive outcome, i.e. that one state from the pre-agreed set of states will be identified. The second parameter $p_{corr}$ gives the probability that this outcome is correct, i.e. that the identified state is indeed the one that was sent. Fully conclusive measurements have $p_{conc}=1$, but usually display a non-unit probability of being correct $p_{corr}<1$ (depending on the set of states). Other measurements will increase $p_{corr}$ by allowing some probability of the measurement being inconclusive, i.e. $p_{conc}<1$. 

Of course, Bob's optimal discrimination strategy will involve optimizing over these parameters for all $N$ states, which is beyond the scope of this paper. Nevertheless, we show here how one powerful attack exploiting inconclusive measurement operators, known as unambiguous state discrimination (USD) \cite{RST:PRA03,Lu:PRA21}, can already provide Dishonest Bob with a cheating advantage over states which do not exhibit photon number coherence under XX pumping. Intuitively, a USD POVM will return an outcome that is always correct ($p_{corr}=1$), at the risk of getting an inconclusive outcome with some probability $p_{conc}<1$. Since this attack yields inconclusive outcomes instead of erroneous ones, Bob can repeatedly perform the same attack on each state until the outcome is conclusive, in which case he has identified Alice's state without any error. If, after $(N-1)$ states, the attack is still inconclusive, he may then perform a conclusive minimum-error discrimination measurement on the last state, with $p_{corr}<1$ and $p_{conc}=1$, whose maximum success probability is given by the Helstrom bound \cite{BK:AOP09}.

To derive Bob's cheating probability in this case, we must first justify that USD is possible for our set of states (i.e., that there exists a USD attack which yields a non-zero probability of successfully identifying which state was sent by Alice). For this, it suffices to note that the kernels associated with the states $\sigma_0^{\left(y,\eta,\phi\right)}$ and $\sigma_1^{\left(y,\eta,\phi\right)}$ living in the full photon space are both non-zero \cite{RST:PRA03}. Although this is enough to derive an upper bound on the success of the USD measurement, i.e. on the value of $p_{conc}$, we must find a bound that is \textit{tight} in order to provide a meaningful comparison with the over-estimated bounds from Eq. (\ref{eq:bobbias}). For this, we once again use the semidefinite programming techniques introduced in Section \ref{sec:tools}. We extend the pure state treatment from \cite{E:IEEE03} and the discrimination problem from \cite{EMV:IEEE03} to recast Dishonest Bob's search for the optimal USD strategy:
\begin{equation}
\begin{aligned}
\max\quad& \frac{1}{2}\left[\Tr\left(M_0 \sigma_0^{\left(y,\eta,\phi\right)} \right)+\Tr\left(M_1 \sigma_1^{\left(y,\eta,\phi\right)}\right) \right] \\
\text{s.t. }&\Tr\left(M_0 \sigma_1^{\left(y,\eta,\phi\right)} \right) = 0 \\
    &\Tr\left(M_1 \sigma_0^{\left(y,\eta,\phi\right)} \right) = 0\\
    &M_0 + M_1 + M_{inc} = \mathbb{1}\\
    &M_0, M_1,  M_{inc} \geqslant 0,
\end{aligned}\label{eq:maxBob}
\end{equation}
where $M_0$ and $M_1$ are the POVM operators which identify the states $\sigma_0^{\left(y,\eta,\phi\right)}$ and $\sigma_1^{\left(y,\eta,\phi\right)}$, respectively, and  $M_{inc}$ is the POVM operator yielding an inconclusive outcome. All three operators serve as the optimization variables. The first two constraints ensure that the optimal POVM operators identify the correct state with zero error probability. Similarly to unforgeable quantum tokens, the tightness of this upper bound can be shown by using strong duality (see Section \ref{sec:tools} and \cite{EMV:IEEE03}).

%we use the necessary and sufficient conditions derived in \cite{RL:PRA05}, that provide the following optimal bound on the probability of successful USD:

%\begin{equation}
%    \mathcal{P}_{USD} = 1-\frac{1}{2}\left[\frac{F^2\left(\sigma_0^{\left(y,\eta,\phi\right)},\sigma_1^{\left(y,\eta,\phi\right)}\right)}{\Tr\left(\Pi_1\sigma_0^{\left(y,\eta,\phi\right)}\right)}+\Tr\left(\Pi_1\sigma_0^{\left(y,\eta,\phi\right)}\right)\right],
%\end{equation}
%
%where $F(a,b) = \Tr\left(\sqrt{a}\sqrt{b}\right)$ is the quantum state fidelity for states $a$ and $b$, and $\Pi_k$ is the projector onto the support of $\sigma_k^{\left(y,\eta,\phi\right)}$. 

We label the optimal value to problem (\ref{eq:maxBob}), i.e. Bob's optimal USD cheating probability, as $\mathcal{P}_{USD}$. On the other hand, the probability of a successful Helstrom measurement is given by \cite{BK:AOP09}:

\begin{equation}
    \mathcal{P}_{HEL} = \frac{1}{2}+\frac{1}{4}\norm{\sigma_0^{\left(y,\eta,\phi\right)}-\sigma_1^{\left(y,\eta,\phi\right)}}_1,
\end{equation}
where $\norm{\odot}_1$ denotes the Schatten $1$-norm.

Since Bob performs the same attack on each of the $(N-1)$ states sent by Alice, followed by a Helstrom measurement on the $N$th state, we can finally upper-bound his cheating probability as:

\begin{equation}
    \mathcal{P}_B^{bias} \leqslant \left[1-\left(1- \mathcal{P}_{USD}\right)^{N-1}\right]\times 1 + \left(1- \mathcal{P}_{USD}\right)^{N-1}\times \mathcal{P}_{HEL}  .
\end{equation}

\section{Security of quantum bit commitment}\label{sec:bitcommitment}

 \noindent We focus on the quantum protocol from \cite{NJC:NatComms12}, under the bounded storage assumption. This assumption circumvents the no-go theorem for two-party computations \cite{M:PRL97,LC:PRL97} by placing restrictions on Dishonest Bob's storage capabilities: he may store only $S$ of the $N$ quantum states sent by Honest Alice, over a time duration no longer than $\Delta t$. Similarly to BB84 QKD in Section \ref{sec:decoy}, the practical security analysis assumes that states emitted by Honest Alice bear no coherence in the photon number basis. We therefore consider only phase-randomized PDS, and LA/TPE QDS. We adapt here the conditions provided in the Appendix of \cite{NJC:NatComms12} for a $3\epsilon$-secure quantum bit commitment implementation. Here, $\epsilon$ is a fixed parameter which upper-bounds the occurrence of bad events, governed by the Hoeffding inequality.
 
 In a nutshell, the practical bit commitment protocol is constructed from a weak string erasure sub-routine with errors (WSEE) \cite{KWW:IEEE12}. An $\left(N,\lambda,\epsilon,e\right)$-WSEE provides Alice with a string $X_N$ and Bob with a randomly chosen subset $I\in [N]$, as well as a substring $\widetilde{X}_I$. This substring is given by the substring $X_I$ (the bits of $X_N$ corresponding to the indices in $I$) passed through a binary symmetric channel that flips each bit of $X_I$ with probability $e$. The security statements then read as follows:
 
 \begin{itemize}
     \item If Alice is honest, then the amount of information a Dishonest Bob holds about $X_N$
is limited, i.e. the $\epsilon$-smooth min entropy of $X_N$ conditioned on a Dishonest Bob’s information is lower bounded by a value $\lambda$.
\item If Bob is honest, then Alice does not have any information $I$. That is, Alice does not learn
which bits of $X_N$ are known to Bob.
 \end{itemize}
 
Essentially, losses allow Dishonest Bob to discard a fraction of single-photon detection events, and keep more multiphoton events so that his chance of guessing $X_N$ correctly is increased. The resulting min-entropy rate $\lambda$ can thereby be calculated as a function
of experimental parameters $\{P_x(k)\}$ and $\{P_{\left(x\eta_c\right)}(k)\}$, where $\{P_x(k)\}$ are the PDS coefficients from Eq. (\ref{eq:PDScoeffs}) for $x=\mu$ or the QDS coefficients from Eq. (\ref{eq:dotss}) for $x=\eta$, and $\{P_{\left(x\eta_c\right)}(k)\}$  are defined similarly, only with an extra channel transmission factor $\eta_c$ multiplying $x$ to account for honest losses.
 
 Following Lemma $12$ from the Supplementary Information of \cite{NJC:NatComms12}, given an experimental error rate $e$, channel losses $\eta_c$, fixed parameters $\epsilon$ and $\left(\beta,\gamma\right)\in\left(0,0.01\right]$, and considering a Dishonest Bob with bounded storage size $S$, we define the following parameters:
 
 \begin{equation}
     \begin{aligned}
     &m_2 = P_x(1)-P_{\left(x\eta_c\right)}(0)+P_x(0)-3\gamma \\
     &m_3 = 1 -P_{\left(x\eta_c\right)}(k)\\ 
     &L' = \max_{s\in\left(0,1\right]} -\frac{1}{s}\left[\log\left(1+2^s\right)-1-s\right]-\frac{3\epsilon}{s}\\
          &\delta= 2\:\frac{e+\frac{\beta}{\sqrt{1-2\beta}}}{1-4\sqrt{5}\beta}\\
     &\lambda = H_2(\delta) + 3\beta^2\\
     &M_1 = \frac{1}{2\gamma^2}\log\frac{2}{\epsilon}\\
     &M_2 = \frac{\log\frac{1}{\epsilon}}{\epsilon\: m_2}\\
     &M_3 = \frac{\log\frac{2}{\epsilon}}{\left(m_3-\gamma\right)\beta^2}\\
     &M_4 = \frac{S}{m_2 L'-m_3\lambda}
     \end{aligned}
 \end{equation}
where $H_2(x)=-x\log_2(x)-(1-x)\log_2(1-x)$ is the binary entropy function for $0<x\leqslant 1$. For security to hold in a practical implementation, the following condition must hold:

\begin{equation}
    m_2 L'-m_3\lambda > 0.
    \label{eq:bitcommcondition}
\end{equation}
If Eq. (\ref{eq:bitcommcondition}) is true, then quantum bit commitment under the bounded storage model can be implemented $3\epsilon$-securely by using a randomly constructed error-correcting code, whenever the number of states sent by Alice:

\begin{equation}
    N > \max\{M_1,M_2,M_3,M_4\}.
    \label{eq:bitcommcondition2}
\end{equation}

\end{appendix}
  

\medskip

\begin{thebibliography}{10}
\expandafter\ifx\csname url\endcsname\relax
  \def\url#1{\texttt{#1}}\fi
\expandafter\ifx\csname urlprefix\endcsname\relax\def\urlprefix{URL }\fi
\providecommand{\bibinfo}[2]{#2}
\providecommand{\eprint}[2][]{\url{#2}}

\bibitem{GS:PRL21}
\bibinfo{author}{Gouzien, E.} \& \bibinfo{author}{Sangouard, N.}
\newblock \bibinfo{title}{Factoring 2048-bit rsa integers in 177 days with 13
  436 qubits and a multimode memory}.
\newblock \emph{\bibinfo{journal}{Phys. Rev. Lett.}}
  \textbf{\bibinfo{volume}{127}}, \bibinfo{pages}{140503}
  (\bibinfo{year}{2021}).
\newblock
  \urlprefix\url{https://link.aps.org/doi/10.1103/PhysRevLett.127.140503}.

\bibitem{MLL:NatPhot12}
\bibinfo{author}{Martín-López, E.} \emph{et~al.}
\newblock \bibinfo{title}{Experimental realization of shor's quantum factoring
  algorithm using qubit recycling}.
\newblock \emph{\bibinfo{journal}{Nat. Photonics}}
  \textbf{\bibinfo{volume}{6}}, \bibinfo{pages}{773--776}
  (\bibinfo{year}{2012}).
\newblock \urlprefix\url{https://www.nature.com/articles/nphoton.2012.259}.

\bibitem{S:SIAM97}
\bibinfo{author}{Shor, P.~W.}
\newblock \bibinfo{title}{Polynomial-time algorithms for prime factorization
  and discrete logarithms on a quantum computer}.
\newblock \emph{\bibinfo{journal}{SIAM Journal on Computing}}
  \textbf{\bibinfo{volume}{26}}, \bibinfo{pages}{1484--1509}
  (\bibinfo{year}{1997}).
\newblock \urlprefix\url{https://doi.org/10.1137/S0097539795293172}.

\bibitem{BB84}
\bibinfo{author}{Bennett, C.~H.} \& \bibinfo{author}{Brassard, G.}
\newblock \bibinfo{title}{Quantum cryptography: Public key distribution and
  coin tossing}.
\newblock In \emph{\bibinfo{booktitle}{Proc. IEEE International Conference on
  Computers, Systems and Signal Processing}}, vol.~\bibinfo{volume}{1},
  \bibinfo{pages}{175--179} (\bibinfo{address}{Bangalore, India},
  \bibinfo{year}{1984}).
\newblock
  \urlprefix\url{https://researcher.watson.ibm.com/researcher/files/us-bennetc/BB84highest.pdf}.
  
 \bibitem{Pan:RevMod20}
\bibinfo{author}{Xu, F.}, \bibinfo{author}{Ma, X.}, \bibinfo{author}{Zhang,
  Q.}, \bibinfo{author}{Lo, H.-K.} \& \bibinfo{author}{Pan, J.-W.}
\newblock \bibinfo{title}{Secure quantum key distribution with realistic
  devices}.
\newblock \emph{\bibinfo{journal}{Rev. Mod. Phys.}}
  \textbf{\bibinfo{volume}{92}}, \bibinfo{pages}{025002}
  (\bibinfo{year}{2020}).
\newblock
  \urlprefix\url{https://link.aps.org/doi/10.1103/RevModPhys.92.025002}.

\bibitem{WEH:Sci18}
\bibinfo{author}{Wehner, S.}, \bibinfo{author}{Elkouss, D.} \&
  \bibinfo{author}{Hanson, R.}
\newblock \bibinfo{title}{Quantum internet: A vision for the road ahead}.
\newblock \emph{\bibinfo{journal}{Science}} \textbf{\bibinfo{volume}{362}}
  (\bibinfo{year}{2018}).

\bibitem{BS:dcc16}
\bibinfo{author}{Broadbent, A.} \& \bibinfo{author}{Schaffner, C.}
\newblock \bibinfo{title}{Quantum cryptography beyond quantum key
  distribution}.
\newblock \emph{\bibinfo{journal}{Designs, Codes and Cryptography}}
  \textbf{\bibinfo{volume}{78}}, \bibinfo{pages}{351--382}
  (\bibinfo{year}{2016}).

\bibitem{BCD:PRX21}
\bibinfo{author}{Bozzio, M.}, \bibinfo{author}{Cavaill\`es, A.},
  \bibinfo{author}{Diamanti, E.}, \bibinfo{author}{Kent, A.} \&
  \bibinfo{author}{Pital\'ua-Garc\'{\i}a, D.}
\newblock \bibinfo{title}{Multiphoton and side-channel attacks in mistrustful
  quantum cryptography}.
\newblock \emph{\bibinfo{journal}{PRX Quantum}} \textbf{\bibinfo{volume}{2}},
  \bibinfo{pages}{030338} (\bibinfo{year}{2021}).
\newblock \urlprefix\url{https://link.aps.org/doi/10.1103/PRXQuantum.2.030338}.

\bibitem{Wie:acm83}
\bibinfo{author}{Wiesner, S.}
\newblock \bibinfo{title}{Conjugate coding}.
\newblock \emph{\bibinfo{journal}{ACM Sigact News}}
  \textbf{\bibinfo{volume}{15}}, \bibinfo{pages}{78} (\bibinfo{year}{1983}).

\bibitem{Michler17}
\bibinfo{author}{Michler, P.}
\newblock \emph{\bibinfo{title}{Quantum Dots for Quantum Information
  Technologies}} (\bibinfo{publisher}{Springer International Publishing},
  \bibinfo{year}{2017}).
\newblock \urlprefix\url{https://www.springer.com/de/book/9783319563770#}.

\bibitem{SSW:NNT17}
\bibinfo{author}{Senellart, P.}, \bibinfo{author}{Solomon, G.} \&
  \bibinfo{author}{White, A.}
\newblock \bibinfo{title}{High-performance semiconductor quantum-dot
  single-photon sources}.
\newblock \emph{\bibinfo{journal}{Nat. Nanotechnol.}}
  \textbf{\bibinfo{volume}{12}}, \bibinfo{pages}{1026--1039}
  (\bibinfo{year}{2017}).

\bibitem{BBR:PRL18}
\bibinfo{author}{Boaron, A.} \emph{et~al.}
\newblock \bibinfo{title}{Secure quantum key distribution over 421 km of
  optical fiber}.
\newblock \emph{\bibinfo{journal}{Phys. Rev. Lett.}}
  \textbf{\bibinfo{volume}{121}}, \bibinfo{pages}{190502}
  (\bibinfo{year}{2018}).
\newblock
  \urlprefix\url{https://link.aps.org/doi/10.1103/PhysRevLett.121.190502}.

\bibitem{AMS:PRApp20}
\bibinfo{author}{Anderson, M.} \emph{et~al.}
\newblock \bibinfo{title}{Gigahertz-clocked teleportation of time-bin qubits
  with a quantum dot in the telecommunication $c$ band}.
\newblock \emph{\bibinfo{journal}{Phys. Rev. Appl.}}
  \textbf{\bibinfo{volume}{13}}, \bibinfo{pages}{054052}
  (\bibinfo{year}{2020}).
\newblock
  \urlprefix\url{https://link.aps.org/doi/10.1103/PhysRevApplied.13.054052}.

\bibitem{BLM:PRL00}
\bibinfo{author}{Brassard, G.}, \bibinfo{author}{L\"utkenhaus, N.},
  \bibinfo{author}{Mor, T.} \& \bibinfo{author}{Sanders, B.~C.}
\newblock \bibinfo{title}{Limitations on practical quantum cryptography}.
\newblock \emph{\bibinfo{journal}{Phys. Rev. Lett.}}
  \textbf{\bibinfo{volume}{85}}, \bibinfo{pages}{1330--1333}
  (\bibinfo{year}{2000}).
\newblock \urlprefix\url{https://link.aps.org/doi/10.1103/PhysRevLett.85.1330}.

\bibitem{SKB:JO19}
\bibinfo{author}{Schneeloch, J.} \emph{et~al.}
\newblock \bibinfo{title}{Introduction to the absolute brightness and number
  statistics in spontaneous parametric down-conversion}.
\newblock \emph{\bibinfo{journal}{J. Opt.}} \textbf{\bibinfo{volume}{21}},
  \bibinfo{pages}{043501} (\bibinfo{year}{2019}).
\newblock \urlprefix\url{https://doi.org/10.1088/2040-8986/ab05a8}.

\bibitem{LMC:PRL05}
\bibinfo{author}{Lo, H.-K.}, \bibinfo{author}{Ma, X.} \& \bibinfo{author}{Chen,
  K.}
\newblock \bibinfo{title}{Decoy state quantum key distribution}.
\newblock \emph{\bibinfo{journal}{Phys. Rev. Lett.}}
  \textbf{\bibinfo{volume}{94}}, \bibinfo{pages}{230504}
  (\bibinfo{year}{2005}).
\newblock
  \urlprefix\url{https://link.aps.org/doi/10.1103/PhysRevLett.94.230504}.

\bibitem{LP:QIC07}
\bibinfo{author}{Lo, H.-K.} \& \bibinfo{author}{Preskill, J.}
\newblock \bibinfo{title}{Security of quantum key distribution using weak
  coherent states with nonrandom phases}.
\newblock \emph{\bibinfo{journal}{Quantum Info. Comput.}}
  \textbf{\bibinfo{volume}{7}}, \bibinfo{pages}{431–458}
  (\bibinfo{year}{2007}).

\bibitem{VRG:AQT22}
\bibinfo{author}{Vajner, D.~A.}, \bibinfo{author}{Rickert, L.},
  \bibinfo{author}{Gao, T.}, \bibinfo{author}{Kaymazlar, K.} \&
  \bibinfo{author}{Heindel, T.}
\newblock \bibinfo{title}{Quantum communication using semiconductor quantum
  dots}.
\newblock \emph{\bibinfo{journal}{Adv. Quantum Technol.}}
  \textbf{\bibinfo{volume}{5}}, \bibinfo{pages}{2100116}
  (\bibinfo{year}{2022}).
\newblock
  \urlprefix\url{https://onlinelibrary.wiley.com/doi/abs/10.1002/qute.202100116}.

\bibitem{BVM:SciAdv21}
\bibinfo{author}{Basset, F.~B.} \emph{et~al.}
\newblock \bibinfo{title}{Quantum key distribution with entangled photons
  generated on demand by a quantum dot}.
\newblock \emph{\bibinfo{journal}{Sci. Adv.}} \textbf{\bibinfo{volume}{7}},
  \bibinfo{pages}{12} (\bibinfo{year}{2021}).

\bibitem{SRH:SciAdv21}
\bibinfo{author}{Schimpf, C.} \emph{et~al.}
\newblock \bibinfo{title}{Quantum cryptography with highly entangled photons
  from semiconductor quantum dots}.
\newblock \emph{\bibinfo{journal}{Sci. Adv.}} \textbf{\bibinfo{volume}{7}},
  \bibinfo{pages}{16} (\bibinfo{year}{2021}).

\bibitem{TNM:SciRep15}
\bibinfo{author}{Takemoto, K.} \emph{et~al.}
\newblock \emph{\bibinfo{journal}{Sci. Rep.}} \textbf{\bibinfo{volume}{5}},
  \bibinfo{pages}{14383} (\bibinfo{year}{2015}).

\bibitem{HKR:NJP12}
\bibinfo{author}{Heindel, T.} \emph{et~al.}
\newblock \bibinfo{title}{Quantum key distribution using quantum dot
  single-photon emitting diodes in the red and near infrared spectral range}.
\newblock \emph{\bibinfo{journal}{New J. Phys.}} \textbf{\bibinfo{volume}{14}},
  \bibinfo{pages}{083001} (\bibinfo{year}{2012}).
\newblock \urlprefix\url{https://doi.org/10.1088/1367-2630/14/8/083001}.

\bibitem{CCF:JAP10}
\bibinfo{author}{Collins, R.~J.} \emph{et~al.}
\newblock \bibinfo{title}{Quantum key distribution system in standard
  telecommunications fiber using a short wavelength single photon source}.
\newblock \emph{\bibinfo{journal}{J. Appl. Phys.}}
  \textbf{\bibinfo{volume}{107}}, \bibinfo{pages}{073102}
  (\bibinfo{year}{2010}).
\newblock \urlprefix\url{https://doi.org/10.1063/1.3327427}.
\newblock \eprint{https://doi.org/10.1063/1.3327427}.

\bibitem{IWK:JOA09}
\bibinfo{author}{Intallura, P.} \emph{et~al.}
\newblock \bibinfo{title}{Quantum communication using single photons from a
  semiconductor quantum dot emitting at a telecommunication wavelength}.
\newblock \emph{\bibinfo{journal}{Journal of Optics A: Pure and Applied
  Optics}} \textbf{\bibinfo{volume}{11}}, \bibinfo{pages}{5}
  (\bibinfo{year}{2009}).

\bibitem{Y:Nat02}
\bibinfo{author}{Waks, E.} \emph{et~al.}
\newblock \bibinfo{title}{Quantum cryptography with a photon turnstile}.
\newblock \emph{\bibinfo{journal}{Nature}} \textbf{\bibinfo{volume}{420}},
  \bibinfo{pages}{762} (\bibinfo{year}{2002}).

\bibitem{WCX:PRL08}
\bibinfo{author}{Wang, Q.} \emph{et~al.}
\newblock \bibinfo{title}{Experimental decoy-state quantum key distribution
  with a sub-poissionian heralded single-photon source}.
\newblock \emph{\bibinfo{journal}{Phys. Rev. Lett.}}
  \textbf{\bibinfo{volume}{100}}, \bibinfo{pages}{090501}
  (\bibinfo{year}{2008}).
\newblock
  \urlprefix\url{https://link.aps.org/doi/10.1103/PhysRevLett.100.090501}.

\bibitem{MCH:Arx22}
\bibinfo{author}{Murtaza, G.} \emph{et~al.}
\newblock \bibinfo{title}{Efficient room-temperature molecular single-photon
  sources for quantum key distribution. preprint at
  https://arxiv.org/abs/2202.12635} (\bibinfo{year}{2022}).
\newblock \urlprefix\url{https://arxiv.org/abs/2202.12635}.

\bibitem{L:NatPhot19}
\bibinfo{author}{Loredo, J.~C.} \emph{et~al.}
\newblock \bibinfo{title}{Generation of non-classical light in a photon-number
  superposition}.
\newblock \emph{\bibinfo{journal}{Nat. Photonics}}
  \textbf{\bibinfo{volume}{13}}, \bibinfo{pages}{803--808}
  (\bibinfo{year}{2019}).
\newblock \urlprefix\url{http://dx.doi.org/10.1038/s41566-019-0506-3}.

\bibitem{CZL:NJP15}
\bibinfo{author}{Cao, Z.}, \bibinfo{author}{Zhang, Z.}, \bibinfo{author}{Lo,
  H.-K.} \& \bibinfo{author}{Ma, X.}
\newblock \bibinfo{title}{Discrete-phase-randomized coherent state source and
  its application in quantum key distribution}.
\newblock \emph{\bibinfo{journal}{New J. Phys.}} \textbf{\bibinfo{volume}{17}},
  \bibinfo{pages}{053014} (\bibinfo{year}{2015}).
\newblock \urlprefix\url{https://doi.org/10.1088/1367-2630/17/5/053014}.

\bibitem{WYH:NatPhot21}
\bibinfo{author}{Wang, S.} \emph{et~al.}
\newblock \bibinfo{title}{Twin-field quantum key distribution over 830-km
  fibre}.
\newblock \emph{\bibinfo{journal}{Nat. Photonics}}
  \textbf{\bibinfo{volume}{16}}, \bibinfo{pages}{154--161}
  (\bibinfo{year}{2022}).

\bibitem{LYD:Nature18}
\bibinfo{author}{Lucamarini, M.}, \bibinfo{author}{Yuan, Z.~L.},
  \bibinfo{author}{Dynes, J.~F.} \& \bibinfo{author}{Shields, A.~J.}
\newblock \bibinfo{title}{Overcoming the rate–distance limit of quantum key
  distribution without quantum repeaters}.
\newblock \emph{\bibinfo{journal}{Nature}} \textbf{\bibinfo{volume}{557}},
  \bibinfo{pages}{400–403} (\bibinfo{year}{2018}).
\newblock \urlprefix\url{http://dx.doi.org/10.1038/s41586-018-0066-6}.

\bibitem{BOV:npj18}
\bibinfo{author}{Bozzio, M.} \emph{et~al.}
\newblock \bibinfo{title}{Experimental investigation of practical unforgeable
  quantum money}.
\newblock \emph{\bibinfo{journal}{Npj Quantum Inf.}}
  \textbf{\bibinfo{volume}{4}}, \bibinfo{pages}{5} (\bibinfo{year}{2018}).
\newblock \urlprefix\url{https://doi.org/10.1038/s41534-018-0058-2}.

\bibitem{GAA:pra18}
\bibinfo{author}{Guan, J.-Y.} \emph{et~al.}
\newblock \bibinfo{title}{Experimental preparation and verification of quantum
  money}.
\newblock \emph{\bibinfo{journal}{Phys. Rev. A}} \textbf{\bibinfo{volume}{97}},
  \bibinfo{pages}{032338} (\bibinfo{year}{2018}).
\newblock \eprint{1709.05882}.

\bibitem{Kent:npjQI22}
\bibinfo{author}{Kent, A.}, \bibinfo{author}{Lowndes, D.},
  \bibinfo{author}{Pital{\'{u}}a-Garc{\'{\i}}a, D.} \& \bibinfo{author}{Rarity,
  J.}
\newblock \emph{\bibinfo{journal}{Npj Quantum Inf.}}
  \textbf{\bibinfo{volume}{8}} (\bibinfo{year}{2022}).
\newblock \urlprefix\url{https://doi.org/10.1038%2Fs41534-022-00524-4}.

\bibitem{PJ+:natcomm14}
\bibinfo{author}{Pappa, A.} \emph{et~al.}
\newblock \bibinfo{title}{Experimental plug and play quantum coin flipping}.
\newblock \emph{\bibinfo{journal}{Nat. Commun.}} \textbf{\bibinfo{volume}{5}},
  \bibinfo{pages}{3717} (\bibinfo{year}{2014}).

\bibitem{BBB:NC11}
\bibinfo{author}{Berl{\'\i}n, G.} \emph{et~al.}
\newblock \bibinfo{title}{Experimental loss-tolerant quantum coin flipping}.
\newblock \emph{\bibinfo{journal}{Nat. Commun.}} \textbf{\bibinfo{volume}{2}},
  \bibinfo{pages}{561} (\bibinfo{year}{2011}).

\bibitem{BCK:PRA20}
\bibinfo{author}{Bozzio, M.}, \bibinfo{author}{Chabaud, U.},
  \bibinfo{author}{Kerenidis, I.} \& \bibinfo{author}{Diamanti, E.}
\newblock \bibinfo{title}{Quantum weak coin flipping with a single photon}.
\newblock \emph{\bibinfo{journal}{Phys. Rev. A}}
  \textbf{\bibinfo{volume}{102}}, \bibinfo{pages}{022414}
  (\bibinfo{year}{2020}).
\newblock \urlprefix\url{https://link.aps.org/doi/10.1103/PhysRevA.102.022414}.

\bibitem{NJC:NatComms12}
\bibinfo{author}{Ng, S.~K., N. Huei Y.and~Joshi}, \bibinfo{author}{Chen~Ming,
  C.}, \bibinfo{author}{Kurtsiefer, C.} \& \bibinfo{author}{Wehner, S.}
\newblock \bibinfo{title}{Experimental implementation of bit commitment in the
  noisy-storage model}.
\newblock \emph{\bibinfo{journal}{Nat. Commun.}} \textbf{\bibinfo{volume}{3}}
  (\bibinfo{year}{2012}).

\bibitem{Zbind:PRL13}
\bibinfo{author}{Lunghi, T.} \emph{et~al.}
\newblock \bibinfo{title}{Experimental bit commitment based on quantum
  communication and special relativity}.
\newblock \emph{\bibinfo{journal}{Phys. Rev. Lett.}}
  \textbf{\bibinfo{volume}{111}}, \bibinfo{pages}{180504}
  (\bibinfo{year}{2013}).
\newblock
  \urlprefix\url{https://link.aps.org/doi/10.1103/PhysRevLett.111.180504}.

\bibitem{Pan:PRL14}
\bibinfo{author}{Liu, Y.} \emph{et~al.}
\newblock \bibinfo{title}{Experimental unconditionally secure bit commitment}.
\newblock \emph{\bibinfo{journal}{Phys. Rev. Lett.}}
  \textbf{\bibinfo{volume}{112}}, \bibinfo{pages}{010504}
  (\bibinfo{year}{2014}).
\newblock
  \urlprefix\url{https://link.aps.org/doi/10.1103/PhysRevLett.112.010504}.

\bibitem{UPW:SciAdv20}
\bibinfo{author}{Uppu, R.} \emph{et~al.}
\newblock \bibinfo{title}{Scalable integrated single-photon source}.
\newblock \emph{\bibinfo{journal}{Sci. Adv.}} \textbf{\bibinfo{volume}{6}},
  \bibinfo{pages}{eabc8268} (\bibinfo{year}{2020}).
\newblock
  \urlprefix\url{https://www.science.org/doi/abs/10.1126/sciadv.abc8268}.

\bibitem{SGS:NP16}
\bibinfo{author}{Somaschi, N.} \emph{et~al.}
\newblock \bibinfo{title}{Near-optimal single-photon sources in the solid
  state}.
\newblock \emph{\bibinfo{journal}{Nat. Photonics}}
  \textbf{\bibinfo{volume}{10}}, \bibinfo{pages}{340--345}
  (\bibinfo{year}{2016}).
\newblock \urlprefix\url{http://dx.doi.org/10.1038/nphoton.2016.23}.

\bibitem{Pan:NNano2013}
\bibinfo{author}{He, Y.-M.} \emph{et~al.}
\newblock \bibinfo{title}{On-demand semiconductor single-photon source with
  near-unity indistinguishability}.
\newblock \emph{\bibinfo{journal}{Nat. Nanotechnol.}}
  \textbf{\bibinfo{volume}{8}}, \bibinfo{pages}{213--217}
  (\bibinfo{year}{2013}).
\newblock \urlprefix\url{http://dx.doi.org/10.1038/nnano.2012.262}.

\bibitem{Lodahl:PRL2014}
\bibinfo{author}{Arcari, M.} \emph{et~al.}
\newblock \bibinfo{title}{Near-unity coupling efficiency of a quantum emitter
  to a photonic crystal waveguide}.
\newblock \emph{\bibinfo{journal}{Phys. Rev. Lett.}}
  \textbf{\bibinfo{volume}{113}}, \bibinfo{pages}{093603}
  (\bibinfo{year}{2014}).
\newblock
  \urlprefix\url{https://link.aps.org/doi/10.1103/PhysRevLett.113.093603}.

\bibitem{Pan:NatPhys19}
\bibinfo{author}{He, Y.-M.} \emph{et~al.}
\newblock \bibinfo{title}{Coherently driving a single quantum two-level system
  with dichromatic laser pulses}.
\newblock \emph{\bibinfo{journal}{Nat. Phys.}} \textbf{\bibinfo{volume}{15}},
  \bibinfo{pages}{941–946} (\bibinfo{year}{2019}).
\newblock \urlprefix\url{http://dx.doi.org/10.1038/s41567-019-0585-6}.

\bibitem{TJA:NatNan21}
\bibinfo{author}{Tomm, N.} \emph{et~al.}
\newblock \bibinfo{title}{A bright and fast source of coherent single photons}.
\newblock \emph{\bibinfo{journal}{Nat. Nanotechnol.}}
  \textbf{\bibinfo{volume}{16}}, \bibinfo{pages}{399–403}
  (\bibinfo{year}{2021}).
\newblock \urlprefix\url{http://dx.doi.org/10.1038/s41565-020-00831-x}.

\bibitem{Armando:ACS17}
\bibinfo{author}{Reindl, M.} \emph{et~al.}
\newblock \bibinfo{title}{Phonon-assisted two-photon interference from remote
  quantum emitters}.
\newblock \emph{\bibinfo{journal}{Nano Lett.}} \textbf{\bibinfo{volume}{17}},
  \bibinfo{pages}{4090--4095} (\bibinfo{year}{2017}).
\newblock \urlprefix\url{http://dx.doi.org/10.1021/acs.nanolett.7b00777}.

\bibitem{Fischer:IOP17}
\bibinfo{author}{Fischer, K.~A.} \emph{et~al.}
\newblock \bibinfo{title}{Pulsed rabi oscillations in quantum two-level
  systems: beyond the area theorem}.
\newblock \emph{\bibinfo{journal}{Quantum Sci. Technol.}}
  \textbf{\bibinfo{volume}{3}}, \bibinfo{pages}{014006} (\bibinfo{year}{2017}).
\newblock \urlprefix\url{http://dx.doi.org/10.1088/2058-9565/aa9269}.

\bibitem{Fisher:NPJ18}
\bibinfo{author}{Hanschke, L.} \emph{et~al.}
\newblock \bibinfo{title}{Quantum dot single-photon sources with ultra-low
  multi-photon probability}.
\newblock \emph{\bibinfo{journal}{Npj Quantum Inf.}}
  \textbf{\bibinfo{volume}{4}} (\bibinfo{year}{2018}).
\newblock \urlprefix\url{http://dx.doi.org/10.1038/s41534-018-0092-0}.

\bibitem{LR:IOP19}
\bibinfo{author}{L{\"u}ker, S.} \& \bibinfo{author}{Reiter, D.~E.}
\newblock \bibinfo{title}{A review on optical excitation of semiconductor
  quantum dots under the influence of phonons}.
\newblock \emph{\bibinfo{journal}{Semicond. Sci. Technol.}}
  \textbf{\bibinfo{volume}{34}}, \bibinfo{pages}{063002}
  (\bibinfo{year}{2019}).
\newblock \urlprefix\url{https://doi.org/10.1088/1361-6641/ab1c14}.

\bibitem{Axt:PRL19}
\bibinfo{author}{Cosacchi, M.}, \bibinfo{author}{Ungar, F.},
  \bibinfo{author}{Cygorek, M.}, \bibinfo{author}{Vagov, A.} \&
  \bibinfo{author}{Axt, V.~M.}
\newblock \bibinfo{title}{Emission-frequency separated high quality
  single-photon sources enabled by phonons}.
\newblock \emph{\bibinfo{journal}{Phys. Rev. Lett.}}
  \textbf{\bibinfo{volume}{123}}, \bibinfo{pages}{017403}
  (\bibinfo{year}{2019}).
\newblock
  \urlprefix\url{https://link.aps.org/doi/10.1103/PhysRevLett.123.017403}.

\bibitem{LAParis:21}
\bibinfo{author}{Thomas, S.~E.} \emph{et~al.}
\newblock \bibinfo{title}{Bright polarized single-photon source based on a
  linear dipole}.
\newblock \emph{\bibinfo{journal}{Phys. Rev. Lett.}}
  \textbf{\bibinfo{volume}{126}}, \bibinfo{pages}{233601}
  (\bibinfo{year}{2021}).
\newblock
  \urlprefix\url{https://link.aps.org/doi/10.1103/PhysRevLett.126.233601}.

\bibitem{Ding2016}
\bibinfo{author}{Ding, X.} \emph{et~al.}
\newblock \bibinfo{title}{On-demand single photons with high extraction
  efficiency and near-unity indistinguishability from a resonantly driven
  quantum dot in a micropillar}.
\newblock \emph{\bibinfo{journal}{Phys. Rev. Lett.}}
  \textbf{\bibinfo{volume}{116}}, \bibinfo{pages}{020401}
  (\bibinfo{year}{2016}).
\newblock
  \urlprefix\url{https://link.aps.org/doi/10.1103/PhysRevLett.116.020401}.

\bibitem{SJZ:APL18}
\bibinfo{author}{Schweickert, L.} \emph{et~al.}
\newblock \bibinfo{title}{On-demand generation of background-free single
  photons from a solid-state source}.
\newblock \emph{\bibinfo{journal}{Appl. Phys. Lett.}}
  \textbf{\bibinfo{volume}{112}}, \bibinfo{pages}{093106}
  (\bibinfo{year}{2018}).
\newblock \urlprefix\url{http://dx.doi.org/10.1063/1.5020038}.

\bibitem{Mueller2014}
\bibinfo{author}{M{\"u}ller, M.}, \bibinfo{author}{Bounouar, S.},
  \bibinfo{author}{J{\"o}ns, K.~D.}, \bibinfo{author}{Gl{\"a}ssl, M.} \&
  \bibinfo{author}{Michler, P.}
\newblock \bibinfo{title}{On-demand generation of indistinguishable
  polarization-entangled photon pairs}.
\newblock \emph{\bibinfo{journal}{Nat. Photonics}}
  \textbf{\bibinfo{volume}{8}}, \bibinfo{pages}{224} (\bibinfo{year}{2014}).
\newblock \urlprefix\url{http://dx.doi.org/10.1038/nphoton.2013.377}.

\bibitem{SHS:PRL22}
\bibinfo{author}{Sbresny, F.} \emph{et~al.}
\newblock \bibinfo{title}{Stimulated generation of indistinguishable single
  photons from a quantum ladder system}.
\newblock \emph{\bibinfo{journal}{Phys. Rev. Lett.}}
  \textbf{\bibinfo{volume}{128}}, \bibinfo{pages}{093603}
  (\bibinfo{year}{2022}).
\newblock
  \urlprefix\url{https://link.aps.org/doi/10.1103/PhysRevLett.128.093603}.

\bibitem{WLL:NatNano22}
\bibinfo{author}{Wei, Y.} \emph{et~al.}
\newblock \emph{\bibinfo{journal}{Nat. Nanotechnol.}}
  \textbf{\bibinfo{volume}{17}}, \bibinfo{pages}{470--476}
  (\bibinfo{year}{2022}).
\newblock \urlprefix\url{https://doi.org/10.1038%2Fs41565-022-01092-6}.

\bibitem{YLL:NanoLett22}
\bibinfo{author}{Yan, J.} \emph{et~al.}
\newblock \emph{\bibinfo{journal}{Nano Lett.}} \textbf{\bibinfo{volume}{22}},
  \bibinfo{pages}{1483--1490} (\bibinfo{year}{2022}).
\newblock \urlprefix\url{https://doi.org/10.1021%2Facs.nanolett.1c03543}.

\bibitem{GLLP04}
\bibinfo{author}{Gottesman, D.}, \bibinfo{author}{Lo, H.-K.},
  \bibinfo{author}{L\"{u}tkenhaus, N.} \& \bibinfo{author}{Preskill, J.}
\newblock \bibinfo{title}{Security of quantum key distribution with imperfect
  devices}.
\newblock \emph{\bibinfo{journal}{Quantum Info. Comput.}}
  \textbf{\bibinfo{volume}{4}}, \bibinfo{pages}{325--360}
  (\bibinfo{year}{2004}).

\bibitem{KGR:PRL05}
\bibinfo{author}{Kraus, B.}, \bibinfo{author}{Gisin, N.} \&
  \bibinfo{author}{Renner, R.}
\newblock \bibinfo{title}{Lower and upper bounds on the secret-key rate for
  quantum key distribution protocols using one-way classical communication}.
\newblock \emph{\bibinfo{journal}{Phys. Rev. Lett.}}
  \textbf{\bibinfo{volume}{95}}, \bibinfo{pages}{080501}
  (\bibinfo{year}{2005}).
\newblock
  \urlprefix\url{https://link.aps.org/doi/10.1103/PhysRevLett.95.080501}.

\bibitem{BDG:PRA19}
\bibinfo{author}{Bozzio, M.}, \bibinfo{author}{Diamanti, E.} \&
  \bibinfo{author}{Grosshans, F.}
\newblock \bibinfo{title}{Semi-device-independent quantum money with coherent
  states}.
\newblock \emph{\bibinfo{journal}{Phys. Rev. A}} \textbf{\bibinfo{volume}{99}},
  \bibinfo{pages}{022336} (\bibinfo{year}{2019}).
\newblock \urlprefix\url{https://link.aps.org/doi/10.1103/PhysRevA.99.022336}.

\bibitem{Wang:2019XX}
\bibinfo{author}{Wang, H.} \emph{et~al.}
\newblock \bibinfo{title}{On-demand semiconductor source of entangled photons
  which simultaneously has high fidelity, efficiency, and
  indistinguishability}.
\newblock \emph{\bibinfo{journal}{Phys. Rev. Lett.}}
  \textbf{\bibinfo{volume}{122}}, \bibinfo{pages}{113602}
  (\bibinfo{year}{2019}).
\newblock
  \urlprefix\url{https://link.aps.org/doi/10.1103/PhysRevLett.122.113602}.

\bibitem{Ding:2016}
\bibinfo{author}{Ding, X.} \emph{et~al.}
\newblock \bibinfo{title}{On-demand single photons with high extraction
  efficiency and near-unity indistinguishability from a resonantly driven
  quantum dot in a micropillar}.
\newblock \emph{\bibinfo{journal}{Phys. Rev. Lett.}}
  \textbf{\bibinfo{volume}{116}}, \bibinfo{pages}{020401}
  (\bibinfo{year}{2016}).
\newblock
  \urlprefix\url{https://link.aps.org/doi/10.1103/PhysRevLett.116.020401}.

\bibitem{KTO:PRA14}
\bibinfo{author}{Kobayashi, T.}, \bibinfo{author}{Tomita, A.} \&
  \bibinfo{author}{Okamoto, A.}
\newblock \bibinfo{title}{Evaluation of the phase randomness of a light source
  in quantum-key-distribution systems with an attenuated laser}.
\newblock \emph{\bibinfo{journal}{Phys. Rev. A}} \textbf{\bibinfo{volume}{90}},
  \bibinfo{pages}{032320} (\bibinfo{year}{2014}).
\newblock \urlprefix\url{https://link.aps.org/doi/10.1103/PhysRevA.90.032320}.

\bibitem{TYM:PRA13}
\bibinfo{author}{Tang, Y.-L.} \emph{et~al.}
\newblock \bibinfo{title}{Source attack of decoy-state quantum key distribution
  using phase information}.
\newblock \emph{\bibinfo{journal}{Phys. Rev. A}} \textbf{\bibinfo{volume}{88}},
  \bibinfo{pages}{022308} (\bibinfo{year}{2013}).
\newblock \urlprefix\url{https://link.aps.org/doi/10.1103/PhysRevA.88.022308}.

\bibitem{DJL:pra00}
\bibinfo{author}{Duvsek, M.}, \bibinfo{author}{Jahma, M.} \&
  \bibinfo{author}{N.Lütkenhaus}.
\newblock \bibinfo{title}{Unambiguous state discrimination in quantum
  cryptography with weak coherent states}.
\newblock \emph{\bibinfo{journal}{Phys. Rev. A}} \textbf{\bibinfo{volume}{62}},
  \bibinfo{pages}{022306} (\bibinfo{year}{2000}).
\newblock \urlprefix\url{https://doi.org/10.1103/PhysRevA.62.022306}.

\bibitem{LBO:NatNanotech18}
\bibinfo{author}{Liu, F.} \emph{et~al.}
\newblock \bibinfo{title}{High purcell factor generation of indistinguishable
  on-chip single photons}.
\newblock \emph{\bibinfo{journal}{Nat. Nanotechnol.}}
  \textbf{\bibinfo{volume}{13}}, \bibinfo{pages}{835--840}
  (\bibinfo{year}{2018}).
\newblock \urlprefix\url{http://dx.doi.org/10.1038/s41565-018-0188-x}.

\bibitem{GG:PRL02}
\bibinfo{author}{Grosshans, F.} \& \bibinfo{author}{Grangier, P.}
\newblock \bibinfo{title}{Continuous variable quantum cryptography using
  coherent states}.
\newblock \emph{\bibinfo{journal}{Phys. Rev. Lett.}}
  \textbf{\bibinfo{volume}{88}}, \bibinfo{pages}{057902}
  (\bibinfo{year}{2002}).
\newblock
  \urlprefix\url{https://link.aps.org/doi/10.1103/PhysRevLett.88.057902}.

\bibitem{JKL:NatPhot13}
\bibinfo{author}{Jouguet, P.}, \bibinfo{author}{Kunz-Jacques, S.},
  \bibinfo{author}{Leverrier, A.}, \bibinfo{author}{Grangier, P.} \&
  \bibinfo{author}{Diamanti, E.}
\newblock \bibinfo{title}{Experimental demonstration of long-distance
  continuous-variable quantum key distribution}.
\newblock \emph{\bibinfo{journal}{Nat. Photonics}}
  \textbf{\bibinfo{volume}{7}}, \bibinfo{pages}{378--381}
  (\bibinfo{year}{2013}).
\newblock \urlprefix\url{https://doi.org/10.1038%2Fnphoton.2013.63}.

\bibitem{VEA:OptExp20}
\bibinfo{author}{Valivarthi, R.}, \bibinfo{author}{Etcheverry, S.},
  \bibinfo{author}{Aldama, J.}, \bibinfo{author}{Zwiehoff, F.} \&
  \bibinfo{author}{Pruneri, V.}
\newblock \bibinfo{title}{Plug-and-play continuous-variable quantum key
  distribution for metropolitan networks}.
\newblock \emph{\bibinfo{journal}{Opt. Express}} \textbf{\bibinfo{volume}{28}},
  \bibinfo{pages}{14547--14559} (\bibinfo{year}{2020}).
\newblock
  \urlprefix\url{http://opg.optica.org/oe/abstract.cfm?URI=oe-28-10-14547}.

\bibitem{KNB:Nano21}
\bibinfo{author}{Kolatschek, S.} \emph{et~al.}
\newblock \bibinfo{title}{Bright purcell enhanced single-photon source in the
  telecom o-band based on a quantum dot in a circular bragg grating}.
\newblock \emph{\bibinfo{journal}{Nano Lett.}}
  \textbf{\bibinfo{volume}{21,18}}, \bibinfo{pages}{7740–7745}
  (\bibinfo{year}{2021}).
\newblock \urlprefix\url{https://doi.org/10.1021/acs.nanolett.1c02647}.

\bibitem{NVF:APL21}
\bibinfo{author}{Nawrath, C.} \emph{et~al.}
\newblock \bibinfo{title}{Resonance fluorescence of single in(ga)as quantum
  dots emitting in the telecom c-band}.
\newblock \emph{\bibinfo{journal}{Appl. Phys. Lett.}}
  \textbf{\bibinfo{volume}{118}}, \bibinfo{pages}{244002}
  (\bibinfo{year}{2021}).
\newblock \urlprefix\url{https://doi.org/10.1063/5.0048695}.
\newblock \eprint{https://doi.org/10.1063/5.0048695}.

\bibitem{TTH:JAP07}
\bibinfo{author}{Takemoto, K.}, \bibinfo{author}{Takatsu, M.},
  \bibinfo{author}{Hirose, S.} \& \bibinfo{author}{Yokoyama, N.}
\newblock \bibinfo{title}{An optical horn structure for single-photon source
  using quantum dots at telecommunication wavelength}.
\newblock \emph{\bibinfo{journal}{J. Appl. Phys.}}
  \textbf{\bibinfo{volume}{101}}, \bibinfo{pages}{081720}
  (\bibinfo{year}{2007}).

\bibitem{SNK:Nano22}
\bibinfo{author}{Sittig, R.} \emph{et~al.}
\newblock \emph{\bibinfo{journal}{Nanophotonics}}
  \textbf{\bibinfo{volume}{11}}, \bibinfo{pages}{1109--1116}
  (\bibinfo{year}{2022}).
\newblock \urlprefix\url{https://doi.org/10.1515/nanoph-2021-0552}.

\bibitem{MTN:APL16}
\bibinfo{author}{Miyazawa, T.} \emph{et~al.}
\newblock \emph{\bibinfo{journal}{Appl. Phys. Lett.}}
  \textbf{\bibinfo{volume}{109}}, \bibinfo{pages}{132106}
  (\bibinfo{year}{2016}).
\newblock \urlprefix\url{https://doi.org/10.1063/1.4961888}.

\bibitem{VBG:PRL20}
\bibinfo{author}{van Leent, T.} \emph{et~al.}
\newblock \bibinfo{title}{Long-distance distribution of atom-photon
  entanglement at telecom wavelength}.
\newblock \emph{\bibinfo{journal}{Phys. Rev. Lett.}}
  \textbf{\bibinfo{volume}{124}}, \bibinfo{pages}{010510}
  (\bibinfo{year}{2020}).
\newblock\urlprefix\url{https://link.aps.org/doi/10.1103/PhysRevLett.124.010510}.

\bibitem{BMH:NatComms14}
\bibinfo{author}{Bell, B.~A.} \emph{et~al.}
\newblock \bibinfo{title}{Experimental demonstration of graph-state quantum
  secret sharing}.
\newblock \emph{\bibinfo{journal}{Nat. Commun.}} \textbf{\bibinfo{volume}{5}}
  (\bibinfo{year}{2014}).
\newblock \urlprefix\url{http://dx.doi.org/10.1038/ncomms6480}.

\bibitem{UMY:PRL19}
\bibinfo{author}{Unnikrishnan, A.} \emph{et~al.}
\newblock \bibinfo{title}{Anonymity for practical quantum networks}.
\newblock \emph{\bibinfo{journal}{Phys. Rev. Lett.}}
  \textbf{\bibinfo{volume}{122}}, \bibinfo{pages}{240501}
  (\bibinfo{year}{2019}).
\newblock
  \urlprefix\url{https://link.aps.org/doi/10.1103/PhysRevLett.122.240501}.

\bibitem{Besombes2001}
\bibinfo{author}{Besombes, L.}, \bibinfo{author}{Kheng, K.},
  \bibinfo{author}{Marsal, L.} \& \bibinfo{author}{Mariette, H.}
\newblock \bibinfo{title}{Acoustic phonon broadening mechanism in single
  quantum dot emission}.
\newblock \emph{\bibinfo{journal}{Phys. Rev. B}} \textbf{\bibinfo{volume}{63}},
  \bibinfo{pages}{155307} (\bibinfo{year}{2001}).
\newblock \urlprefix\url{https://link.aps.org/doi/10.1103/PhysRevB.63.155307}.

\bibitem{Borri2001}
\bibinfo{author}{Borri, P.} \emph{et~al.}
\newblock \bibinfo{title}{Ultralong dephasing time in {I}n{G}a{A}s quantum
  dots}.
\newblock \emph{\bibinfo{journal}{Phys. Rev. Lett.}}
  \textbf{\bibinfo{volume}{87}}, \bibinfo{pages}{157401}
  (\bibinfo{year}{2001}).
\newblock
  \urlprefix\url{https://link.aps.org/doi/10.1103/PhysRevLett.87.157401}.

\bibitem{Axt2005}
\bibinfo{author}{Axt, V.~M.}, \bibinfo{author}{Kuhn, T.},
  \bibinfo{author}{Vagov, A.} \& \bibinfo{author}{Peeters, F.~M.}
\newblock \bibinfo{title}{Phonon-induced pure dephasing in exciton-biexciton
  quantum dot systems driven by ultrafast laser pulse sequences}.
\newblock \emph{\bibinfo{journal}{Phys. Rev. B}} \textbf{\bibinfo{volume}{72}},
  \bibinfo{pages}{125309} (\bibinfo{year}{2005}).
\newblock \urlprefix\url{https://link.aps.org/doi/10.1103/PhysRevB.72.125309}.

\bibitem{Reiter2019}
\bibinfo{author}{Reiter, D.~E.}, \bibinfo{author}{Kuhn, T.} \&
  \bibinfo{author}{Axt, V.~M.}
\newblock \bibinfo{title}{Distinctive characteristics of carrier-phonon
  interactions in optically driven semiconductor quantum dots}.
\newblock \emph{\bibinfo{journal}{Advances in Physics: X}}
  \textbf{\bibinfo{volume}{4}}, \bibinfo{pages}{1655478}
  (\bibinfo{year}{2019}).
\newblock \urlprefix\url{https://doi.org/10.1080/23746149.2019.1655478}.

\bibitem{Vagov2011}
\bibinfo{author}{Vagov, A.}, \bibinfo{author}{Croitoru, M.~D.},
  \bibinfo{author}{Gl\"assl, M.}, \bibinfo{author}{Axt, V.~M.} \&
  \bibinfo{author}{Kuhn, T.}
\newblock \bibinfo{title}{Real-time path integrals for quantum dots: {Q}uantum
  dissipative dynamics with superohmic environment coupling}.
\newblock \emph{\bibinfo{journal}{Phys. Rev. B}} \textbf{\bibinfo{volume}{83}},
  \bibinfo{pages}{094303} (\bibinfo{year}{2011}).
\newblock \urlprefix\url{https://link.aps.org/doi/10.1103/PhysRevB.83.094303}.

\bibitem{Cygorek2017}
\bibinfo{author}{Cygorek, M.}, \bibinfo{author}{Barth, A.~M.},
  \bibinfo{author}{Ungar, F.}, \bibinfo{author}{Vagov, A.} \&
  \bibinfo{author}{Axt, V.~M.}
\newblock \bibinfo{title}{Nonlinear cavity feeding and unconventional photon
  statistics in solid-state cavity {QED} revealed by many-level real-time
  path-integral calculations}.
\newblock \emph{\bibinfo{journal}{Phys. Rev. B}} \textbf{\bibinfo{volume}{96}},
  \bibinfo{pages}{201201(R)} (\bibinfo{year}{2017}).
\newblock \urlprefix\url{https://link.aps.org/doi/10.1103/PhysRevB.96.201201}.

\bibitem{Cosacchi2018}
\bibinfo{author}{Cosacchi, M.} \emph{et~al.}
\newblock \bibinfo{title}{Path-integral approach for nonequilibrium multitime
  correlation functions of open quantum systems coupled to {M}arkovian and
  non-{M}arkovian environments}.
\newblock \emph{\bibinfo{journal}{Phys. Rev. B}} \textbf{\bibinfo{volume}{98}},
  \bibinfo{pages}{125302} (\bibinfo{year}{2018}).
\newblock \urlprefix\url{https://link.aps.org/doi/10.1103/PhysRevB.98.125302}.


\end{thebibliography}

\begin{thebibliography}{58}
\expandafter\ifx\csname natexlab\endcsname\relax\def\natexlab#1{#1}\fi
\expandafter\ifx\csname bibnamefont\endcsname\relax
  \def\bibnamefont#1{#1}\fi
\expandafter\ifx\csname bibfnamefont\endcsname\relax
  \def\bibfnamefont#1{#1}\fi
\expandafter\ifx\csname citenamefont\endcsname\relax
  \def\citenamefont#1{#1}\fi
\expandafter\ifx\csname url\endcsname\relax
  \def\url#1{\texttt{#1}}\fi
\expandafter\ifx\csname urlprefix\endcsname\relax\def\urlprefix{URL }\fi
\providecommand{\bibinfo}[2]{#2}
\providecommand{\eprint}[2][]{\url{#2}}

\bibitem[{\citenamefont{Krummheuer et~al.}(2002)\citenamefont{Krummheuer, Axt,
  and Kuhn}}]{Krummheuer2002}
\bibinfo{author}{\bibfnamefont{B.}~\bibnamefont{Krummheuer}},
  \bibinfo{author}{\bibfnamefont{V.~M.} \bibnamefont{Axt}}, \bibnamefont{and}
  \bibinfo{author}{\bibfnamefont{T.}~\bibnamefont{Kuhn}},
  \bibinfo{journal}{Phys. Rev. B} \textbf{\bibinfo{volume}{65}},
  \bibinfo{pages}{195313} (\bibinfo{year}{2002}),
  \urlprefix\url{https://link.aps.org/doi/10.1103/PhysRevB.65.195313}.


\bibitem[{\citenamefont{Krummheuer et~al.}(2005)\citenamefont{Krummheuer, Axt,
  Kuhn, D'Amico, and Rossi}}]{Krummheuer2005}
\bibinfo{author}{\bibfnamefont{B.}~\bibnamefont{Krummheuer}},
  \bibinfo{author}{\bibfnamefont{V.~M.} \bibnamefont{Axt}},
  \bibinfo{author}{\bibfnamefont{T.}~\bibnamefont{Kuhn}},
  \bibinfo{author}{\bibfnamefont{I.}~\bibnamefont{D'Amico}}, \bibnamefont{and}
  \bibinfo{author}{\bibfnamefont{F.}~\bibnamefont{Rossi}},
  \bibinfo{journal}{Phys. Rev. B} \textbf{\bibinfo{volume}{71}},
  \bibinfo{pages}{235329} (\bibinfo{year}{2005}),
  \urlprefix\url{https://link.aps.org/doi/10.1103/PhysRevB.71.235329}.


\bibitem[{\citenamefont{Bounouar et~al.}(2015)\citenamefont{Bounouar, M\"uller,
  Barth, Gl\"assl, Axt, and Michler}}]{Bounouar2015}
\bibinfo{author}{\bibfnamefont{S.}~\bibnamefont{Bounouar}},
  \bibinfo{author}{\bibfnamefont{M.}~\bibnamefont{M\"uller}},
  \bibinfo{author}{\bibfnamefont{A.~M.} \bibnamefont{Barth}},
  \bibinfo{author}{\bibfnamefont{M.}~\bibnamefont{Gl\"assl}},
  \bibinfo{author}{\bibfnamefont{V.~M.} \bibnamefont{Axt}}, \bibnamefont{and}
  \bibinfo{author}{\bibfnamefont{P.}~\bibnamefont{Michler}},
  \bibinfo{journal}{Phys. Rev. B} \textbf{\bibinfo{volume}{91}},
  \bibinfo{pages}{161302} (\bibinfo{year}{2015}),
  \urlprefix\url{https://link.aps.org/doi/10.1103/PhysRevB.91.161302}.

\bibitem[{\citenamefont{Quilter et~al.}(2015)\citenamefont{Quilter, Brash, Liu,
  Gl\"assl, Barth, Axt, Ramsay, Skolnick, and Fox}}]{Quilter2015}
\bibinfo{author}{\bibfnamefont{J.~H.} \bibnamefont{Quilter}},
  \bibinfo{author}{\bibfnamefont{A.~J.} \bibnamefont{Brash}},
  \bibinfo{author}{\bibfnamefont{F.}~\bibnamefont{Liu}},
  \bibinfo{author}{\bibfnamefont{M.}~\bibnamefont{Gl\"assl}},
  \bibinfo{author}{\bibfnamefont{A.~M.} \bibnamefont{Barth}},
  \bibinfo{author}{\bibfnamefont{V.~M.} \bibnamefont{Axt}},
  \bibinfo{author}{\bibfnamefont{A.~J.} \bibnamefont{Ramsay}},
  \bibinfo{author}{\bibfnamefont{M.~S.} \bibnamefont{Skolnick}},
  \bibnamefont{and} \bibinfo{author}{\bibfnamefont{A.~M.} \bibnamefont{Fox}},
  \bibinfo{journal}{Phys. Rev. Lett.} \textbf{\bibinfo{volume}{114}},
  \bibinfo{pages}{137401} (\bibinfo{year}{2015}),
  \urlprefix\url{https://link.aps.org/doi/10.1103/PhysRevLett.114.137401}.

\bibitem[{\citenamefont{Kaldewey et~al.}(2017)\citenamefont{Kaldewey,
  L{\"u}ker, Kuhlmann, Valentin, Chauveau, Ludwig, Wieck, Reiter, Kuhn, and
  Warburton}}]{kaldewey2017demonstrating}
\bibinfo{author}{\bibfnamefont{T.}~\bibnamefont{Kaldewey}},
  \bibinfo{author}{\bibfnamefont{S.}~\bibnamefont{L{\"u}ker}},
  \bibinfo{author}{\bibfnamefont{A.~V.} \bibnamefont{Kuhlmann}},
  \bibinfo{author}{\bibfnamefont{S.~R.} \bibnamefont{Valentin}},
  \bibinfo{author}{\bibfnamefont{J.-M.} \bibnamefont{Chauveau}},
  \bibinfo{author}{\bibfnamefont{A.}~\bibnamefont{Ludwig}},
  \bibinfo{author}{\bibfnamefont{A.~D.} \bibnamefont{Wieck}},
  \bibinfo{author}{\bibfnamefont{D.~E.} \bibnamefont{Reiter}},
  \bibinfo{author}{\bibfnamefont{T.}~\bibnamefont{Kuhn}}, \bibnamefont{and}
  \bibinfo{author}{\bibfnamefont{R.~J.} \bibnamefont{Warburton}},
  \bibinfo{journal}{Phys. Rev. B} \textbf{\bibinfo{volume}{95}},
  \bibinfo{pages}{241306} (\bibinfo{year}{2017}).

\bibitem[{\citenamefont{L\"uker et~al.}(2017)\citenamefont{L\"uker, Kuhn, and
  Reiter}}]{Luker2017}
\bibinfo{author}{\bibfnamefont{S.}~\bibnamefont{L\"uker}},
  \bibinfo{author}{\bibfnamefont{T.}~\bibnamefont{Kuhn}}, \bibnamefont{and}
  \bibinfo{author}{\bibfnamefont{D.~E.} \bibnamefont{Reiter}},
  \bibinfo{journal}{Phys. Rev. B} \textbf{\bibinfo{volume}{96}},
  \bibinfo{pages}{245306} (\bibinfo{year}{2017}),
  \urlprefix\url{https://link.aps.org/doi/10.1103/PhysRevB.96.245306}.

\bibitem[{\citenamefont{Ding et~al.}(2010)\citenamefont{Ding, Singh, Plumhof,
  Zander, Křápek, Chen, Benyoucef, Zwiller, Dörr, Bester et~al.}}]{Ding2010}
\bibinfo{author}{\bibfnamefont{F.}~\bibnamefont{Ding}},
  \bibinfo{author}{\bibfnamefont{R.}~\bibnamefont{Singh}},
  \bibinfo{author}{\bibfnamefont{J.~D.} \bibnamefont{Plumhof}},
  \bibinfo{author}{\bibfnamefont{T.}~\bibnamefont{Zander}},
  \bibinfo{author}{\bibfnamefont{V.}~\bibnamefont{Křápek}},
  \bibinfo{author}{\bibfnamefont{Y.~H.} \bibnamefont{Chen}},
  \bibinfo{author}{\bibfnamefont{M.}~\bibnamefont{Benyoucef}},
  \bibinfo{author}{\bibfnamefont{V.}~\bibnamefont{Zwiller}},
  \bibinfo{author}{\bibfnamefont{K.}~\bibnamefont{Dörr}},
  \bibinfo{author}{\bibfnamefont{G.}~\bibnamefont{Bester}},
  \bibnamefont{et~al.}, \bibinfo{journal}{Phys. Rev. Lett.}
  \textbf{\bibinfo{volume}{104}}, \bibinfo{pages}{067405}
  (\bibinfo{year}{2010}), ISSN \bibinfo{issn}{0031-9007, 1079-7114},
  \urlprefix\url{https://link.aps.org/doi/10.1103/PhysRevLett.104.067405}.


\bibitem[{\citenamefont{Barth et~al.}(2016)\citenamefont{Barth, Vagov, and
  Axt}}]{Barth2016}
\bibinfo{author}{\bibfnamefont{A.~M.} \bibnamefont{Barth}},
  \bibinfo{author}{\bibfnamefont{A.}~\bibnamefont{Vagov}}, \bibnamefont{and}
  \bibinfo{author}{\bibfnamefont{V.~M.} \bibnamefont{Axt}},
  \bibinfo{journal}{Phys. Rev. B} \textbf{\bibinfo{volume}{94}},
  \bibinfo{pages}{125439} (\bibinfo{year}{2016}),
  \urlprefix\url{https://link.aps.org/doi/10.1103/PhysRevB.94.125439}.


\bibitem[{\citenamefont{Lo and Preskill}(2005)}]{LP:calt05}
\bibinfo{author}{\bibfnamefont{H.-K.} \bibnamefont{Lo}} \bibnamefont{and}
  \bibinfo{author}{\bibfnamefont{J.}~\bibnamefont{Preskill}}
  (\bibinfo{year}{2005}), \eprint{quant-ph/0504209}.

\bibitem[{\citenamefont{Miller}(1882)}]{M:CM82}
\bibinfo{author}{\bibfnamefont{F.}~\bibnamefont{Miller}},
  \emph{\bibinfo{title}{Telegraphic code to ensure privacy and secrecy in the
  transmission of telegrams}} (\bibinfo{publisher}{C.M. Cornwell},
  \bibinfo{year}{1882}).

\bibitem[{\citenamefont{Deutsch et~al.}(1996)\citenamefont{Deutsch, Ekert,
  Jozsa, Macchiavello, Popescu, and Sanpera}}]{DEJ:PRL96}
\bibinfo{author}{\bibfnamefont{D.}~\bibnamefont{Deutsch}},
  \bibinfo{author}{\bibfnamefont{A.}~\bibnamefont{Ekert}},
  \bibinfo{author}{\bibfnamefont{R.}~\bibnamefont{Jozsa}},
  \bibinfo{author}{\bibfnamefont{C.}~\bibnamefont{Macchiavello}},
  \bibinfo{author}{\bibfnamefont{S.}~\bibnamefont{Popescu}}, \bibnamefont{and}
  \bibinfo{author}{\bibfnamefont{A.}~\bibnamefont{Sanpera}},
  \bibinfo{journal}{Phys. Rev. Lett.} \textbf{\bibinfo{volume}{77}},
  \bibinfo{pages}{2818} (\bibinfo{year}{1996}),
  \urlprefix\url{https://link.aps.org/doi/10.1103/PhysRevLett.77.2818}.

\bibitem[{\citenamefont{Kitaev}(2003)}]{K:PC03}
\bibinfo{author}{\bibfnamefont{A.}~\bibnamefont{Kitaev}}, \bibinfo{journal}{6th
  Workshop on Quantum Information Processing}  (\bibinfo{year}{2003}).

\bibitem[{\citenamefont{Chailloux and Kerenidis}(2009)}]{IEEE:CK09}
\bibinfo{author}{\bibfnamefont{A.}~\bibnamefont{Chailloux}} \bibnamefont{and}
  \bibinfo{author}{\bibfnamefont{I.}~\bibnamefont{Kerenidis}},
  \bibinfo{journal}{50th Annual IEEE Symposium on Foundations of Computer
  Science} pp. \bibinfo{pages}{527--533} (\bibinfo{year}{2009}).

\bibitem[{\citenamefont{Berlin et~al.}(2009)\citenamefont{Berlin, Brassard,
  Bussieres, and Godbout}}]{BBB:PRA09}
\bibinfo{author}{\bibfnamefont{G.}~\bibnamefont{Berlin}},
  \bibinfo{author}{\bibfnamefont{G.}~\bibnamefont{Brassard}},
  \bibinfo{author}{\bibfnamefont{F.}~\bibnamefont{Bussieres}},
  \bibnamefont{and} \bibinfo{author}{\bibfnamefont{N.}~\bibnamefont{Godbout}},
  \bibinfo{journal}{Phys. Rev. A} \textbf{\bibinfo{volume}{80}},
  \bibinfo{pages}{062321} (\bibinfo{year}{2009}).
  \urlprefix\url{https://doi.org/10.1103/PhysRevA.80.062321}.
  

\bibitem[{\citenamefont{Hänggi and Wullschleger}(2011)}]{HW:TCC11}
\bibinfo{author}{\bibfnamefont{E.}~\bibnamefont{Hänggi}} \bibnamefont{and}
  \bibinfo{author}{\bibfnamefont{J.}~\bibnamefont{Wullschleger}},
  \bibinfo{journal}{Proceedings of TCC} pp. \bibinfo{pages}{468--485}
  (\bibinfo{year}{2011}).
  \urlprefix\url{https://iacr.org/archive/tcc2011/65970464/65970464.pdf}.
  

\bibitem[{\citenamefont{Molina-Terriza
  et~al.}(2005)\citenamefont{Molina-Terriza, Vaziri, Ursin, and
  Zeilinger}}]{MTV:PRL05}
\bibinfo{author}{\bibfnamefont{G.}~\bibnamefont{Molina-Terriza}},
  \bibinfo{author}{\bibfnamefont{A.}~\bibnamefont{Vaziri}},
  \bibinfo{author}{\bibfnamefont{R.}~\bibnamefont{Ursin}}, \bibnamefont{and}
  \bibinfo{author}{\bibfnamefont{A.}~\bibnamefont{Zeilinger}},
  \bibinfo{journal}{Phys. Rev. Lett.} \textbf{\bibinfo{volume}{94}},
  \bibinfo{pages}{040501} (\bibinfo{year}{2005}).
  \urlprefix\url{https://doi.org/10.1103/PhysRevLett.94.040501}.
  
\bibitem[{\citenamefont{Mayers}(1997)}]{M:PRL97}
\bibinfo{author}{\bibfnamefont{D.}~\bibnamefont{Mayers}},
  \bibinfo{journal}{Phys. Rev. Lett.} \textbf{\bibinfo{volume}{78}},
  \bibinfo{pages}{3414} (\bibinfo{year}{1997}),
  \urlprefix\url{https://link.aps.org/doi/10.1103/PhysRevLett.78.3414}.

\bibitem[{\citenamefont{Lo and Chau}(1997)}]{LC:PRL97}
\bibinfo{author}{\bibfnamefont{H.-K.} \bibnamefont{Lo}} \bibnamefont{and}
  \bibinfo{author}{\bibfnamefont{H.~F.} \bibnamefont{Chau}},
  \bibinfo{journal}{Phys. Rev. Lett.} \textbf{\bibinfo{volume}{78}},
  \bibinfo{pages}{3410} (\bibinfo{year}{1997}),
  \urlprefix\url{https://link.aps.org/doi/10.1103/PhysRevLett.78.3410}.


\bibitem[{\citenamefont{Molina et~al.}(2013)\citenamefont{Molina, Vidick, and
  Watrous}}]{MVW:tqc12}
\bibinfo{author}{\bibfnamefont{A.}~\bibnamefont{Molina}},
  \bibinfo{author}{\bibfnamefont{T.}~\bibnamefont{Vidick}}, \bibnamefont{and}
  \bibinfo{author}{\bibfnamefont{J.}~\bibnamefont{Watrous}}, in
  \emph{\bibinfo{booktitle}{{TQC} 2012: Theory of Quantum Computation,
  Communication, and Cryptography}}, edited by
  \bibinfo{editor}{\bibfnamefont{K.}~\bibnamefont{Iwama}},
  \bibinfo{editor}{\bibfnamefont{Y.}~\bibnamefont{Kawano}}, \bibnamefont{and}
  \bibinfo{editor}{\bibfnamefont{M.}~\bibnamefont{Murao}}
  (\bibinfo{publisher}{Springer}, \bibinfo{year}{2013}), vol.
  \bibinfo{volume}{7582} of \emph{\bibinfo{series}{Lecture Notes in Computer
  Science}}, \eprint{1202.4010}.

\bibitem[{\citenamefont{Watrous}(2011)}]{W:LN11}
\bibinfo{author}{\bibfnamefont{J.}~\bibnamefont{Watrous}},
  \emph{\bibinfo{title}{Semidefinite Programming}}
  (\bibinfo{publisher}{University of Waterloo}, \bibinfo{year}{2011}),
  chap.~\bibinfo{chapter}{7},
  \urlprefix\url{https://cs.uwaterloo.ca/~watrous/LectureNotes.html}.

\bibitem[{\citenamefont{Vandenberghe and Boyd}(1996)}]{VB:SIAM96}
\bibinfo{author}{\bibfnamefont{L.}~\bibnamefont{Vandenberghe}}
  \bibnamefont{and} \bibinfo{author}{\bibfnamefont{S.}~\bibnamefont{Boyd}},
  \bibinfo{journal}{SIAM Review} \textbf{\bibinfo{volume}{38}},
  \bibinfo{pages}{49} (\bibinfo{year}{1996}).


\bibitem[{\citenamefont{Ma}(2008)}]{Ma:Thesis08}
\bibinfo{author}{\bibfnamefont{X.}~\bibnamefont{Ma}},
  \emph{\bibinfo{title}{Quantum cryptography: theory and practice}}
  (\bibinfo{year}{2008}), \eprint{0808.1385}.
  \urlprefix\url{https://doi.org/10.48550/arXiv.0808.1385}.

\bibitem[{\citenamefont{Cui et~al.}(2019)\citenamefont{Cui, Yin, Wang, Chen,
  Wang, Guo, and Han}}]{Cui_2019}
\bibinfo{author}{\bibfnamefont{C.}~\bibnamefont{Cui}},
  \bibinfo{author}{\bibfnamefont{Z.-Q.} \bibnamefont{Yin}},
  \bibinfo{author}{\bibfnamefont{R.}~\bibnamefont{Wang}},
  \bibinfo{author}{\bibfnamefont{W.}~\bibnamefont{Chen}},
  \bibinfo{author}{\bibfnamefont{S.}~\bibnamefont{Wang}},
  \bibinfo{author}{\bibfnamefont{G.-C.} \bibnamefont{Guo}}, \bibnamefont{and}
  \bibinfo{author}{\bibfnamefont{Z.-F.} \bibnamefont{Han}},
  \bibinfo{journal}{Physical Review Applied} \textbf{\bibinfo{volume}{11}}
  (\bibinfo{year}{2019}), ISSN \bibinfo{issn}{2331-7019},
  \urlprefix\url{http://dx.doi.org/10.1103/PhysRevApplied.11.034053}.

\bibitem[{\citenamefont{Curty et~al.}(2019)\citenamefont{Curty, Azuma, and
  Lo}}]{CAL:npjQI19}
\bibinfo{author}{\bibfnamefont{M.}~\bibnamefont{Curty}},
  \bibinfo{author}{\bibfnamefont{K.}~\bibnamefont{Azuma}}, \bibnamefont{and}
  \bibinfo{author}{\bibfnamefont{H.-K.} \bibnamefont{Lo}},
  \bibinfo{journal}{npj Quantum Information} \textbf{\bibinfo{volume}{5}},
  \bibinfo{pages}{64} (\bibinfo{year}{2019}).

\bibitem[{\citenamefont{Beaudry et~al.}(2008)\citenamefont{Beaudry, Moroder,
  and L\"utkenhaus}}]{BML:prl08}
\bibinfo{author}{\bibfnamefont{N.~J.} \bibnamefont{Beaudry}},
  \bibinfo{author}{\bibfnamefont{T.}~\bibnamefont{Moroder}}, \bibnamefont{and}
  \bibinfo{author}{\bibfnamefont{N.}~\bibnamefont{L\"utkenhaus}},
  \bibinfo{journal}{Phys. Rev. Lett.} \textbf{\bibinfo{volume}{101}},
  \bibinfo{pages}{093601} (\bibinfo{year}{2008}), \eprint{0804.3082}.


\bibitem[{\citenamefont{Rudolph et~al.}(2003)\citenamefont{Rudolph, Spekkens,
  and Turner}}]{RST:PRA03}
\bibinfo{author}{\bibfnamefont{T.}~\bibnamefont{Rudolph}},
  \bibinfo{author}{\bibfnamefont{R.~W.} \bibnamefont{Spekkens}},
  \bibnamefont{and} \bibinfo{author}{\bibfnamefont{P.~S.}
  \bibnamefont{Turner}}, \bibinfo{journal}{Phys. Rev. A}
  \textbf{\bibinfo{volume}{68}}, \bibinfo{pages}{010301}
  (\bibinfo{year}{2003}),
  \urlprefix\url{https://link.aps.org/doi/10.1103/PhysRevA.68.010301}.

\bibitem[{\citenamefont{L\"u}(2021)}]{Lu:PRA21}
\bibinfo{author}{\bibfnamefont{X.}~\bibnamefont{L\"u}}, \bibinfo{journal}{Phys.
  Rev. A} \textbf{\bibinfo{volume}{103}}, \bibinfo{pages}{022216}
  (\bibinfo{year}{2021}),
  \urlprefix\url{https://link.aps.org/doi/10.1103/PhysRevA.103.022216}.

\bibitem[{\citenamefont{Barnett and Croke}(2009)}]{BK:AOP09}
\bibinfo{author}{\bibfnamefont{S.~M.} \bibnamefont{Barnett}} \bibnamefont{and}
  \bibinfo{author}{\bibfnamefont{S.}~\bibnamefont{Croke}},
  \bibinfo{journal}{Adv. Opt. Photon.} \textbf{\bibinfo{volume}{1}},
  \bibinfo{pages}{238} (\bibinfo{year}{2009}),
  \urlprefix\url{http://www.osapublishing.org/aop/abstract.cfm?URI=aop-1-2-238}.

\bibitem[{\citenamefont{Eldar}(2003)}]{E:IEEE03}
\bibinfo{author}{\bibfnamefont{Y.}~\bibnamefont{Eldar}}, \bibinfo{journal}{IEEE
  Transactions on Information Theory} \textbf{\bibinfo{volume}{49}},
  \bibinfo{pages}{446} (\bibinfo{year}{2003}).

\bibitem[{\citenamefont{Eldar et~al.}(2003)\citenamefont{Eldar, Megretski, and
  Verghese}}]{EMV:IEEE03}
\bibinfo{author}{\bibfnamefont{Y.}~\bibnamefont{Eldar}},
  \bibinfo{author}{\bibfnamefont{A.}~\bibnamefont{Megretski}},
  \bibnamefont{and} \bibinfo{author}{\bibfnamefont{G.}~\bibnamefont{Verghese}},
  \bibinfo{journal}{IEEE Transactions on Information Theory}
  \textbf{\bibinfo{volume}{49}}, \bibinfo{pages}{1007} (\bibinfo{year}{2003}).
  \urlprefix\url{https://ieeexplore.ieee.org/document/1193807}.

\bibitem[{\citenamefont{Konig et~al.}(2012)\citenamefont{Konig, Wehner, and
  Wullschleger}}]{KWW:IEEE12}
\bibinfo{author}{\bibfnamefont{R.}~\bibnamefont{Konig}},
  \bibinfo{author}{\bibfnamefont{S.}~\bibnamefont{Wehner}}, \bibnamefont{and}
  \bibinfo{author}{\bibfnamefont{J.}~\bibnamefont{Wullschleger}},
  \bibinfo{journal}{IEEE Transactions on Information Theory}
  \textbf{\bibinfo{volume}{58}}, \bibinfo{pages}{1962} (\bibinfo{year}{2012}).
   \urlprefix\url{https://ieeexplore.ieee.org/document/6157089}.
  

\end{thebibliography}
\end{document}